\documentclass[11pt]{article}
\usepackage{amsthm}
\usepackage{amsmath}   
\usepackage{amsfonts}    
\usepackage{mathrsfs} 
\usepackage{xfrac} 
\usepackage{tabu}
\usepackage{color}
\usepackage{tikz}
\usetikzlibrary{trees,snakes}
\usepackage{verbatim}
\usepackage{amssymb}
\usepackage{graphicx}
\usepackage{tikz}
\usetikzlibrary{calc}
\usepackage{pgfplots}
\usepackage{cleveref}
\usepackage{graphicx}
\usepackage{subcaption}
\usepackage{bbm}

\definecolor{myred}{RGB}{255, 0, 0}
\definecolor{myblue}{RGB}{0, 0, 255}
\newtheorem{theorem}{Theorem}
\newtheorem{lemma}{Lemma}
\newtheorem{proposition}{Proposition}

\newtheorem{remark}{Remark}

\newcommand{\tr}{\mathsf{T}}
\newcommand {\nn} {\nonumber}
\newcommand {\E} {\mathbb{E}}
\newcommand {\pr} {\mathbb{P}}
\newcommand {\prob} {\mathbb{P}}
\newcommand{\IND}{\mathbbm{1}}

\newcommand{\dfn}{\stackrel{\triangle}{=}}
\newcommand{\dint}{\mathrm{d}}

\newcommand{\txz}{\mbox{\small xz}}

\newcommand {\reals} {{\rm I\!R}}
\newcommand {\naturals} {\mathbb{N}}

\newcommand {\bi} {\mbox{\boldmath $i$}}
\newcommand {\bj} {\mbox{\boldmath $j$}}
\newcommand {\bk} {\mbox{\boldmath $k$}}

\newcommand {\bx} {\boldsymbol{x}}
\newcommand {\by} {\boldsymbol{y}}

\newcommand {\bI} {\mbox{\boldmath $I$}}

\newcommand {\bX}{\boldsymbol{X}}

\newcommand {\bY}{\boldsymbol{Y}}
\newcommand {\bZ} {\boldsymbol{Z}}
\newcommand{\calA}{{\cal A}}
\newcommand{\calB}{{\cal B}}

\newcommand{\calE}{{\cal E}}
\newcommand{\calF}{{\cal F}}

\newcommand{\calI}{{\cal I}}

\newcommand{\calK}{{\cal K}}

\newcommand{\calN}{{\cal N}}
\newcommand{\calO}{{\cal O}}
\newcommand{\calP}{{\cal P}}

\newcommand{\calS}{{\cal S}}

\newcommand{\calV}{{\cal V}}

\newcommand{\calX}{{\cal X}}
\newcommand{\calY}{{\cal Y}}

\newcommand{\sL}{\mathsf{L}}

\newcommand{\sR}{\mathsf{R}}

\newcommand{\HypA}{\mathsf{H}_{0}}
\newcommand{\HypB}{\mathsf{H}_{1}}

\begin{document}
\thispagestyle{empty}
\title{On Correlation Detection and Alignment Recovery of Gaussian Databases\footnote{
		This research has been funded in part by ETH Foundations of Data Science (ETH-FDS).}\\}
\author{\\Ran Tamir\\}
\maketitle
\begin{center}
Signal and Information Processing Laboratory \\
ETH Zurich, 8092 Zurich, Switzerland \\ 
Email: tamir@isi.ee.ethz.ch
\end{center}

\vspace{1.5\baselineskip}
\setlength{\baselineskip}{1.5\baselineskip}

\begin{abstract}
In this work, we propose an efficient two-stage algorithm solving a joint problem of correlation detection and partial alignment recovery between two Gaussian databases. Correlation detection is a hypothesis testing problem; under the null hypothesis, the databases are independent, and under the alternate hypothesis, they are correlated, under an unknown row permutation. We develop bounds on the type-I and type-II error probabilities, and show that the analyzed detector performs better than a recently proposed detector, at least for some specific parameter choices. 
Since the proposed detector relies on a statistic, which is a sum of dependent indicator random variables, then in order to bound the type-I probability of error, we develop a novel graph-theoretic technique for bounding the $k$-th order moments of such statistics.
When the databases are accepted as correlated, the algorithm also recovers some partial alignment between the given databases.
We also propose two more algorithms: (i) One more algorithm for partial alignment recovery, whose reliability and computational complexity are both higher than those of the first proposed algorithm. (ii) An algorithm for full alignment recovery, which has a reduced amount of calculations and a not much lower error probability, when compared to the optimal recovery procedure.         
\\

\end{abstract}

\clearpage
\section{Introduction}
Two fundamental problems in the statistical analysis of Gaussian databases are correlation detection (or testing) and alignment recovery. Correlation detection of two databases is basically a hypothesis testing problem; under the null hypothesis, the databases are independent, and under the alternate hypothesis, there exists a permutation, for which the databases are correlated. In this task, the main objective is to attain the best trade-off between the type-I and type-II error probabilities. In the problem of database alignment recovery, we make an assumption that the two databases are correlated, and want to estimate the underlying permutation. The objective is to minimize the probability of alignment error. 

While alignment recovery of databases with $n$ sequences, each containing $d$ Gaussian entries has been recently studied in \cite{DCK2019} and correlation detection of such Gaussian databases has been lately explored in \cite{KN2022}, it seems very natural to tackle the two individual problems together as a joint problem of correlation detection and alignment recovery. In addition, we also refer to the problem of {\it partial alignment recovery}, in which one would like to estimate only part of the underlying alignment between the databases. The main reasons for preferring partial alignment recovery instead of full alignment recovery are as follows.
\begin{enumerate}
	\item In some practical scenarios, the requirement on high reliability may be more crucial than the need for having a full alignment recovery. In such cases, one may prefer to use algorithms that provide a partial alignment recovery with lower guaranteed error probabilities, compared to the error probability of the optimal full alignment recovery.  
	\item Empirical studies show that if the correlation coefficient between two correlated databases is lower than some critical value, then the probability of error regarding full alignment recovery equals one. Hence, when analyzing correlated databases that are governed by relatively low correlation coefficients, only partial alignment recovery may still be feasible.
	\item Practically, in many real-world scenarios, two databases may not contain data on exactly the same subjects, i.e., only some fraction of the rows will contain data on a common set of subjects. In such cases, it obviously makes no sense to execute a full alignment recovery procedure, but only an algorithm that aims to recover a partial alignment.          
\end{enumerate}   

In this manuscript, we propose an efficient two-stage algorithm that solves jointly the two problems of correlation detection and partial alignment recovery of correlated Gaussian databases. In its first stage, the algorithm first calculates a statistic which is based on local decisions for all $n^{2}$ pairs of sequences, and then takes a final decision in favor of one of the hypotheses based on this statistic. 
Since the proposed statistical test is based on a sum of indicator random variables, which are weakly dependent, the analysis of the two error probabilities, especially the type-I probability of error, turns out to be highly non-trivial, and this is one of the main contribution of this work.
If the algorithm accepts the alternate hypothesis (i.e., that the databases are probably correlated), then in his second stage, it outputs an estimation of a partial alignment. The proposed algorithm can be tuned to trade-off between the size of the estimated partial alignment and the probability of error (in terms that at least one of the estimated pairs of feature vectors is wrong).

The main problem in \cite{KN2022} is identifying the lowest possible correlation coefficient between each consecutive entries in the databases, as a function of $n$ and $d$, such that the sum of the type-I and type-II error probabilities can be driven to zero as $n$ and $d$ grow to infinity. While the detector that was proposed in \cite{KN2022} can be used also for non-asymptotic correlation values, we show in Section \ref{Sec7} that, at least for some parameter choices, its performance is inferior to the performance of the new proposed detector, which relies on many local decisions and a final global decision. Moreover, these local decisions that are made in the detection stage of the proposed algorithm, are in fact the basis for estimating the underlying permutation (in case the alternate hypothesis is accepted). As opposed to the proposed correlation detector, the detector in \cite{KN2022}, which is merely based on summing up all $n^{2}$ inner products between the pairs of sequences, obviously cannot be used as a basis for permutation recovery.               

More contributions of this work are as follows.
\begin{enumerate}
	\item While the computational complexity of the first partial alignment recovery algorithm is relatively low (it grows only quadratically in $n$), we propose in Section \ref{Sec5} a second recovery algorithm, whose computational complexity grows cubically in $n$, but its probability of error turns out to be much smaller than the error probability of the first algorithm.
	\item Regarding full alignment recovery, we propose in Section \ref{Sec6} an algorithm for this task, which has a reduced computational complexity, relative to the computational complexity of the optimal maximum-likelihood estimator. We show by computer simulations, that in some cases, the probability of error of the proposed algorithm is not much smaller than the error probability of the optimal estimator.       
\end{enumerate}

\subsection{Related Work}
The database alignment problem was originally introduced by Cullina et al.\ \cite{Cullina2018}. The discrete case was studied in \cite{Cullina2018}, which derived achievability and converse bounds in terms of mutual information. Exact recovery of the underlying permutation for correlated Gaussian databases was studied in \cite{DCK2019}, and follow-up work extended the results to partial recovery \cite{Cullina2020}. 
A typicality-based framework for permutation estimation was investigated in \cite{Erkip2019}. Random feature deletions and repetitions were researched respectively in \cite{Erkip2021} and \cite{Erkip2022}.
The Gaussian database alignment recovery problem is equivalent to a certain idealized tracking problem studied in \cite{Chertkov2010, Kunisky2022}. The recovery problem between two correlated random graphs has also been investigated in the past few years. A starting point on this problem proposed the correlated Erd\H{o}s--R\'enyi graph model with dependent Bernoulli edge pairs \cite{Pedarsani2011}. A subsequent work studied the recovery problem under the Gaussian setting \cite{Ding2021}. More recent papers have investigated the corresponding detection problem for correlation between graphs \cite{Yu2020, Yu2021}.   
It has been lately proved \cite{Sophie2022, Ganassali2022} that detecting whether Gaussian graphs are correlated is as difficult as recovering the node labeling. 

\subsection{Outline}
The remaining part of the paper is organized as follows. 
In Section \ref{Sec2}, we establish notation conventions. 
In Section \ref{Sec3}, we provide some motivation, formalize the model, formulate the problem, and state some results from previous work. 
In Section \ref{Sec4} we state the new proposed algorithm, and provide and discuss its theoretical guarantees.
In Sections \ref{Sec5} and \ref{Sec6}, we propose algorithms regarding partial and full alignment recovery, respectively. In Section \ref{Sec7} we compare our results to a previous work. 
All the results in this work are proved in the Appendixes.

\section{Notation Conventions} \label{Sec2}

Throughout the paper, random variables will be denoted by capital letters and specific values they may take will be denoted by the corresponding lower case letters. Random vectors and their realizations will be denoted, respectively, by capital letters and the corresponding lower case letters, both in the bold face font. For example, the random vector $\bX = (X_1,X_2,\ldots,X_d)$, ($d$ -- positive integer) may take a specific vector value $\bx = (x_1,x_2,\ldots,x_d)$ in $\reals^{d}$. When used in the linear-algebraic context, these vectors should be thought of as column vectors, and so, when they appear
with superscript $\mathsf{T}$, they will be transformed into row vectors by transposition. Thus, $\bx^{\mathsf{T}}\by$ is
understood as the inner product of $\bx$ and $\by$. The notation $\|\bx\|$ will stand for the Euclidean norm of vector $\bx$.   
As customary in probability theory, we write $\bX=(X_{1}, \ldots, X_{d}) \sim \calN(\boldsymbol{0}_{d},\bI_{d})$ ($\boldsymbol{0}_{d}$ being a vector of $d$ zeros and $\bI_{d}$ being the $d \times d$ identity matrix) to denote that the probability density function of $\bX$ is 
\begin{align}
P_{\bX}(\bx) = (2\pi)^{-d/2} \cdot \exp\left\{-\frac{1}{2} \|\bx\|^{2} \right\}.
\end{align} 
For $\bX=(X_{1}, \ldots, X_{d})$ and $\bY=(Y_{1}, \ldots, Y_{d})$, we write
\begin{align}
(\bX,\bY) \sim \calN^{\otimes d}\left(\left[\begin{matrix} 0 \\ 0
\end{matrix}\right], \left[\begin{matrix} 1 & \rho \\ \rho & 1 \end{matrix}\right] \right)
\end{align}
to denote the fact that $\bX$ and $\bY$ are jointly Gaussian with IID pairs, where each pair $(X_i,Y_i)$, $i \in \{1,\ldots,d\}$, is a Gaussian vector with a zero mean and the specified covariance matrix.

Logarithms are taken to the natural base.
The probability of an event $\calE$ will be denoted by 
$\pr\{\calE\}$ and the indicator function by $\IND\{\calE\}$. 
The expectation operator will be denoted by $\mathbb{E}[\cdot]$.

\section{Motivations, Settings, Problem Formulation, and Prior Work}
\label{Sec3}
\subsection{Motivations}

To get started, assume that one has access to some set of databases. Each database is given by a matrix; each row contains information on a different subject, while each column contains information for all subjects regarding a specific feature (e.g., height, weight, etc.). We assume that different databases contain information concerning similar features, but also regarding different features for the same subjects. 
Hence, it follows that by merging together two (or more) databases, one can gather more data on each given subject.  
One of the main difficulties in database analysis is that a database must not contain the specific identities of its subjects; the identities may have been erased due to data-privacy issues \cite{Pedarsani2011}. It follows that gathering data from two databases is nontrivial; one has to recover the alignment between the databases.
The main goal is to detect groups of databases that contain data regarding similar subjects, and then to gather the maximum amount of data for each subject.

\begin{figure}[h!]
	\centering
	\begin{tikzpicture}[scale=0.8,cross/.style={path picture={ 
			\draw[red]
			(path picture bounding box.south east) -- (path picture bounding box.north west) (path picture bounding box.south west) -- (path picture bounding box.north east);}}]
	\filldraw[draw=black,fill=white,rounded corners=3, ultra thick] (-0.5,6.5) rectangle (8.5,11.6);
	\filldraw[draw=black,fill=gray!10, ultra thick] (0,7) rectangle (8,7.8);
	\filldraw[draw=black,fill=pink!20] (1.1,7.1) rectangle (4.35,7.7);
	\filldraw[draw=black,fill=yellow!40] (4.55,7.1) rectangle (7.8,7.7);
	\node at (0.5,7.4) {\includegraphics[width=5mm]{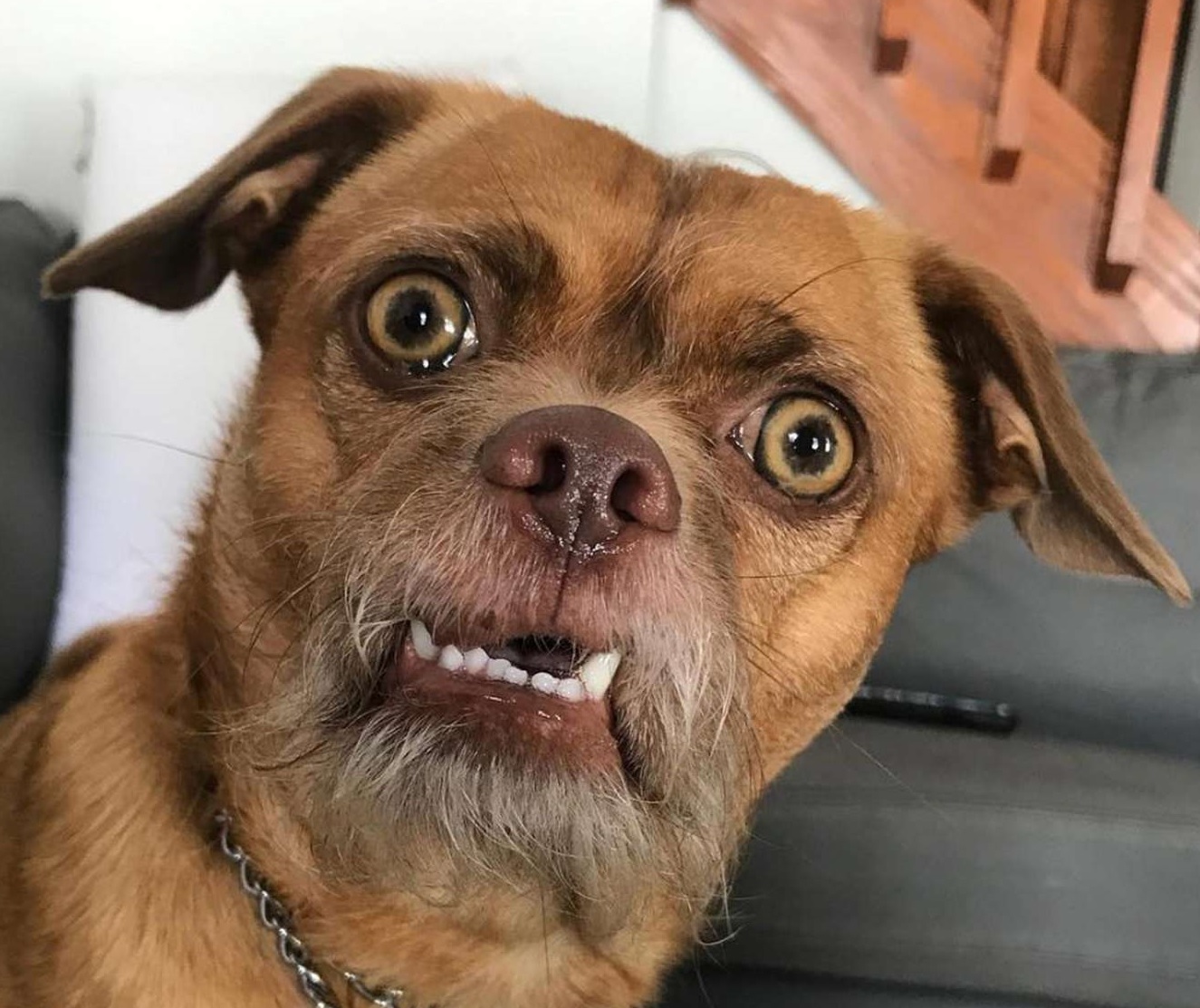}};
	\filldraw[draw=black,fill=gray!10, ultra thick] (0,8.1) rectangle (8,8.9);
	\filldraw[draw=black,fill=pink!20] (1.1,8.2) rectangle (4.35,8.8);
	\filldraw[draw=black,fill=red!40] (4.55,8.2) rectangle (7.8,8.8);
	\node at (0.5,8.5) {\includegraphics[width=5mm]{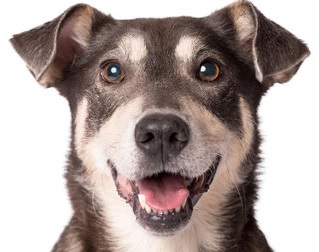}};
	\filldraw[draw=black,fill=gray!10, ultra thick] (0,9.2) rectangle (8,10.0);
	\filldraw[draw=black,fill=pink!20] (1.1,9.3) rectangle (4.35,9.9);
	\filldraw[draw=black,fill=green!40] (4.55,9.3) rectangle (7.8,9.9);
	\node at (0.5,9.6) {\includegraphics[width=5mm]{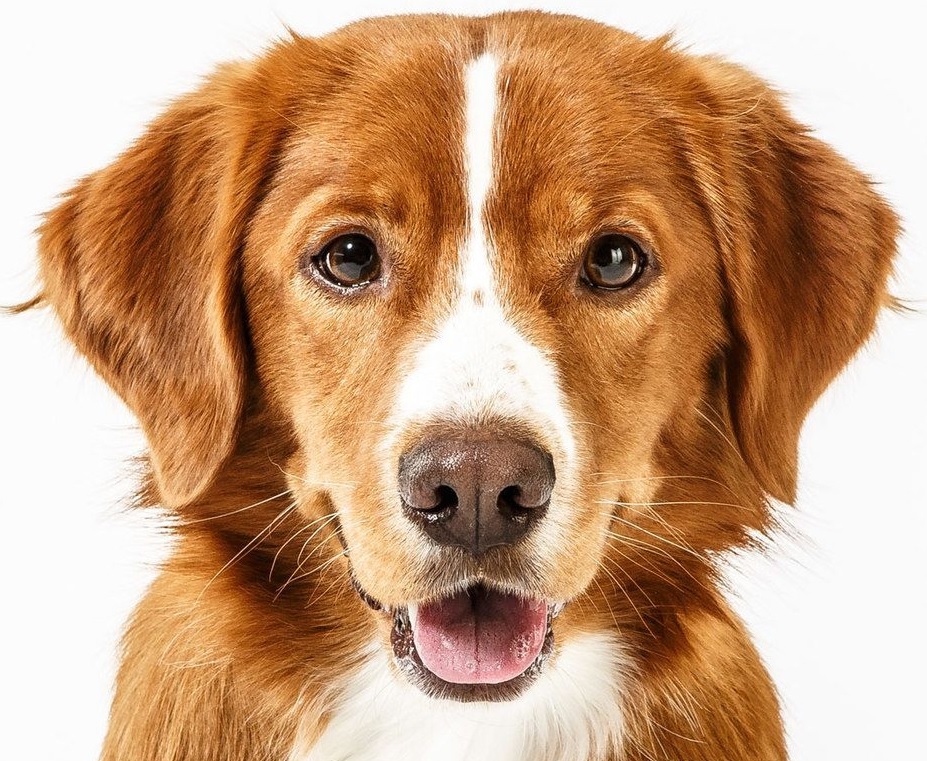}};
	\filldraw[draw=black,fill=gray!10, ultra thick] (0,10.3) rectangle (8,11.1);
	\filldraw[draw=black,fill=pink!20] (1.1,10.4) rectangle (4.35,11);
	\filldraw[draw=black,fill=blue!40] (4.55,10.4) rectangle (7.8,11);
	\node at (0.5,10.7) {\includegraphics[width=5mm]{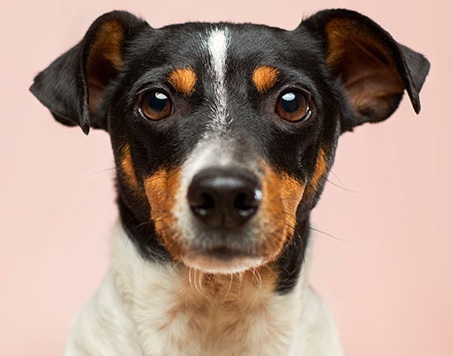}};
	
	\draw[line width = 0.4mm] (2.725,7.4) circle [radius=0.25];
	\draw[line width = 0.4mm] (2.725,8.5) circle [radius=0.25];
	\draw[line width = 0.4mm] (2.725,9.6) circle [radius=0.25];
	\draw[line width = 0.4mm] (2.725,10.7) circle [radius=0.25];

	\filldraw[draw=black,fill=white,rounded corners=3, ultra thick] (-0.5,-0.5) rectangle (8.5,4.6);
	\filldraw[draw=black,fill=gray!10, ultra thick] (0,0) rectangle (8,0.8);
	\filldraw[draw=black,fill=pink!20] (1.1,0.1) rectangle (4.35,0.7);
	\filldraw[draw=black,fill=purple!40] (4.55,0.1) rectangle (7.8,0.7);
	\node at (0.5,0.4) {\includegraphics[width=5mm]{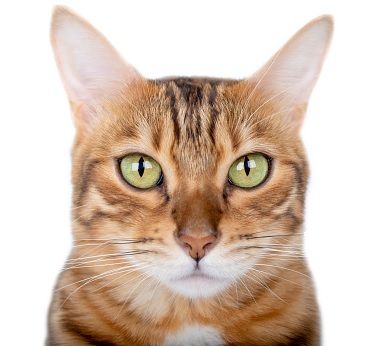}};
	\filldraw[draw=black,fill=gray!10, ultra thick] (0,1.1) rectangle (8,1.9);
	\filldraw[draw=black,fill=pink!20] (1.1,1.2) rectangle (4.35,1.8);
	\filldraw[draw=black,fill=cyan!40] (4.55,1.2) rectangle (7.8,1.8);
	\node at (0.5,1.5) {\includegraphics[width=5mm]{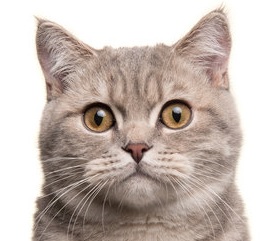}};
	\filldraw[draw=black,fill=gray!10, ultra thick] (0,2.2) rectangle (8,3.0);
	\filldraw[draw=black,fill=pink!20] (1.1,2.3) rectangle (4.35,2.9);
	\filldraw[draw=black,fill=orange!40] (4.55,2.3) rectangle (7.8,2.9);
	\node at (0.5,2.6) {\includegraphics[width=5mm]{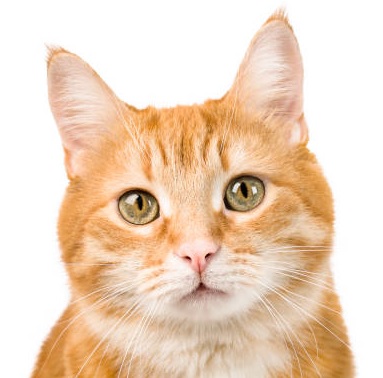}};
	\filldraw[draw=black,fill=gray!10, ultra thick] (0,3.3) rectangle (8,4.1);
	\filldraw[draw=black,fill=pink!20] (1.1,3.4) rectangle (4.35,4.0);
	\filldraw[draw=black,fill=olive!40] (4.55,3.4) rectangle (7.8,4.0);
	\node at (0.5,3.7) {\includegraphics[width=5mm]{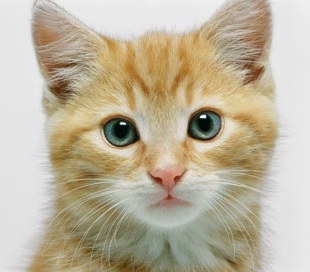}};
	
	\draw[line width = 0.4mm] (2.725,0.4) circle [radius=0.25];
	\draw[line width = 0.4mm] (2.725,1.5) circle [radius=0.25];
	\draw[line width = 0.4mm] (2.725,2.6) circle [radius=0.25];
	\draw[line width = 0.4mm] (2.725,3.7) circle [radius=0.25];

	\filldraw[draw=black,fill=white,rounded corners=3, ultra thick] (9.5,6.5) rectangle (18.5,11.6);
	\filldraw[draw=black,fill=gray!10, ultra thick] (10,7) rectangle (18,7.8);
	\filldraw[draw=black,fill=pink!20] (11.1,7.1) rectangle (14.35,7.7);
	\filldraw[draw=black,fill=olive!70] (14.55,7.1) rectangle (17.8,7.7);
	\node at (10.5,7.4) {\includegraphics[width=5mm]{Cat4.jpg}};
	\node [draw,circle,red,cross,minimum width=0.5 cm] at (10.5,7.4){};
	\filldraw[draw=black,fill=gray!10, ultra thick] (10,8.1) rectangle (18,8.9);
	\filldraw[draw=black,fill=pink!20] (11.1,8.2) rectangle (14.35,8.8);
	\filldraw[draw=black,fill=cyan!70] (14.55,8.2) rectangle (17.8,8.8);
	\node at (10.5,8.5) {\includegraphics[width=5mm]{Cat2.jpg}};
	\node [draw,circle,red,cross,minimum width=0.5 cm] at (10.5,8.5){};
	\filldraw[draw=black,fill=gray!10, ultra thick] (10,9.2) rectangle (18,10.0);
	\filldraw[draw=black,fill=pink!20] (11.1,9.3) rectangle (14.35,9.9);
	\filldraw[draw=black,fill=purple!70] (14.55,9.3) rectangle (17.8,9.9);
	\node at (10.5,9.6) {\includegraphics[width=5mm]{Cat1.jpg}};
	\node [draw,circle,red,cross,minimum width=0.5 cm] at (10.5,9.6){};
	\filldraw[draw=black,fill=gray!10, ultra thick] (10,10.3) rectangle (18,11.1);
	\filldraw[draw=black,fill=pink!20] (11.1,10.4) rectangle (14.35,11);
	\filldraw[draw=black,fill=orange!70] (14.55,10.4) rectangle (17.8,11);
	\node at (10.5,10.7) {\includegraphics[width=5mm]{Cat3.jpg}};
	\node [draw,circle,red,cross,minimum width=0.5 cm] at (10.5,10.7){};
	
	\draw[line width = 0.4mm] (12.725,7.4) circle [radius=0.25];
	\draw[line width = 0.4mm] (12.725,8.5) circle [radius=0.25];
	\draw[line width = 0.4mm] (12.725,9.6) circle [radius=0.25];
	\draw[line width = 0.4mm] (12.725,10.7) circle [radius=0.25];

	\filldraw[draw=black,fill=white,rounded corners=3, ultra thick] (9.5,-0.5) rectangle (18.5,4.6);
	\filldraw[draw=black,fill=gray!10, ultra thick] (10,0) rectangle (18,0.8);
	\filldraw[draw=black,fill=pink!20] (11.1,0.1) rectangle (14.35,0.7);
	\filldraw[draw=black,fill=green!70] (14.55,0.1) rectangle (17.8,0.7);
	\node at (10.5,0.4) {\includegraphics[width=5mm]{Dog3.jpg}};
	\node [draw,circle,red,cross,minimum width=0.5 cm] at (10.5,0.4){};
	\filldraw[draw=black,fill=gray!10, ultra thick] (10,1.1) rectangle (18,1.9);
	\filldraw[draw=black,fill=pink!20] (11.1,1.2) rectangle (14.35,1.8);
	\filldraw[draw=black,fill=yellow!70] (14.55,1.2) rectangle (17.8,1.8);
	\node at (10.5,1.5) {\includegraphics[width=5mm]{Dog1.jpg}};
	\node [draw,circle,red,cross,minimum width=0.5 cm] at (10.5,1.5){};
	\filldraw[draw=black,fill=gray!10, ultra thick] (10,2.2) rectangle (18,3.0);
	\filldraw[draw=black,fill=pink!20] (11.1,2.3) rectangle (14.35,2.9);
	\filldraw[draw=black,fill=blue!70] (14.55,2.3) rectangle (17.8,2.9);
	\node at (10.5,2.6) {\includegraphics[width=5mm]{Dog4.jpg}};
	\node [draw,circle,red,cross,minimum width=0.5 cm] at (10.5,2.6){};
	\filldraw[draw=black,fill=gray!10, ultra thick] (10,3.3) rectangle (18,4.1);
	\filldraw[draw=black,fill=pink!20] (11.1,3.4) rectangle (14.35,4.0);
	\filldraw[draw=black,fill=red!70] (14.55,3.4) rectangle (17.8,4.0);
	\node at (10.5,3.7) {\includegraphics[width=5mm]{Dog2.jpg}};
	\node [draw,circle,red,cross,minimum width=0.5 cm] at (10.5,3.7){};
	
	\draw[line width = 0.4mm] (12.725,0.4) circle [radius=0.25];
	\draw[line width = 0.4mm] (12.725,1.5) circle [radius=0.25];
	\draw[line width = 0.4mm] (12.725,2.6) circle [radius=0.25];
	\draw[line width = 0.4mm] (12.725,3.7) circle [radius=0.25];
	\end{tikzpicture}
	\caption{In the above four databases, the two left-hand contain the specific identities, while the two right-hand have been anonymized.}
	\label{fig:Database_Problem}
\end{figure}

\begin{figure}[h!]
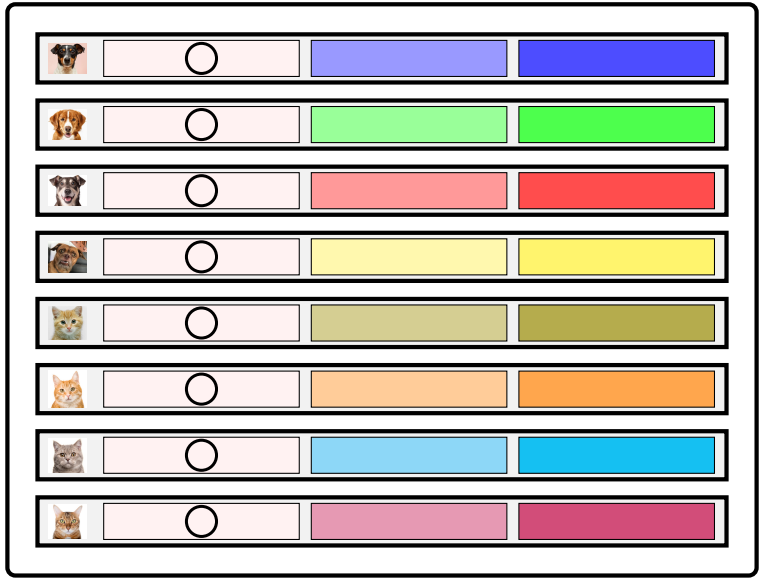

	\centering
	\begin{tikzpicture}[scale=0.8,cross/.style={path picture={ 
			\draw[red]
			(path picture bounding box.south east) -- (path picture bounding box.north west) (path picture bounding box.south west) -- (path picture bounding box.north east);}}]
	
	\filldraw[draw=black,fill=white,rounded corners=3, ultra thick] (-0.5,-0.5) rectangle (11.95,9.0);
	
	\filldraw[draw=black,fill=gray!10, ultra thick] (0,4.4) rectangle (11.45,5.2);
	\filldraw[draw=black,fill=pink!20] (1.1,4.5) rectangle (4.35,5.1);
	\filldraw[draw=black,fill=yellow!40] (4.55,4.5) rectangle (7.8,5.1);
	\filldraw[draw=black,fill=yellow!70] (8.0,4.5) rectangle (11.25,5.1);
	\node at (0.5,4.8) {\includegraphics[width=5mm]{Dog1.jpg}};
	\filldraw[draw=black,fill=gray!10, ultra thick] (0,5.5) rectangle (11.45,6.3);
	\filldraw[draw=black,fill=pink!20] (1.1,5.6) rectangle (4.35,6.2);
	\filldraw[draw=black,fill=red!40] (4.55,5.6) rectangle (7.8,6.2);
	\filldraw[draw=black,fill=red!70] (8.0,5.6) rectangle (11.25,6.2);
	\node at (0.5,5.9) {\includegraphics[width=5mm]{Dog2.jpg}};
	\filldraw[draw=black,fill=gray!10, ultra thick] (0,6.6) rectangle (11.45,7.4);
	\filldraw[draw=black,fill=pink!20] (1.1,6.7) rectangle (4.35,7.3);
	\filldraw[draw=black,fill=green!40] (4.55,6.7) rectangle (7.8,7.3);
	\filldraw[draw=black,fill=green!70] (8.0,6.7) rectangle (11.25,7.3);
	\node at (0.5,7.0) {\includegraphics[width=5mm]{Dog3.jpg}};
	\filldraw[draw=black,fill=gray!10, ultra thick] (0,7.7) rectangle (11.45,8.5);
	\filldraw[draw=black,fill=pink!20] (1.1,7.8) rectangle (4.35,8.4);
	\filldraw[draw=black,fill=blue!40] (4.55,7.8) rectangle (7.8,8.4);
	\filldraw[draw=black,fill=blue!70] (8.0,7.8) rectangle (11.25,8.4);
	\node at (0.5,8.1) {\includegraphics[width=5mm]{Dog4.jpg}};
	
	\draw[line width = 0.4mm] (2.725,4.8) circle [radius=0.25];
	\draw[line width = 0.4mm] (2.725,5.9) circle [radius=0.25];
	\draw[line width = 0.4mm] (2.725,7.0) circle [radius=0.25];
	\draw[line width = 0.4mm] (2.725,8.1) circle [radius=0.25];

	\filldraw[draw=black,fill=gray!10, ultra thick] (0,0) rectangle (11.45,0.8);
	\filldraw[draw=black,fill=pink!20] (1.1,0.1) rectangle (4.35,0.7);
	\filldraw[draw=black,fill=purple!40] (4.55,0.1) rectangle (7.8,0.7);
	\filldraw[draw=black,fill=purple!70] (8.0,0.1) rectangle (11.25,0.7);
	\node at (0.5,0.4) {\includegraphics[width=5mm]{Cat1.jpg}};
	\filldraw[draw=black,fill=gray!10, ultra thick] (0,1.1) rectangle (11.45,1.9);
	\filldraw[draw=black,fill=pink!20] (1.1,1.2) rectangle (4.35,1.8);
	\filldraw[draw=black,fill=cyan!40] (4.55,1.2) rectangle (7.8,1.8);
	\filldraw[draw=black,fill=cyan!70] (8.0,1.2) rectangle (11.25,1.8);
	\node at (0.5,1.5) {\includegraphics[width=5mm]{Cat2.jpg}};
	\filldraw[draw=black,fill=gray!10, ultra thick] (0,2.2) rectangle (11.45,3.0);
	\filldraw[draw=black,fill=pink!20] (1.1,2.3) rectangle (4.35,2.9);
	\filldraw[draw=black,fill=orange!40] (4.55,2.3) rectangle (7.8,2.9);
	\filldraw[draw=black,fill=orange!70] (8.0,2.3) rectangle (11.25,2.9);
	\node at (0.5,2.6) {\includegraphics[width=5mm]{Cat3.jpg}};
	\filldraw[draw=black,fill=gray!10, ultra thick] (0,3.3) rectangle (11.45,4.1);
	\filldraw[draw=black,fill=pink!20] (1.1,3.4) rectangle (4.35,4.0);
	\filldraw[draw=black,fill=olive!40] (4.55,3.4) rectangle (7.8,4.0);
	\filldraw[draw=black,fill=olive!70] (8.0,3.4) rectangle (11.25,4.0);
	\node at (0.5,3.7) {\includegraphics[width=5mm]{Cat4.jpg}};
	
	\draw[line width = 0.4mm] (2.725,0.4) circle [radius=0.25];
	\draw[line width = 0.4mm] (2.725,1.5) circle [radius=0.25];
	\draw[line width = 0.4mm] (2.725,2.6) circle [radius=0.25];
	\draw[line width = 0.4mm] (2.725,3.7) circle [radius=0.25];
	
	\end{tikzpicture}
\caption{The resulted unified database after processing the four databases in Figure \ref{fig:Database_Problem}.}
\label{fig:Database_Problem_output}
\end{figure}

For example, consider the four databases given in Figure \ref{fig:Database_Problem} above. As can be seen, only the two left-hand databases contain the identities of eight different subjects, while the right-hand databases have been anonymized. The key idea is that all four databases contain data regarding some common features -- these are denoted by the rings. Based on these common features, one would like to: (i) match between databases that contain data for the same subjects (dogs and cats in this case), (ii) for each matched pair of databases, to recover the alignment between the rows. The resulted output from this process is given in Figure \ref{fig:Database_Problem_output}; the unified database contains data regarding the eight subjects, where each subject has more data than in the original databases.    

\begin{figure}[h!]
	\centering
	\begin{tikzpicture}[scale=0.8,cross/.style={path picture={ 
			\draw[red]
			(path picture bounding box.south east) -- (path picture bounding box.north west) (path picture bounding box.south west) -- (path picture bounding box.north east);}}]
	
	\filldraw[draw=black,fill=white,rounded corners=3, ultra thick] (9.5,-0.5) rectangle (18.5,5.7);
	\filldraw[draw=black,fill=gray!10, ultra thick] (10,0) rectangle (18,0.8);
	\filldraw[draw=black,fill=pink!20] (11.1,0.1) rectangle (14.35,0.7);
	\filldraw[draw=black,fill=green!70] (14.55,0.1) rectangle (17.8,0.7);
	\node at (10.5,0.4) {\includegraphics[width=5mm]{Dog3.jpg}};
	\node [draw,circle,red,cross,minimum width=0.5 cm] at (10.5,0.4){};
	\filldraw[draw=black,fill=gray!10, ultra thick] (10,1.1) rectangle (18,1.9);
	\filldraw[draw=black,fill=pink!20] (11.1,1.2) rectangle (14.35,1.8);
	\filldraw[draw=black,fill=yellow!70] (14.55,1.2) rectangle (17.8,1.8);
	\node at (10.5,1.5) {\includegraphics[width=5mm]{Dog1.jpg}};
	\node [draw,circle,red,cross,minimum width=0.5 cm] at (10.5,1.5){};
	\filldraw[draw=black,fill=gray!10, ultra thick] (10,2.2) rectangle (18,3.0);
	\filldraw[draw=black,fill=pink!20] (11.1,2.3) rectangle (14.35,2.9);
	\filldraw[draw=black,fill=blue!70] (14.55,2.3) rectangle (17.8,2.9);
	\node at (10.5,2.6) {\includegraphics[width=5mm]{Dog4.jpg}};
	\node [draw,circle,red,cross,minimum width=0.5 cm] at (10.5,2.6){};
	\filldraw[draw=black,fill=gray!10, ultra thick] (10,3.3) rectangle (18,4.1);
	\filldraw[draw=black,fill=pink!20] (11.1,3.4) rectangle (14.35,4.0);
	\filldraw[draw=black,fill=red!70] (14.55,3.4) rectangle (17.8,4.0);
	\node at (10.5,3.7) {\includegraphics[width=5mm]{Dog2.jpg}};
	\node [draw,circle,red,cross,minimum width=0.5 cm] at (10.5,3.7){};
	
	\filldraw[draw=black,fill=gray!10, ultra thick] (10,4.4) rectangle (18,5.2);
	\filldraw[draw=black,fill=pink!20] (11.1,4.5) rectangle (14.35,5.1);
	\filldraw[draw=black,fill=orange!70] (14.55,4.5) rectangle (17.8,5.1);
	\node at (10.5,4.8) {\includegraphics[width=5mm]{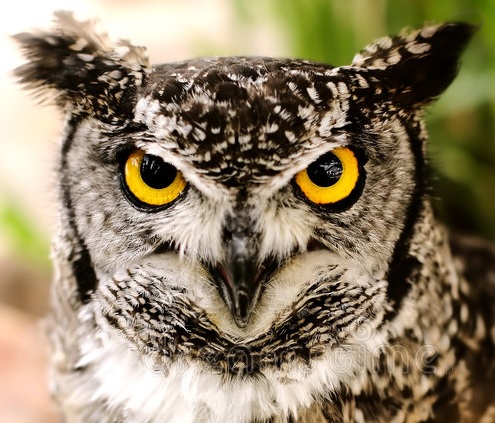}};
	\node [draw,circle,red,cross,minimum width=0.5 cm] at (10.5,4.8){};
	
	\draw[line width = 0.4mm] (12.725,0.4) circle [radius=0.25];
	\draw[line width = 0.4mm] (12.725,1.5) circle [radius=0.25];
	\draw[line width = 0.4mm] (12.725,2.6) circle [radius=0.25];
	\draw[line width = 0.4mm] (12.725,3.7) circle [radius=0.25];
	\draw[line width = 0.4mm] (12.725,4.8) circle [radius=0.25];

	\filldraw[draw=black,fill=white,rounded corners=3, ultra thick] (-0.5,-0.5) rectangle (8.5,5.7);
	\filldraw[draw=black,fill=gray!10, ultra thick] (0,0) rectangle (8,0.8);
	\filldraw[draw=black,fill=pink!20] (1.1,0.1) rectangle (4.35,0.7);
	\filldraw[draw=black,fill=yellow!40] (4.55,0.1) rectangle (7.8,0.7);
	\node at (0.5,0.4) {\includegraphics[width=5mm]{Dog1.jpg}};
	\filldraw[draw=black,fill=gray!10, ultra thick] (0,1.1) rectangle (8,1.9);
	\filldraw[draw=black,fill=pink!20] (1.1,1.2) rectangle (4.35,1.8);
	\filldraw[draw=black,fill=red!40] (4.55,1.2) rectangle (7.8,1.8);
	\node at (0.5,1.5) {\includegraphics[width=5mm]{Dog2.jpg}};
	\filldraw[draw=black,fill=gray!10, ultra thick] (0,2.2) rectangle (8,3.0);
	\filldraw[draw=black,fill=pink!20] (1.1,2.3) rectangle (4.35,2.9);
	\filldraw[draw=black,fill=green!40] (4.55,2.3) rectangle (7.8,2.9);
	\node at (0.5,2.6) {\includegraphics[width=5mm]{Dog3.jpg}};
	\filldraw[draw=black,fill=gray!10, ultra thick] (0,3.3) rectangle (8,4.1);
	\filldraw[draw=black,fill=pink!20] (1.1,3.4) rectangle (4.35,4.0);
	\filldraw[draw=black,fill=blue!40] (4.55,3.4) rectangle (7.8,4.0);
	\node at (0.5,3.7) {\includegraphics[width=5mm]{Dog4.jpg}};
	
	\filldraw[draw=black,fill=gray!10, ultra thick] (0,4.4) rectangle (8,5.2);
	\filldraw[draw=black,fill=pink!20] (1.1,4.5) rectangle (4.35,5.1);
	\filldraw[draw=black,fill=purple!40] (4.55,4.5) rectangle (7.8,5.1);
	\node at (0.5,4.8) {\includegraphics[width=5mm]{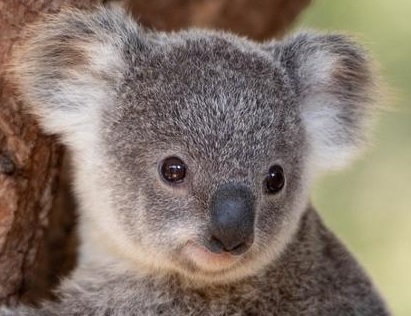}};
	
	\draw[line width = 0.4mm] (2.725,0.4) circle [radius=0.25];
	\draw[line width = 0.4mm] (2.725,1.5) circle [radius=0.25];
	\draw[line width = 0.4mm] (2.725,2.6) circle [radius=0.25];
	\draw[line width = 0.4mm] (2.725,3.7) circle [radius=0.25];
	\draw[line width = 0.4mm] (2.725,4.8) circle [radius=0.25];
	\end{tikzpicture} 
	\caption{An example for two databases with a partial overlap.} 
	\label{fig:Database_Problem_PartialAlignment}
\end{figure}

Yet another realistic assumption motivates this study, which is the fact that two databases may only have a {\it partial overlap} between the subjects, i.e., the two databases contain information regarding a common set of subjects, but each database also contain data on more subjects that are not included in the other database. For example, consider the two databases in Figure \ref{fig:Database_Problem_PartialAlignment}; four dog subjects are in common, while the left-hand-side database has a Koala and the right-hand-side database has an Owl. 

\begin{figure}[h!] 
	\centering 
	\begin{tikzpicture}[scale=0.8,cross/.style={path picture={ 
			\draw[red]
			(path picture bounding box.south east) -- (path picture bounding box.north west) (path picture bounding box.south west) -- (path picture bounding box.north east);}}]
	
	\filldraw[draw=black,fill=white,rounded corners=3, ultra thick] (-0.5,2.8) rectangle (11.95,9.0);
	
	\filldraw[draw=black,fill=gray!10, ultra thick] (0,3.3) rectangle (11.45,4.1);
	\filldraw[draw=black,fill=pink!20] (1.1,3.4) rectangle (4.35,4.0);
	\filldraw[draw=black,fill=purple!40] (4.55,3.4) rectangle (7.8,4.0);
	\filldraw[draw=black,fill=orange!70] (8.0,3.4) rectangle (11.25,4.0);
	\node at (0.5,3.7) {\includegraphics[width=5mm]{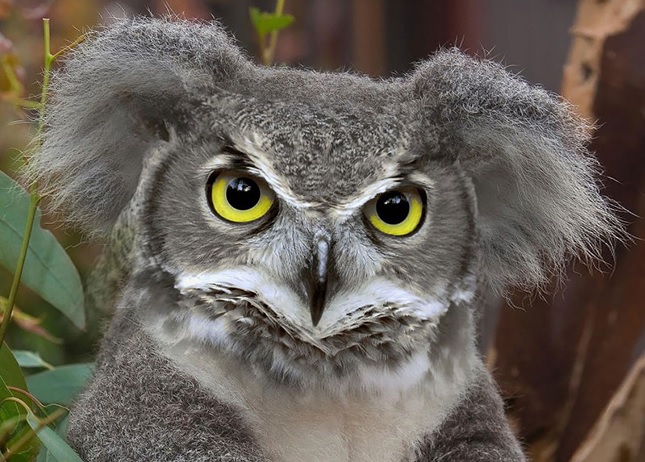}};
	\filldraw[draw=black,fill=gray!10, ultra thick] (0,4.4) rectangle (11.45,5.2);
	\filldraw[draw=black,fill=pink!20] (1.1,4.5) rectangle (4.35,5.1);
	\filldraw[draw=black,fill=yellow!40] (4.55,4.5) rectangle (7.8,5.1);
	\filldraw[draw=black,fill=yellow!70] (8.0,4.5) rectangle (11.25,5.1);
	\node at (0.5,4.8) {\includegraphics[width=5mm]{Dog1.jpg}};
	\filldraw[draw=black,fill=gray!10, ultra thick] (0,5.5) rectangle (11.45,6.3);
	\filldraw[draw=black,fill=pink!20] (1.1,5.6) rectangle (4.35,6.2);
	\filldraw[draw=black,fill=red!40] (4.55,5.6) rectangle (7.8,6.2);
	\filldraw[draw=black,fill=red!70] (8.0,5.6) rectangle (11.25,6.2);
	\node at (0.5,5.9) {\includegraphics[width=5mm]{Dog2.jpg}};
	\filldraw[draw=black,fill=gray!10, ultra thick] (0,6.6) rectangle (11.45,7.4);
	\filldraw[draw=black,fill=pink!20] (1.1,6.7) rectangle (4.35,7.3);
	\filldraw[draw=black,fill=green!40] (4.55,6.7) rectangle (7.8,7.3);
	\filldraw[draw=black,fill=green!70] (8.0,6.7) rectangle (11.25,7.3);
	\node at (0.5,7.0) {\includegraphics[width=5mm]{Dog3.jpg}};
	\filldraw[draw=black,fill=gray!10, ultra thick] (0,7.7) rectangle (11.45,8.5);
	\filldraw[draw=black,fill=pink!20] (1.1,7.8) rectangle (4.35,8.4);
	\filldraw[draw=black,fill=blue!40] (4.55,7.8) rectangle (7.8,8.4);
	\filldraw[draw=black,fill=blue!70] (8.0,7.8) rectangle (11.25,8.4);
	\node at (0.5,8.1) {\includegraphics[width=5mm]{Dog4.jpg}};
	
	\draw[line width = 0.4mm] (2.725,3.7) circle [radius=0.25];
	\draw[line width = 0.4mm] (2.725,4.8) circle [radius=0.25];
	\draw[line width = 0.4mm] (2.725,5.9) circle [radius=0.25];
	\draw[line width = 0.4mm] (2.725,7.0) circle [radius=0.25];
	\draw[line width = 0.4mm] (2.725,8.1) circle [radius=0.25];
	\end{tikzpicture}
	\caption{An example for the resulted unified database after operating a full recovery algorithm over two databases with partial overlap as in Figure \ref{fig:Database_Problem_PartialAlignment}.} 
	\label{fig:Database_Problem_PartialAlignment_Error!}
\end{figure}

Operating an algorithm that aims on a full alignment recovery between the databases is obviously a wrong idea; the four dogs will probably be matched together, but also the Koala with the Owl, which is an error (Figure \ref{fig:Database_Problem_PartialAlignment_Error!}).  

Indeed, one of the main goals of this work is to propose algorithms that are able to handle databases with partial overlap. We will refer to this problem by partial alignment recovery. Apart from recovering the (partial) alignment between two databases with a partial overlap, we mention another motivation for studying such algorithms: even when two databases have a full overlap, the probability of error of the optimal ML estimator may be too high in some scenarios that require a relatively high level of reliability. 
Thus, relying on algorithms that are able to estimate only part of the alignment but with lower guaranteed error probability may be a good compromise in such cases.

\subsection{Settings and Problem Formulation}

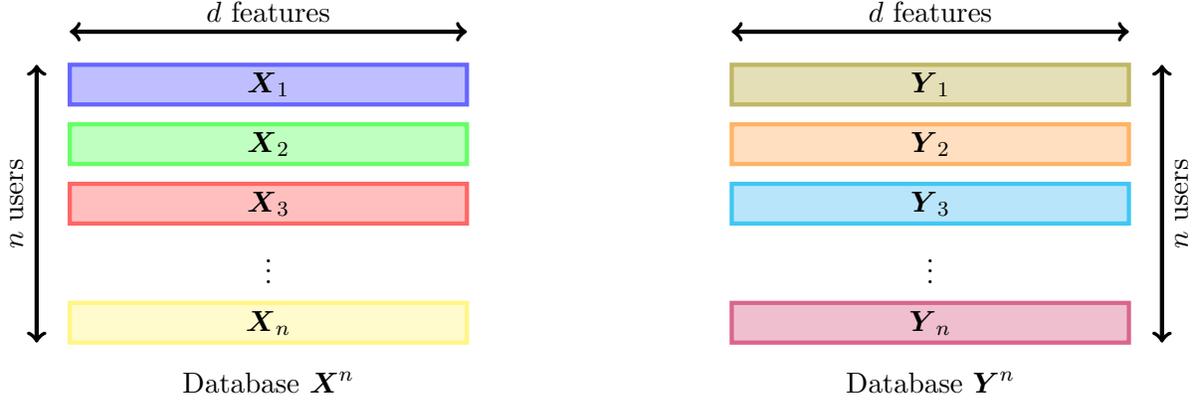
\begin{figure}[h!]
	\centering
	\begin{tikzpicture}[scale=0.88]
	\node [anchor=center] at (3,-0.9) {Database $\bX^{n}$};
	\filldraw[color=yellow!60, fill=yellow!25, ultra thick] (0,-0.3) rectangle (6,0.3);
	\node [anchor=center] at (3,0.9) {$\vdots$};
	\filldraw[color=red!60, fill=red!25, ultra thick] (0,1.5) rectangle (6,2.1);
	\filldraw[color=green!60, fill=green!25, ultra thick] (0,2.4) rectangle (6,3.0);
	\filldraw[color=blue!60, fill=blue!25, ultra thick] (0,3.3) rectangle (6,3.9);
	\node [anchor=center] at (3,3.6) {$\bX_{1}$};
	\node [anchor=center] at (3,2.7) {$\bX_{2}$};
	\node [anchor=center] at (3,1.8) {$\bX_{3}$};
	\node [anchor=center] at (3,0) {$\bX_{n}$};
	\draw[ultra thick,<->] (0,4.4) -- (6,4.4);
	\node [anchor=center] at (3,4.7) {$d$ features};	
	\draw[ultra thick,<->] (-0.5,-0.3) -- (-0.5,3.9);
	\node[rotate=90] at (-0.8,1.8) {$n$ users};

	\node [anchor=center] at (13,-0.9) {Database $\bY^{n}$};
	\filldraw[color=purple!60, fill=purple!25, ultra thick] (10,-0.3) rectangle (16,0.3);
	\node [anchor=center] at (13,0.9) {$\vdots$};
	\filldraw[color=cyan!60, fill=cyan!25, ultra thick] (10,1.5) rectangle (16,2.1);
	\filldraw[color=orange!60, fill=orange!25, ultra thick] (10,2.4) rectangle (16,3.0);
	\filldraw[color=olive!60, fill=olive!25, ultra thick] (10,3.3) rectangle (16,3.9);
	\node [anchor=center] at (13,3.6) {$\bY_{1}$};
	\node [anchor=center] at (13,2.7) {$\bY_{2}$};
	\node [anchor=center] at (13,1.8) {$\bY_{3}$};
	\node [anchor=center] at (13,0) {$\bY_{n}$};
	\draw[ultra thick,<->] (10,4.4) -- (16,4.4);
	\node [anchor=center] at (13,4.7) {$d$ features};	
	\draw[ultra thick,<->] (16.5,-0.3) -- (16.5,3.9);
	\node[rotate=90] at (16.8,1.8) {$n$ users};
	\end{tikzpicture}
	\caption{Under the null hypothesis $\HypA$ in the detection problem, the feature vectors $\{\bX_{i}\}_{i=1}^{n}$ and $\{\bY_{i}\}_{i=1}^{n}$ are generated independently according to $\calN(\boldsymbol{0}_{d},\bI_{d})$.} 
	\label{fig:NULL_Hypothesis}
\end{figure}

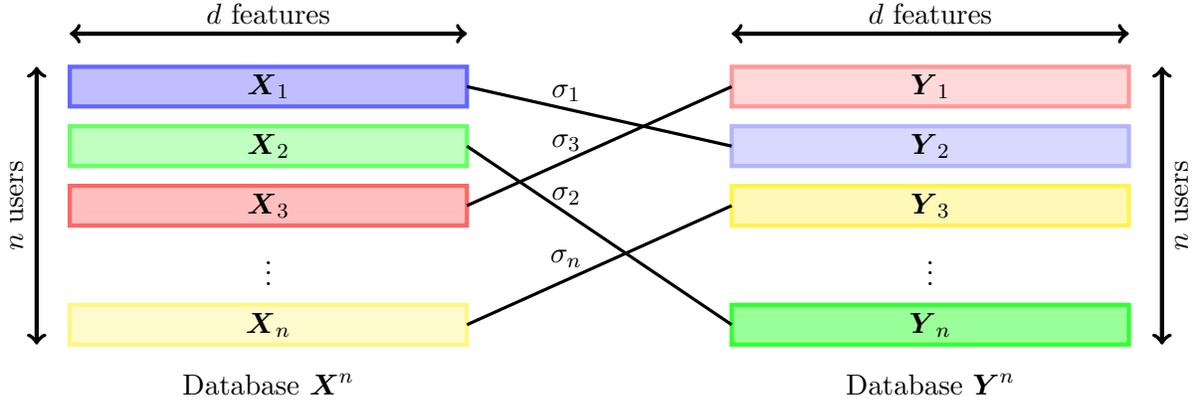
\begin{figure}[h!]
	\centering
	\begin{tikzpicture}[scale=0.88]
	\node [anchor=center] at (3,-0.9) {Database $\bX^{n}$};
	\filldraw[color=yellow!60, fill=yellow!25, ultra thick] (0,-0.3) rectangle (6,0.3);
	\node [anchor=center] at (3,0.9) {$\vdots$};
	\filldraw[color=red!60, fill=red!25, ultra thick] (0,1.5) rectangle (6,2.1);
	\filldraw[color=green!60, fill=green!25, ultra thick] (0,2.4) rectangle (6,3.0);
	\filldraw[color=blue!60, fill=blue!25, ultra thick] (0,3.3) rectangle (6,3.9);
	\node [anchor=center] at (3,3.6) {$\bX_{1}$};
	\node [anchor=center] at (3,2.7) {$\bX_{2}$};
	\node [anchor=center] at (3,1.8) {$\bX_{3}$};
	\node [anchor=center] at (3,0) {$\bX_{n}$};
	\draw[ultra thick,<->] (0,4.4) -- (6,4.4);
	\node [anchor=center] at (3,4.7) {$d$ features};	
	\draw[ultra thick,<->] (-0.5,-0.3) -- (-0.5,3.9);
	\node[rotate=90] at (-0.8,1.8) {$n$ users};

	\node [anchor=center] at (13,-0.9) {Database $\bY^{n}$};
	\filldraw[color=green!80, fill=green!40, ultra thick] (10,-0.3) rectangle (16,0.3);
	\node [anchor=center] at (13,0.9) {$\vdots$};
	\filldraw[color=yellow!80, fill=yellow!35, ultra thick] (10,1.5) rectangle (16,2.1);
	\filldraw[color=blue!30, fill=blue!15, ultra thick] (10,2.4) rectangle (16,3.0);
	\filldraw[color=red!40, fill=red!15, ultra thick] (10,3.3) rectangle (16,3.9);
	\node [anchor=center] at (13,3.6) {$\bY_{1}$};
	\node [anchor=center] at (13,2.7) {$\bY_{2}$};
	\node [anchor=center] at (13,1.8) {$\bY_{3}$};
	\node [anchor=center] at (13,0) {$\bY_{n}$};
	\draw[ultra thick,<->] (10,4.4) -- (16,4.4);
	\node [anchor=center] at (13,4.7) {$d$ features};	
	\draw[ultra thick,<->] (16.5,-0.3) -- (16.5,3.9);
	\node[rotate=90] at (16.8,1.8) {$n$ users};
	
	\draw[very thick] (6,3.6) -- (10,2.7);
	\draw[very thick] (6,2.7) -- (10,0);
	\draw[very thick] (6,1.8) -- (10,3.6);
	\draw[very thick] (6,0) -- (10,1.8);
	
	\node [anchor=center] at (7.5,3.5) {$\sigma_{1}$};
	\node [anchor=center] at (7.5,2.75) {$\sigma_{3}$};
	\node [anchor=center] at (7.5,1.95) {$\sigma_{2}$};
	\node [anchor=center] at (7.5,1) {$\sigma_{n}$};
	\end{tikzpicture}
	\caption{Under the alternate hypothesis $\HypB$ in the detection problem (or the probabilistic model for the alignment recovery problem), the collection of pairs $(\bX_{1},\bY_{\sigma_{1}}),\ldots,(\bX_{n},\bY_{\sigma_{n}})$ is independent. Each pair of feature vectors $(\bX_{i},\bY_{\sigma_{i}})$ are jointly Gaussian with a known correlation coefficient $\rho > 0$, under some unknown permutation $\sigma \in \calS_{n}$.} 
	\label{fig:Alternate_Hypothesis}
\end{figure}

We consider the following binary hypothesis testing problem. Under the {\it null hypothesis} $\HypA$, the Gaussian databases $\bX^{n}$ and $\bY^{n}$ are generated independently with $\bX_{1},\ldots,\bX_{n},\bY_{1},\ldots,\bY_{n} \overset{\text{IID}}{\sim} \calN(\boldsymbol{0}_{d},\bI_{d})$ (Figure \ref{fig:NULL_Hypothesis}). Let us denote by $\pr_{0}$ the probability distribution that governs $(\bX^{n},\bY^{n})$ under $\HypA$. Under the {\it alternate hypothesis} $\HypB$, the databases $\bX^{n}$ and $\bY^{n}$ are correlated with some unknown permutation $\sigma \in \calS_{n}$ and some known correlation coefficient $\rho >0$ (Figure \ref{fig:Alternate_Hypothesis}). 
Let us denote by $\pr_{1|\sigma}$ the probability distribution that governs $(\bX^{n},\bY^{n})$ under $\HypB$, for some permutation $\sigma \in \calS_{n}$. Summarizing:
\begin{equation} \label{Binary_HT_Problem}
\begin{aligned}
&\HypA:~~~(\bX_{1},\bY_{1}),\ldots,(\bX_{n},\bY_{n}) \overset{\text{IID}}{\sim} \calN^{\otimes d}\left(\left[\begin{matrix} 0 \\ 0
\end{matrix}\right], \left[\begin{matrix} 1 & 0 \\ 0 & 1 \end{matrix}\right] \right) \\
&\HypB:~~~(\bX_{1},\bY_{\sigma_{1}}),\ldots,(\bX_{n},\bY_{\sigma_{n}}) \overset{\text{IID}}{\sim} \calN^{\otimes d}\left(\left[\begin{matrix} 0 \\ 0
\end{matrix}\right], \left[\begin{matrix} 1 & \rho \\ \rho & 1 \end{matrix}\right] \right),
\end{aligned} 
\end{equation}
for some permutation $\sigma \in \calS_{n}$.

We would like to consider a joint problem of correlation detection and low-complexity permutation recovery. 
Let us denote the set 
\begin{align} \label{Def_Ln}
\sL_{n} = \Big\{\{(k_{1},\ell_{1}), (k_{2},\ell_{2}), \ldots, (k_{m},\ell_{m})\} \Big| m \in \{1,2,\ldots,n\}, \forall i \neq j: k_{i} \neq k_{j}, \ell_{i} \neq \ell_{j}   \Big\}.
\end{align}

Given the databases $\bX^{n}$ and $\bY^{n}$ (and not the permutation $\sigma$), a {\it joint detector-recover} $\phi:\reals^{d \times n} \times \reals^{d \times n} \to \{0,\sL_{n}\}$ decides whether the null hypothesis or the alternate hypothesis occurred, and if it accepts the latter, it outputs a member $\psi(\bX^{n},\bY^{n})$ of $\sL_{n}$, i.e., a list of all probably matched pairs. Note that $\psi(\bX^{n},\bY^{n})$ must not be a permutation.    

For a given $n$ and $d$, the type-I probability of error of a test $\phi$ is 
\begin{align}
P_{\mbox{\tiny FA}}(\phi) \dfn \pr_{0}\left\{\phi(\bX^{n},\bY^{n}) \neq 0 \right\},
\end{align}
and the type-II probability of error is 
\begin{align}
P_{\mbox{\tiny MD}}(\phi) \dfn 
\max_{\sigma \in \calS_{n}} \pr_{1|\sigma}\left\{\phi(\bX^{n},\bY^{n})=0 \right\}.
\end{align}
Assuming that $\bX^{n}$ and $\bY^{n}$ are indeed correlated, we define two different error probabilities. The probability of error that the resulting recovery is not the correct permutation is defined by
\begin{align} \label{ErrorProbability_1}
P_{\mbox{\scriptsize e,1}}(\psi)
\dfn \max_{\sigma \in \calS_{n}} \pr_{1|\sigma}\Big\{\psi(\bX^{n},\bY^{n}) \neq \{(1,\sigma_{1}),(2,\sigma_{2}),\ldots,(n,\sigma_{n})\}\Big\}.
\end{align}  
Also, the probability of error that at least one pair in the recovery is a mismatch is defined by 
\begin{align} \label{ErrorProbability_2}
P_{\mbox{\scriptsize e,2}}(\psi)
\dfn \max_{\sigma \in \calS_{n}} \pr_{1|\sigma}\Big\{\text{At least one match in }\psi(\bX^{n},\bY^{n})\text{ is wrong}\Big\}.
\end{align} 

For the specific algorithm that will be defined in Section \ref{Sec4}, which solves the above defined joint detection-recovery problem, our main objective is to find theoretical guarantees on the various error probabilities as functions of the parameters $n$, $d$, and $\rho$.

\subsection{Prior Work}

We present the achievability result for the correlated Gaussian database detection problem that was recently reported in \cite{KN2022}.
The sum-of-inner-products statistic is defined by
\begin{align} \label{Statistic}
T \dfn \sum_{i=1}^{n} \sum_{j=1}^{n} \bX_{i}^{\mathsf{T}} \bY_{j},
\end{align}
and the test itself is defined by comparing $T$ to a threshold as follows:
\begin{align} \label{Test}
\phi_{T,t}(\bX^{n},\bY^{n})
= \left\{   
\begin{array}{l l}
0    & \quad \text{  $T < t$  }   \\
1    & \quad \text{  $T \geq t$  }
\end{array} \right. .
\end{align}

Then, the following result is proved in \cite[Section 4]{KN2022}.

\begin{proposition}[Detection Achievability]
	\label{Theorem Detection Achievability}
	Let $t = \sqrt{\gamma}\tfrac{dn}{2}$ with $\gamma \in (0,4\rho^{2})$. The error probabilities of the sum-of-inner-products test $\phi_{T}$ for the binary hypothesis testing problem \eqref{Binary_HT_Problem} are upper-bounded by
	\begin{align}
	P_{\mbox{\tiny FA}}(\phi_{T}) 
	&\leq \exp\left(-\frac{d}{2}G_{\mbox{\tiny FA}}(\gamma)\right) \\
	P_{\mbox{\tiny MD}}(\phi_{T}) 
	&\leq \exp\left(-\frac{d}{2}G_{\mbox{\tiny MD}}(\gamma)\right), 
	\end{align} 
	where,
	\begin{align}
	G_{\mbox{\tiny FA}}(\gamma) 
	&\dfn \sqrt{1+\gamma} - 1 - \ln\left(\frac{1+\sqrt{1+\gamma}}{2}\right), \\
	G_{\mbox{\tiny MD}}(\gamma)
	&\dfn \frac{1}{1-\rho^{2}}\left(\sqrt{(1-\rho^{2})^{2}+\gamma} - \sqrt{\rho^{2}\gamma}\right) - 1 - \ln\left(\frac{1-\rho^{2}+\sqrt{(1-\rho^{2})^{2}+\gamma}}{2}\right).
	\end{align}
\end{proposition}

To see why the test defined in \eqref{Statistic}-\eqref{Test} may be improved, first note that the double summation in \eqref{Statistic} can also be written by
\begin{align} \label{Test2}
T = \sum_{i=1}^{n} \bX_{i}^{\mathsf{T}} \bY_{\sigma_{i}} + \sum_{i=1}^{n} \sum_{j \neq \sigma_{i}} \bX_{i}^{\mathsf{T}} \bY_{j}
\end{align} 
for some permutation $\sigma \in \calS_{n}$, while the optimal test is based on the statistic
\begin{align}
T_{\mbox{\tiny opt}} = \max_{\sigma \in \calS_{n}} \sum_{i=1}^{n} \bX_{i}^{\mathsf{T}} \bY_{\sigma_{i}}.
\end{align} 
If the two databases are indeed correlated for some $\sigma \in \calS_{n}$, then the expectation of the first term on the right-hand-side of \eqref{Test2} is given by $nd\rho$, but the second term, which is a sum over $n(n-1)$ relatively small terms, acts as an effective noise that increases the probability of missed detection. Alternatively, if the two databases are independent, then the sum $\sum_{i=1}^{n} \bX_{i}^{\mathsf{T}} \bY_{\sigma_{i}}$ is expected to be relatively small for every $\sigma \in \calS_{n}$, so the second term on the right-hand-side of \eqref{Test2} increases the probability of false alarm.

%

\section{Joint Detection and Partial Alignment Recovery} \label{Sec4}

\subsection{The Proposed ``Threshold-and-Clean'' Algorithm}
Denote the normalized vectors
\begin{align}
\tilde{\bX}_{i} = \frac{\bX_{i}}{\|\bX_{i}\|},~~i \in \{1,2,\ldots,n\},
\end{align}
and
\begin{align}
\tilde{\bY}_{j} = \frac{\bY_{j}}{\|\bX_{j}\|},~~j \in \{1,2,\ldots,n\},
\end{align}
and consider the normalized databases $\tilde{\bX^{n}}$ and $\tilde{\bY^{n}}$.
For $\theta \in (0,1)$ (possibly depending on $\rho$), define the statistic
\begin{align} \label{Statistic_new}
N(\theta) \dfn   \sum_{i=1}^{n} \sum_{j=1}^{n} \IND \left\{\tilde{\bX}_{i}^{\mathsf{T}} \tilde{\bY}_{j} \geq \theta \right\},
\end{align}
such that in the detection phase, the proposed algorithm compares $N(\theta)$ to a threshold:
\begin{align} \label{Test_new}
\phi_{N}(\tilde{\bX^{n}},\tilde{\bY^{n}})
= \left\{   
\begin{array}{l l}
0    & \quad \text{  $N(\theta) < \beta nP$  }   \\
1    & \quad \text{  $N(\theta) \geq \beta nP$  }
\end{array} \right. ,
\end{align}
where $\beta \in (0,1)$ 
and $P$ is the probability of $\tilde{\bX}^{\mathsf{T}} \tilde{\bY} \geq \theta$ when $\bX$ and $\bY$ are a matched pair.

\begin{figure}[h!]
	\centering
	\begin{tikzpicture}[scale=0.3]
	\draw[black,very thick] (0,0) -- (20,0);
	\draw[black,very thick] (0,2) -- (20,2);
	\draw[black,very thick] (0,4) -- (20,4);
	\draw[black,very thick] (0,6) -- (20,6);
	\draw[black,very thick] (0,8) -- (20,8);
	\draw[black,very thick] (0,10) -- (20,10);
	\draw[black,very thick] (0,12) -- (20,12);
	\draw[black,very thick] (0,14) -- (20,14);
	\draw[black,very thick] (0,16) -- (20,16);
	\draw[black,very thick] (0,18) -- (20,18);
	\draw[black,very thick] (0,20) -- (20,20);
	
	\draw[black,very thick] (0,0) -- (0,20);
	\draw[black,very thick] (2,0) -- (2,20);
	\draw[black,very thick] (4,0) -- (4,20);
	\draw[black,very thick] (6,0) -- (6,20);
	\draw[black,very thick] (8,0) -- (8,20);
	\draw[black,very thick] (10,0) -- (10,20);
	\draw[black,very thick] (12,0) -- (12,20);
	\draw[black,very thick] (14,0) -- (14,20);
	\draw[black,very thick] (16,0) -- (16,20);
	\draw[black,very thick] (18,0) -- (18,20);
	\draw[black,very thick] (20,0) -- (20,20);
	
	\foreach \x/\xtext in {1/1, 3/2, 5/3, 7/4, 9/5, 11/6, 13/7, 15/8, 17/9, 19/10}
	\draw[shift={(\x,0)}] (0pt,2pt) -- (0pt,-2pt) node[below] {$\bX_{\xtext}$};
	\foreach \y/\ytext in {1/1, 3/2, 5/3, 7/4, 9/5, 11/6, 13/7, 15/8, 17/9, 19/10}
	\draw[shift={(0,\y)}] (2pt,0pt) -- (-2pt,0pt) node[left] {$\bY_{\ytext}$};
	
	\node at (9,19) [circle,fill=black] {};
	\node at (3,7) [circle,fill=black] {};
	\node at (17,3) [circle,fill=black] {};
	
	\end{tikzpicture}
	\hspace{0.3in}
	\begin{tikzpicture}[scale=0.3]
	\draw[black,very thick] (0,0) -- (20,0);
	\draw[black,very thick] (0,2) -- (20,2);
	\draw[black,very thick] (0,4) -- (20,4);
	\draw[black,very thick] (0,6) -- (20,6);
	\draw[black,very thick] (0,8) -- (20,8);
	\draw[black,very thick] (0,10) -- (20,10);
	\draw[black,very thick] (0,12) -- (20,12);
	\draw[black,very thick] (0,14) -- (20,14);
	\draw[black,very thick] (0,16) -- (20,16);
	\draw[black,very thick] (0,18) -- (20,18);
	\draw[black,very thick] (0,20) -- (20,20);
	
	\draw[black,very thick] (0,0) -- (0,20);
	\draw[black,very thick] (2,0) -- (2,20);
	\draw[black,very thick] (4,0) -- (4,20);
	\draw[black,very thick] (6,0) -- (6,20);
	\draw[black,very thick] (8,0) -- (8,20);
	\draw[black,very thick] (10,0) -- (10,20);
	\draw[black,very thick] (12,0) -- (12,20);
	\draw[black,very thick] (14,0) -- (14,20);
	\draw[black,very thick] (16,0) -- (16,20);
	\draw[black,very thick] (18,0) -- (18,20);
	\draw[black,very thick] (20,0) -- (20,20);
	
	\foreach \x/\xtext in {1/1, 3/2, 5/3, 7/4, 9/5, 11/6, 13/7, 15/8, 17/9, 19/10}
	\draw[shift={(\x,0)}] (0pt,2pt) -- (0pt,-2pt) node[below] {$\bX_{\xtext}$};
	\foreach \y/\ytext in {1/1, 3/2, 5/3, 7/4, 9/5, 11/6, 13/7, 15/8, 17/9, 19/10}
	\draw[shift={(0,\y)}] (2pt,0pt) -- (-2pt,0pt) node[left] {$\bY_{\ytext}$};
	
	\node at (1,1) [circle,fill=black] {};
	\node at (5,5) [circle,fill=black] {};
	\node at (7,7) [circle,fill=black] {};
	\node at (9,9) [circle,fill=black] {};
	\node at (11,11) [circle,fill=black] {};
	\node at (13,13) [circle,fill=black] {};
	\node at (15,15) [circle,fill=black] {};
	\node at (17,17) [circle,fill=black] {};
	\node at (19,19) [circle,fill=black] {};
	
	\node at (7,15) [circle,fill=black] {};
	\node at (13,1) [circle,fill=black] {};
	\end{tikzpicture}
	\caption{Examples of typical tables under $\HypA$ (left) and under $\HypB$ with the identity permutation.} 
	\label{fig:Typical_Tables}
\end{figure}
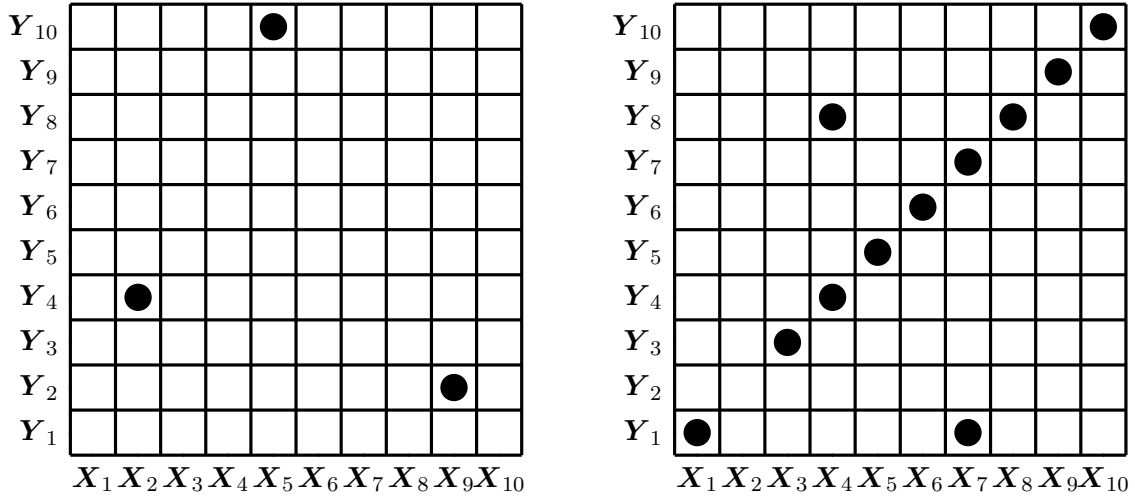

Our proposed algorithm may be presented in a simple graphical way. In order to perform both correlation detection and partial alignment recovery, one has to draw a table with $n$ rows and $n$ columns. In this table, row $i$ stands for $\bX_{i}$ and column $j$ for $\bY_{j}$. The cell $(i,j)$ is filled with a dot only if $\tilde{\bX}_{i}^{\mathsf{T}} \tilde{\bY}_{j} \geq \theta$. The dots are then counted and one declare that the two databases are correlated if the total number of dots exceeds a threshold of $\beta n P$. Typical tables for $n=10$ under $\HypA$ and under $\HypB$ with the identity permutation are given in Figure \ref{fig:Typical_Tables}.   

When the algorithm declares that the databases are correlated (i.e., accepting $\HypB$), it still has to clean the original table before outputting the decoded permutation (more precisely, a member of $\sL_{n}$), since each row and each column must contain at most one dot. 
The algorithm cleans all rows and all columns that contain more than a single dot. Figure \ref{fig:Tables_Expurgation} below presents an example for an original table and a cleaned table, which presents the decoded permutation.     

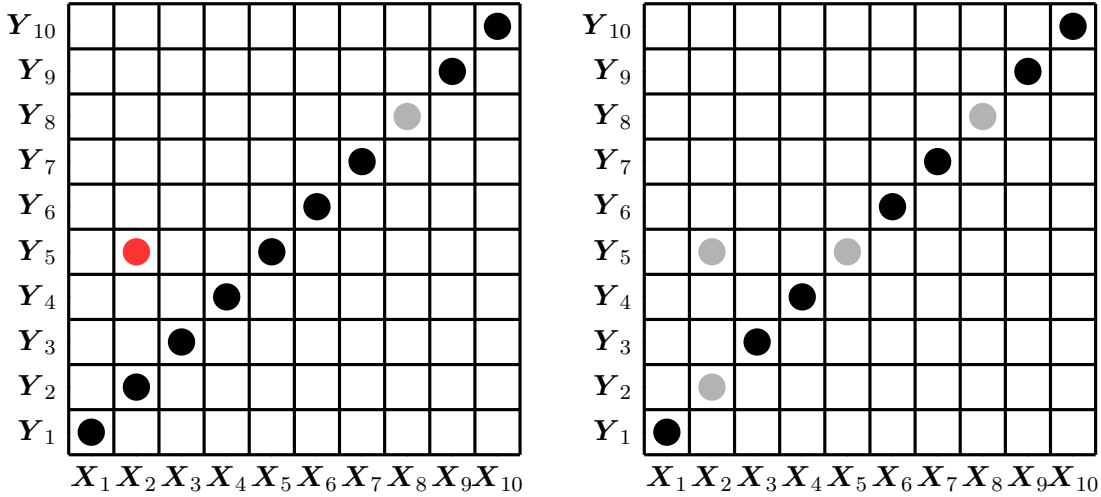
\begin{figure}[h!]
	\centering
	\begin{tikzpicture}[scale=0.3]
	\draw[black,very thick] (0,0) -- (20,0);
	\draw[black,very thick] (0,2) -- (20,2);
	\draw[black,very thick] (0,4) -- (20,4);
	\draw[black,very thick] (0,6) -- (20,6);
	\draw[black,very thick] (0,8) -- (20,8);
	\draw[black,very thick] (0,10) -- (20,10);
	\draw[black,very thick] (0,12) -- (20,12);
	\draw[black,very thick] (0,14) -- (20,14);
	\draw[black,very thick] (0,16) -- (20,16);
	\draw[black,very thick] (0,18) -- (20,18);
	\draw[black,very thick] (0,20) -- (20,20);
	
	\draw[black,very thick] (0,0) -- (0,20);
	\draw[black,very thick] (2,0) -- (2,20);
	\draw[black,very thick] (4,0) -- (4,20);
	\draw[black,very thick] (6,0) -- (6,20);
	\draw[black,very thick] (8,0) -- (8,20);
	\draw[black,very thick] (10,0) -- (10,20);
	\draw[black,very thick] (12,0) -- (12,20);
	\draw[black,very thick] (14,0) -- (14,20);
	\draw[black,very thick] (16,0) -- (16,20);
	\draw[black,very thick] (18,0) -- (18,20);
	\draw[black,very thick] (20,0) -- (20,20);
	
	\foreach \x/\xtext in {1/1, 3/2, 5/3, 7/4, 9/5, 11/6, 13/7, 15/8, 17/9, 19/10}
	\draw[shift={(\x,0)}] (0pt,2pt) -- (0pt,-2pt) node[below] {$\bX_{\xtext}$};
	\foreach \y/\ytext in {1/1, 3/2, 5/3, 7/4, 9/5, 11/6, 13/7, 15/8, 17/9, 19/10}
	\draw[shift={(0,\y)}] (2pt,0pt) -- (-2pt,0pt) node[left] {$\bY_{\ytext}$};
	
	\node at (3,9) [circle,fill=red!80] {};
	
	\node at (1,1) [circle,fill=black] {};
	\node at (3,3) [circle,fill=black] {};
	\node at (5,5) [circle,fill=black] {};
	\node at (7,7) [circle,fill=black] {};
	\node at (9,9) [circle,fill=black] {};
	\node at (11,11) [circle,fill=black] {};
	\node at (13,13) [circle,fill=black] {};
	\node at (15,15) [circle,fill=gray!60] {};
	\node at (17,17) [circle,fill=black] {};
	\node at (19,19) [circle,fill=black] {};
	\end{tikzpicture}
	\hspace{0.15in}
	\begin{tikzpicture}[scale=0.3]
	\draw[black,very thick] (0,0) -- (20,0);
	\draw[black,very thick] (0,2) -- (20,2);
	\draw[black,very thick] (0,4) -- (20,4);
	\draw[black,very thick] (0,6) -- (20,6);
	\draw[black,very thick] (0,8) -- (20,8);
	\draw[black,very thick] (0,10) -- (20,10);
	\draw[black,very thick] (0,12) -- (20,12);
	\draw[black,very thick] (0,14) -- (20,14);
	\draw[black,very thick] (0,16) -- (20,16);
	\draw[black,very thick] (0,18) -- (20,18);
	\draw[black,very thick] (0,20) -- (20,20);
	
	\draw[black,very thick] (0,0) -- (0,20);
	\draw[black,very thick] (2,0) -- (2,20);
	\draw[black,very thick] (4,0) -- (4,20);
	\draw[black,very thick] (6,0) -- (6,20);
	\draw[black,very thick] (8,0) -- (8,20);
	\draw[black,very thick] (10,0) -- (10,20);
	\draw[black,very thick] (12,0) -- (12,20);
	\draw[black,very thick] (14,0) -- (14,20);
	\draw[black,very thick] (16,0) -- (16,20);
	\draw[black,very thick] (18,0) -- (18,20);
	\draw[black,very thick] (20,0) -- (20,20);
	
	\foreach \x/\xtext in {1/1, 3/2, 5/3, 7/4, 9/5, 11/6, 13/7, 15/8, 17/9, 19/10}
	\draw[shift={(\x,0)}] (0pt,2pt) -- (0pt,-2pt) node[below] {$\bX_{\xtext}$};
	\foreach \y/\ytext in {1/1, 3/2, 5/3, 7/4, 9/5, 11/6, 13/7, 15/8, 17/9, 19/10}
	\draw[shift={(0,\y)}] (2pt,0pt) -- (-2pt,0pt) node[left] {$\bY_{\ytext}$};
	
	\node at (3,9) [circle,fill=gray!60] {};
	
	\node at (1,1) [circle,fill=black] {};
	\node at (3,3) [circle,fill=gray!60] {};
	\node at (5,5) [circle,fill=black] {};
	\node at (7,7) [circle,fill=black] {};
	\node at (9,9) [circle,fill=gray!60] {};
	\node at (11,11) [circle,fill=black] {};
	\node at (13,13) [circle,fill=black] {};
	\node at (15,15) [circle,fill=gray!60] {};
	\node at (17,17) [circle,fill=black] {};
	\node at (19,19) [circle,fill=black] {};
	\end{tikzpicture}
	\caption{An example of an original table (left) and a cleaned table. Note that the cleaned table must not represent a permutation, rather a member of $\sL_{n}$.} 
	\label{fig:Tables_Expurgation}
\end{figure}

\subsection{Theoretical Guarantees}

In order to present our main result, we need some definitions. 
For two correlated feature vectors $\bX$ and $\bY$, the {\it local detect probability} is defined by
\begin{align}
P(d,\rho,\theta) \dfn \pr\left\{\tilde{\bX}^{\mathsf{T}} \tilde{\bY} \geq \theta \right\},
\end{align}
while for two independent feature vectors, the {\it local false-alarm probability} is defined by 
\begin{align}
Q(d,\theta) \dfn \pr\left\{\tilde{\bX}^{\mathsf{T}} \tilde{\bY} \geq \theta \right\}.
\end{align}

In the following result, which is proved in Appendix A, we present upper bounds on the type-I and type-II error probabilities for a general set of parameters.
Let us denote the Stirling numbers of the second kind:
\begin{align}
\mathsf{S}(k,\ell) = \frac{1}{\ell!} \sum_{i=0}^{\ell} (-1)^{i} \binom{\ell}{i}(\ell-i)^{k},
\end{align}
and define the numbers
\begin{align}
\label{DEF_Bk}
\mathsf{B}(k) = \sum_{d=1}^{k}  
\mathsf{S}(k,d) \cdot \frac{k^{2d}}{d!}.
\end{align}

\begin{theorem}[type-I and type-II probability bounds] \label{THM_FA_and_MD_probabilities}
	Let $n,d \in \naturals$ and $\rho,\theta \in (0,1]$ be given. Let $P=P(d,\rho,\theta)$ and $Q=Q(d,\theta)$ as defined above. 
	Then, for any $\beta \in (0,1)$,
	\begin{align} \label{FA_upper_bound} 
	P_{\mbox{\tiny FA}}(\phi_{N}) 
	&\leq \inf_{k \in \naturals}  
	\frac{k(k+1)\mathsf{B}(k)\left[(n^{2}Q)^{k} \cdot \IND\{n^{2}Q \geq 1\} + n^{2}Q \cdot \IND\{n^{2}Q < 1\}\right]}{\left(\beta nP\right)^{k}},
	\end{align}
	and,
	\begin{align} \label{MD_upper_bound} 
	P_{\mbox{\tiny MD}}(\phi_{N})
	&\leq \exp \left\{- \min \left( (1-\beta)^{2}  \frac{nP}{16nQ + 2}, (1-\beta) \frac{n}{12} \right) \right\}. 
	\end{align}
\end{theorem}

\subsection*{Discussion}

The threshold $\beta \in (0,1)$ trades-off between the two error probabilities. When $\beta$ is relatively low, then relatively few ``dots'' are required to accept $\HypB$, hence the type-I error probability is high and the type-II error probability is low. When $\beta$ grows towards 1, an increasing number of dots are needed to accept $\HypB$; the type-I error probability gradually decreases while the type-II error probability increases.      

While the two error probabilities of the sum-of-inner-products tester \eqref{Statistic}-\eqref{Test} were upper-bounded using standard large-deviations techniques, i.e., the Chernoff's bound, this cannot be similarly done with the new proposed tester, since calculating the moment generating function (MGF) of the double summation in \eqref{Statistic_new} seems to be hopeless. Thus, alternative tools from large deviations theory have to be invoked. Starting with the type-II probability of error, which is given explicitly by     
\begin{align}
P_{\mbox{\tiny MD}}(\phi_{N}) 
&= \pr_{1|\sigma} \left\{\sum_{i=1}^{n} \sum_{j=1}^{n} \IND\left\{\tilde{\bX}_{i}^{\mathsf{T}} \tilde{\bY}_{j} \geq \theta \right\} \leq  \beta nP \right\}.
\end{align} 
The main difficulty in analyzing this lower tail probability is the statistical dependencies between the indicator random variables. Nevertheless, an existing large deviations result by Janson \cite[Theorem 10]{Janson} provides the appropriate tool to handle the situation in hand. Other recently treated settings where tools from \cite{Janson} proved useful can be found in \cite[Appendixes B and I]{TMWG} and \cite[Appendix K]{TCFG2022}.       

The behavior of the bound in \eqref{MD_upper_bound} depends on $n$, $d$, and $\rho$. As long as $d$ grows sufficiently fast relative to $n$, such that $16nQ(d,\theta) \ll 2$, then the bound decays exponentially with $n$, when the rate function is given by
\begin{align}
\min \left\{ \frac{(1-\beta)^{2}}{2}, \frac{1-\beta}{12} \right\},
\end{align} 
since $P(d,\rho,\theta) \approx 1$ for large $d$ (and fixed $\rho$). On the other hand, if $n$ and $d$ grow in a way such that $16nQ(d,\rho) \gg 2$, then the bound also converges rapidly to zero, but now with an exponent function that depends on $\min\left\{\tfrac{P(d,\rho,\theta)}{Q(d,\theta)},n\right\}$ (note that $P(d,\rho,\theta) \to 1$ and $Q(d,\theta) \to 0$ as $d \to \infty$). 

Regarding the type-I probability of error, the situation is somewhat more complicated, since no general large deviation bounds exist to assess upper tail probabilities of the form 
\begin{align}
P_{\mbox{\tiny FA}}(\phi_{N}) 
&= \pr_{0} \left\{\sum_{i=1}^{n} \sum_{j=1}^{n} \IND\left\{\tilde{\bX}_{i}^{\mathsf{T}} \tilde{\bY}_{j} \geq \theta \right\} \geq \beta nP \right\}.
\end{align} 
Hence, the best we could do is to upper-bound this probability using a generalization of Chebyshev's inequality:  
\begin{align}
\label{FA_Discussion}
P_{\mbox{\tiny FA}}(\phi_{N}) 
\leq \frac{\E_{0}\left[\left(\sum_{i=1}^{n} \sum_{j=1}^{n} \IND\left\{\tilde{\bX}_{i}^{\mathsf{T}} \tilde{\bY}_{j} \geq \theta \right\}\right)^{k}\right]}{\left(\beta nP \right)^{k}},
\end{align}
where $k \in \naturals$ may be optimized to yield the tightest bound. In order to derive (or at least, to upper-bound) the $k$-th moment in \eqref{FA_Discussion}, new techniques, that relies on graph-counting considerations, are developed in Appendix B to prove the general Lemma \ref{Lemma_Square_LD_Integers}, that appears in Appendix A. The relatively interesting fact (at least to the writer of these lines) is the appearing dichotomy between two regions in the parameter space $\{(n,d,\rho)\}$. On the one hand, if $n^{2}Q(d,\theta)$, which is the expected number of dots under $\HypA$, is smaller than 1, then the $k$-th moment in \eqref{FA_Discussion} is proportional to $n^{2}Q(d,\theta)$, for any $k \in \naturals$. In this case, the type-I error probability is relatively small, since the typical number of ``correlated'' pairs of vectors is zero. On the other hand, if $(n,d,\rho)$ are such that $n^{2}Q(d,\theta) \geq 1$, then the $k$-th moment in \eqref{FA_Discussion} is proportional to $[n^{2}Q(d,\theta)]^{k}$. In this case, any increment in $n^{2}Q(d,\theta)$ (due to increasing $n$ or decreasing $\rho$), causes a relatively sharp increment in the type-I error probability.           
Such phenomena, where moments of enumerators (i.e., sums of IID or weakly dependent indicator random variables) undergo phase transitions have already been encountered multiple times, e.g.\ in information theory \cite[Appendix B, Lemma 3]{TMWG}, \cite[p.\ 168]{MERHAV09}, \cite[Appendix A, Lemma 1.1]{BL2014}.

Moving further, we provide theoretical guarantees for the two error probabilities in the permutation recovery part as a function of the parameters $n$, $d$, and $\rho$.
The following result is proved in Appendix C.

\begin{theorem}[Permutation recovery probability bounds]
	\label{Theorem_Alignment_Recovery}
	Let $n,d \in \naturals$ and $\rho,\theta \in (0,1]$ be given. Let $P=P(d,\rho,\theta)$ and $Q=Q(d,\theta)$ as defined above. 
	\begin{enumerate}
		\item Regarding the erroneous recovery error probability in \eqref{ErrorProbability_1},
		\begin{align} \label{ER_Error1}
		P_{\mbox{\scriptsize e,1}}(\psi_{\mbox{\tiny TC}}) \leq \min\left[1,n(1-P) + n(n-1)Q\right],
		\end{align}
		and,
		\begin{align} \label{ER_Error2}
		P_{\mbox{\scriptsize e,1}}(\psi_{\mbox{\tiny TC}})
		\geq \frac{n(1-P)+n(n-1)Q}{\max\{P,1-Q\} + n(1-P) + n(n-1)Q}.
		\end{align}
		\item Regarding the mismatch-undetected error probability in \eqref{ErrorProbability_2},
		\begin{align} \label{MU_Error1}
		P_{\mbox{\scriptsize e,2}}(\psi_{\mbox{\tiny TC}}) \leq \min\left[1,n(n-1) Q (1-P)^{2} (1-Q)^{2n-4}\right].
		\end{align} 
	\end{enumerate}  
\end{theorem}

A few remarks are now in order.

\begin{remark} \normalfont
It is important to note that the proposed algorithm does not necessarily outputs a permutation, since it is primarily based on {\it local} detections of correlated pairs.
In fact, the algorithm outputs a member of the set $\sL_{n}$ (defined in \eqref{Def_Ln}), which can be regarded as a {\it partial-permutation}.  
Although this may be thought of as a major drawback when one would like to have a full alignment recovery of the two databases, we explain why, nonetheless, this is an advantage. 
Assume that the two databases are indeed correlated and that the true permutation is the identity permutation. Furthermore, assume that the two pairs $(\bX_{1},\bY_{1})$ and $(\bX_{2},\bY_{2})$ have been drawn with very low correlation, i.e., that $\tilde{\bX}_{i}^{\mathsf{T}} \tilde{\bY}_{i} \approx 0$, $i=1,2$, and that the rest of the correct pairs have a typical correlation, i.e., $\tilde{\bX}_{i}^{\mathsf{T}} \tilde{\bY}_{i} \approx \rho$, $i=3,\ldots,n$. In such a case, it is likely that an algorithm that aims to recover the entire permutation, will output one out of two possible permutations - the true permutation or an incorrect permutation where $(\bX_{1},\bY_{2})$ and $(\bX_{2},\bY_{1})$ are declared as matching pairs (see Figures \ref{fig:POPULATION1} and \ref{fig:POPULATION2} below). On the other hand, the proposed algorithm will declare only on the $n-2$ pairs with relatively high correlation, while avoiding from ``connecting'' $\bX_{1},\bX_{2}$ to $\bY_{1},\bY_{2}$ in any way (see Figure \ref{fig:POPULATION3}). 
One should pay attention that this event is regarded as an error according to the definition in \eqref{ErrorProbability_1}, because the output is not a full permutation, but on the other hand, this event does not count as an error according to the definition in \eqref{ErrorProbability_2}, since all the estimated vector pairs are correct.  
\end{remark}

\begin{figure}[h!]
	\definecolor{CODEWORD}{rgb}{0,0,0}
	\definecolor{FULL}{rgb}{0.13,0.65,0.13}
	\definecolor{EMPTY}{rgb}{1.0,0,0.22}	
	\begin{subfigure}[b]{0.33\columnwidth}
		\centering 
		\begin{tikzpicture}[scale=0.3]
		\draw[black,very thick] (0,0) -- (12,0);
		\draw[black,very thick] (0,2) -- (12,2);
		\draw[black,very thick] (0,4) -- (12,4);
		\draw[black,very thick] (0,6) -- (12,6);
		\draw[black,very thick] (0,8) -- (12,8);
		\draw[black,very thick] (0,10) -- (12,10);
		\draw[black,very thick] (0,12) -- (12,12);
		
		\draw[black,very thick] (0,0) -- (0,12);
		\draw[black,very thick] (2,0) -- (2,12);
		\draw[black,very thick] (4,0) -- (4,12);
		\draw[black,very thick] (6,0) -- (6,12);
		\draw[black,very thick] (8,0) -- (8,12);
		\draw[black,very thick] (10,0) -- (10,12);
		\draw[black,very thick] (12,0) -- (12,12);
		
		\foreach \x/\xtext in {1/1, 3/2, 5/3, 7/4, 9/5, 11/6}
		\draw[shift={(\x,0)}] (0pt,2pt) -- (0pt,-2pt) node[below] {$\bX_{\xtext}$};
		\foreach \y/\ytext in {1/1, 3/2, 5/3, 7/4, 9/5, 11/6}
		\draw[shift={(0,\y)}] (2pt,0pt) -- (-2pt,0pt) node[left] {$\bY_{\ytext}$};
		
		\node at (1,1) [circle,fill=green!80] {};
		\node at (3,3) [circle,fill=green!80] {};
		\node at (5,5) [circle,fill=black] {};
		\node at (7,7) [circle,fill=black] {};
		\node at (9,9) [circle,fill=black] {};
		\node at (11,11) [circle,fill=black] {};
		\end{tikzpicture}
		\caption{The correct permutation} 
		\label{fig:POPULATION1}
	\end{subfigure}%
	\begin{subfigure}[b]{0.33\columnwidth}
		\centering 
				\begin{tikzpicture}[scale=0.3]
		\draw[black,very thick] (0,0) -- (12,0);
		\draw[black,very thick] (0,2) -- (12,2);
		\draw[black,very thick] (0,4) -- (12,4);
		\draw[black,very thick] (0,6) -- (12,6);
		\draw[black,very thick] (0,8) -- (12,8);
		\draw[black,very thick] (0,10) -- (12,10);
		\draw[black,very thick] (0,12) -- (12,12);
		
		\draw[black,very thick] (0,0) -- (0,12);
		\draw[black,very thick] (2,0) -- (2,12);
		\draw[black,very thick] (4,0) -- (4,12);
		\draw[black,very thick] (6,0) -- (6,12);
		\draw[black,very thick] (8,0) -- (8,12);
		\draw[black,very thick] (10,0) -- (10,12);
		\draw[black,very thick] (12,0) -- (12,12);
		
		\foreach \x/\xtext in {1/1, 3/2, 5/3, 7/4, 9/5, 11/6}
		\draw[shift={(\x,0)}] (0pt,2pt) -- (0pt,-2pt) node[below] {$\bX_{\xtext}$};
		\foreach \y/\ytext in {1/1, 3/2, 5/3, 7/4, 9/5, 11/6}
		\draw[shift={(0,\y)}] (2pt,0pt) -- (-2pt,0pt) node[left] {$\bY_{\ytext}$};
		
		\node at (1,3) [circle,fill=red!80] {};
		\node at (3,1) [circle,fill=red!80] {};
		\node at (5,5) [circle,fill=black] {};
		\node at (7,7) [circle,fill=black] {};
		\node at (9,9) [circle,fill=black] {};
		\node at (11,11) [circle,fill=black] {};
		\end{tikzpicture}
		\caption{An incorrect permutation} 
		\label{fig:POPULATION2}
	\end{subfigure}%
	\begin{subfigure}[b]{0.33\columnwidth}
		\centering 
		\begin{tikzpicture}[scale=0.3]
		\draw[black,very thick] (0,0) -- (12,0);
		\draw[black,very thick] (0,2) -- (12,2);
		\draw[black,very thick] (0,4) -- (12,4);
		\draw[black,very thick] (0,6) -- (12,6);
		\draw[black,very thick] (0,8) -- (12,8);
		\draw[black,very thick] (0,10) -- (12,10);
		\draw[black,very thick] (0,12) -- (12,12);
		
		\draw[black,very thick] (0,0) -- (0,12);
		\draw[black,very thick] (2,0) -- (2,12);
		\draw[black,very thick] (4,0) -- (4,12);
		\draw[black,very thick] (6,0) -- (6,12);
		\draw[black,very thick] (8,0) -- (8,12);
		\draw[black,very thick] (10,0) -- (10,12);
		\draw[black,very thick] (12,0) -- (12,12);
		
		\foreach \x/\xtext in {1/1, 3/2, 5/3, 7/4, 9/5, 11/6}
		\draw[shift={(\x,0)}] (0pt,2pt) -- (0pt,-2pt) node[below] {$\bX_{\xtext}$};
		\foreach \y/\ytext in {1/1, 3/2, 5/3, 7/4, 9/5, 11/6}
		\draw[shift={(0,\y)}] (2pt,0pt) -- (-2pt,0pt) node[left] {$\bY_{\ytext}$};
		
		\node at (1,1) [circle,fill=gray!60] {};
		\node at (3,3) [circle,fill=gray!60] {};
		\node at (5,5) [circle,fill=black] {};
		\node at (7,7) [circle,fill=black] {};
		\node at (9,9) [circle,fill=black] {};
		\node at (11,11) [circle,fill=black] {};
		\end{tikzpicture}
		\caption{A partial alignment} 
		\label{fig:POPULATION3}
	\end{subfigure}%
	\caption{Outputs of different alignment algorithms when two pairs of vectors have been drawn with relatively low correlation, while the rest with a typical correlation. An algorithm that outputs a perfect permutation will output one of the above permutations. In this case, the proposed algorithm will output a member of $\sL_{n}$ which is not a perfect permutation.} 
	\label{fig:POPULATIONS}
\end{figure}
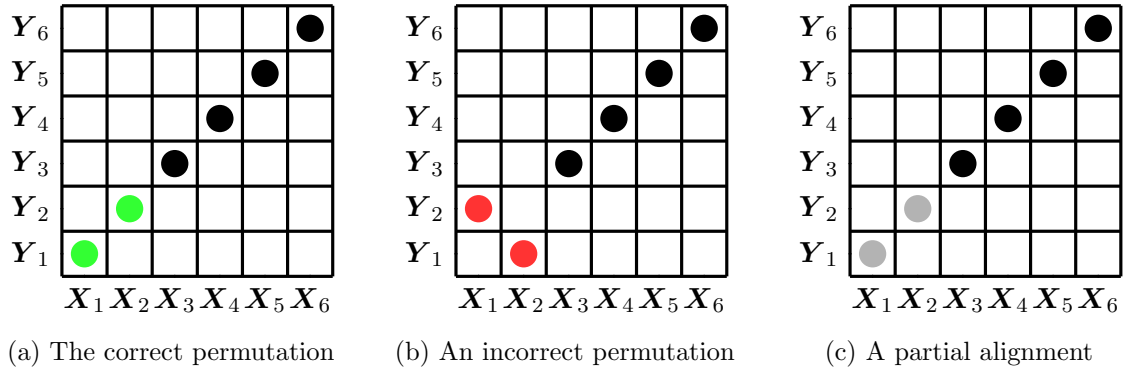

\begin{remark} \normalfont
The conditions for convergence of the error probabilities are different in the two problems; in the hypothesis testing problem and the permutation recovery problem. 
In order to achieve low type-I error probability, we have to require that
\begin{align} \label{Cond}
n \cdot Q(d,\theta) \xrightarrow{d,n \to \infty} 0,
\end{align}
while for the type-II error probability to be low, it is enough to require that $n \cdot Q(d,\theta)$ is only bounded.     
For the alignment error probability to converge to zero, the condition in \eqref{Cond} is not enough, but we must require that  
\begin{align}
n^{2} \cdot Q(d,\theta) \xrightarrow{d,n \to \infty} 0.
\end{align}
This phenomenon that permutation recovery is generally harder than correlation detection between databases has already been pointed out in \cite{KN2022}.	
\end{remark}

\begin{remark} \normalfont
	The design parameter $\theta$ can be tuned in order to trade-off between the reliability and the size of the partial alignment recovery. For a fixed value of $\rho$, if $\theta$ is relatively low, then typically almost all of the dependent vector pairs will cross the threshold $\theta$, but also many independent pairs. Thus, after the cleaning procedure of the Threshold-and-Clean algorithm, many rows and columns will be erased, so the output size will be relatively low, but with a high reliability. 
	On the other extreme, when $\theta$ is relatively high, then some fraction of the dependent vector pairs will cross the threshold $\theta$, and typically almost none of the independent pairs, such that the reliability is also high in this case. 
	In the intermediate range of $\theta$ values, most of the dependent vector pairs cross the threshold, such that the output size is relatively high, but chances are also high that an independent pair will cross the threshold and not be erased in the cleaning procedure, thus causing an undetected error.           	
\end{remark}  

In order to demonstrate this trade-off between the reliability and the output size, we should be able to calculate the error probability bounds of Theorem \ref{Theorem_Alignment_Recovery}. Hence, we first present explicit expressions for $P(d,\rho,\theta)$ and $Q(d,\theta)$.
Regarding $Q(d,\theta)$, which is the probability for an independent pair of vectors to cross a threshold $\theta$, consider the following. Let $d \in \naturals$ and $\theta \in (0,1)$, as well as
\begin{align}
(\bX,\bY) \sim \calN^{\otimes d}\left(\left[\begin{matrix} 0 \\ 0
\end{matrix}\right], \left[\begin{matrix} 1 & 0 \\ 0 & 1 \end{matrix}\right] \right),
\end{align}  
and 
\begin{align}
\tilde{\bX} = \frac{\bX}{\|\bX\|},~~~\tilde{\bY} = \frac{\bY}{\|\bY\|}.
\end{align}
Thanks to symmetry, $\pr\{\tilde{\bX}^{\mathsf{T}} \tilde{\bY} \geq \theta\}$ is equal to the probability that a uniformly distributed vector on a $d$-dimensional hypersphere of unit norm will fall in a $d$-dimensional hyper-spherical cap with half-angle $\varphi = \arccos(\theta)$. This probability is given by \cite{Li}
\begin{align} \label{Q_CALC}
Q(d,\theta)
=\frac{1}{2}\frac{B(\sin^{2}(\varphi); \frac{d-1}{2},\frac{1}{2})}{B(\frac{d-1}{2},\frac{1}{2})}
= \frac{1}{2}\frac{B(1-\theta^{2}; \frac{d-1}{2},\frac{1}{2})}{B(\frac{d-1}{2},\frac{1}{2})},
\end{align}
where the complete beta function 
\begin{align}
B(a,b) = \int_{0}^{1} t^{a-1}(1-t)^{b-1}\dint t,
\end{align}
and the incomplete beta function
\begin{align}
B(x; a,b) = \int_{0}^{x} t^{a-1}(1-t)^{b-1}\dint t.
\end{align}

Regarding $P(d,\rho,\theta)$, which is the probability for a matched pair to cross a threshold, we make a few definitions. For $\rho \in (0,1)$ and $\theta \in (0,1)$, define the constants
\begin{align}
\alpha(\rho) &= \frac{1-\rho^2}{\rho^2}, \\
\beta(\rho,\theta) &= \frac{1-\rho^2}{\rho^2(1-\theta^2)},
\end{align}
and the functions
\begin{align}
F_{1}(u) &= -\frac{\rho  (1-\theta^{2})}{\sqrt{1-\rho^{2}}}\sqrt{u} - \theta\sqrt{1 - \frac{\rho^{2}(1-\theta^{2})}{1-\rho^{2}} u}, \\
F_{2}(u) &= -\frac{\rho  (1-\theta^{2})}{\sqrt{1-\rho^{2}}}\sqrt{u} + \theta\sqrt{1 - \frac{\rho^{2}(1-\theta^{2})}{1-\rho^{2}} u}.
\end{align}
Also, for any $d\in \naturals$, define the two PDFs
\begin{align}
f_{U}(u) = \frac{1}{B\left(\frac{d}{2},\frac{d}{2}\right)}  u^{d/2-1}\left(1+u\right)^{-d},~u\geq 0,
\end{align} 
and 
\begin{align}
g_{S}(s) = \frac{(1-s^{2})^{\frac{d-3}{2}}}{B(\frac{d-1}{2},\frac{1}{2})},~s \in [-1,1].
\end{align}

The following general lemma is proved in Appendix D.

\begin{lemma} \label{Lemma_P_Bound} 
	Let $d \in \naturals$, $\rho \in (0,1)$ and $\theta \in (0,1)$. Let
	\begin{align}
	(\bX,\bY) \sim \calN^{\otimes d}\left(\left[\begin{matrix} 0 \\ 0
	\end{matrix}\right], \left[\begin{matrix} 1 & \rho \\ \rho & 1 \end{matrix}\right] \right),
	\end{align}  
	and 
	\begin{align}
	\tilde{\bX} = \frac{\bX}{\|\bX\|},~~~\tilde{\bY} = \frac{\bY}{\|\bY\|}.
	\end{align}
	Then,
	\begin{align} \label{P_CALC}
	P(d,\rho,\theta) 
	&= \int_{0}^{\alpha(\rho)} \left[\int_{F_{2}(u)}^{1} g_{S}(s) \dint s \right] f_{U}(u) \dint u \nn \\ 
	&~~~+ \int_{\alpha(\rho)}^{\beta(\rho,\theta)} \left[\int_{-1}^{F_{1}(u)} g_{S}(s) \dint s + \int_{F_{2}(u)}^{1} g_{S}(s) \dint s\right] f_{U}(u) \dint u + \int_{\beta(\rho,\theta)}^{\infty} f_{U}(u) \dint u.
	\end{align}
\end{lemma}

In Figure \ref{fig:Trade-off} below, we demonstrate the trade-off between the reliability and the output size for the parameters $n=200$, $d=50$, and $\rho=0.7$. The output size has been evaluated numerically; for each value of $\theta$, we empirically averaged the output size over 50 database realizations and divided by $n$ (we denote this quantity by $\overline{\sR}_{n,d}(\rho,\theta)$ -- the averaged success rate). As can be seen in Figure \ref{fig:Trade-off}, when $\theta$ is relatively low or relatively high, the reliability is high but the success rate is very close to zero. On the other hand, at the intermediate $\theta$ values, the output size is maximized but the reliability is minimized.    

\begin{figure}[h!]	
	\begin{subfigure}[b]{0.5\columnwidth}
		\centering 
		\begin{tikzpicture}[scale=0.9]
		\begin{axis}[
		disabledatascaling,
		scaled x ticks=false,
		xticklabel style={/pgf/number format/fixed,
			/pgf/number format/precision=3},
		scaled y ticks=false,
		yticklabel style={/pgf/number format/fixed,
			/pgf/number format/precision=3},
		xlabel={$\theta$},
		xmin=0.2, xmax=0.95,
		ymin=0, ymax=25,
		legend pos=north west,
		]
		
		\addplot[smooth,color=black!20!orange,thick]
		table[row sep=crcr] 
		{
		0.1 53.938529 \\ 
		0.11 49.173263 \\ 
		0.12 44.771553 \\ 
		0.13 40.716953 \\ 
		0.14 36.992655 \\ 
		0.15 33.581532 \\ 
		0.16 30.466161 \\ 
		0.17 27.628862 \\ 
		0.18 25.051742 \\ 
		0.19 22.71679 \\ 
		0.2 20.606007 \\ 
		0.21 18.701605 \\ 
		0.22 16.986259 \\ 
		0.23 15.443388 \\ 
		0.24 14.057411 \\ 
		0.25 12.813944 \\ 
		0.26 11.699872 \\ 
		0.27 10.703301 \\ 
		0.28 9.813427 \\ 
		0.29 9.020357 \\ 
		0.3 8.314944 \\ 
		0.31 7.688661 \\ 
		0.32 7.133529 \\ 
		0.33 6.64208 \\ 
		0.34 6.207355 \\ 
		0.35 5.82291 \\ 
		0.36 5.482838 \\ 
		0.37 5.181769 \\ 
		0.38 4.914877 \\ 
		0.39 4.677869 \\ 
		0.4 4.466971 \\ 
		0.41 4.278904 \\ 
		0.42 4.110851 \\ 
		0.43 3.960427 \\ 
		0.44 3.825636 \\ 
		0.45 3.704839 \\ 
		0.46 3.596714 \\ 
		0.47 3.500219 \\ 
		0.48 3.414554 \\ 
		0.49 3.339137 \\ 
		0.5 3.273565 \\ 
		0.51 3.217594 \\ 
		0.52 3.171115 \\ 
		0.53 3.134133 \\ 
		0.54 3.10675 \\ 
		0.55 3.089151 \\ 
		0.56 3.081597 \\ 
		0.57 3.08441 \\ 
		0.58 3.097969 \\ 
		0.59 3.122706 \\ 
		0.6 3.159099 \\ 
		0.61 3.207675 \\ 
		0.62 3.269004 \\ 
		0.63 3.343701 \\ 
		0.64 3.432426 \\ 
		0.65 3.535883 \\ 
		0.66 3.654824 \\ 
		0.67 3.790045 \\ 
		0.68 3.942394 \\ 
		0.69 4.112762 \\ 
		0.7 4.302092 \\ 
		0.71 4.511372 \\ 
		0.72 4.741637 \\ 
		0.73 4.993967 \\ 
		0.74 5.269484 \\ 
		0.75 5.569354 \\ 
		0.76 5.894782 \\ 
		0.77 6.247026 \\ 
		0.78 6.627399 \\ 
		0.79 7.037304 \\ 
		0.8 7.47827 \\ 
		0.81 7.952024 \\ 
		0.82 8.460584 \\ 
		0.83 9.006405 \\ 
		0.84 9.592549 \\ 
		0.85 10.222912 \\ 
		0.86 10.902494 \\ 
		0.87 11.637731 \\ 
		0.88 12.436917 \\ 
		0.89 13.310777 \\ 
		0.9 14.273308 \\ 
		0.91 15.343069 \\ 
		0.92 16.545267 \\ 
		0.93 17.915279 \\ 
		0.94 19.504926 \\ 
		0.95 21.394557 \\ 
		0.96 23.71878 \\
		};
		\legend{}
		\addlegendentry{$-\log_{10}P_{\mbox{\small e,2}}^{\mbox{\tiny UB}}$}	
		
		\end{axis}
		\end{tikzpicture}
		\caption{The upper bound on $P_{\mbox{\small e,2}}(\psi_{\mbox{\tiny TC}})$ in \eqref{MU_Error1}.}
	\end{subfigure}%
	\begin{subfigure}[b]{0.5\columnwidth}
		\centering 
		\begin{tikzpicture}[scale=0.9]
		\begin{axis}[
		disabledatascaling,
		scaled x ticks=false,
		xticklabel style={/pgf/number format/fixed,
			/pgf/number format/precision=3},
		scaled y ticks=false,
		yticklabel style={/pgf/number format/fixed,
			/pgf/number format/precision=3},
		xlabel={$\theta$},
		xmin=0.20, xmax=0.95,
		ymin=0, ymax=1.0,
		legend pos=north east
		]		
		
		\addplot[smooth,color=violet,thick,mark=o]
		table[row sep=crcr]
		{ 
		0.2 0.0 \\ 
		0.25 0.0 \\ 
		0.3 0.0019 \\ 
		0.35 0.1085 \\ 
		0.4 0.4842 \\ 
		0.45 0.8409 \\ 
		0.5 0.9527 \\ 
		0.55 0.9575 \\ 
		0.6 0.9027 \\ 
		0.65 0.7566 \\ 
		0.7 0.5256 \\ 
		0.75 0.2438 \\ 
		0.8 0.0624 \\ 
		0.85 0.0036 \\ 
		0.9 0.0 \\ 
		0.95 0.0 \\  		
		};
		\legend{}
		\addlegendentry{$\overline{\sR}_{n,d}(\rho,\theta)$}
		
		\end{axis}
		\end{tikzpicture}
		\caption{The average output size.}
	\end{subfigure}%
\caption{A demonstration for the trade-off between the reliability and the output size.} 
\label{fig:Trade-off}
\end{figure}
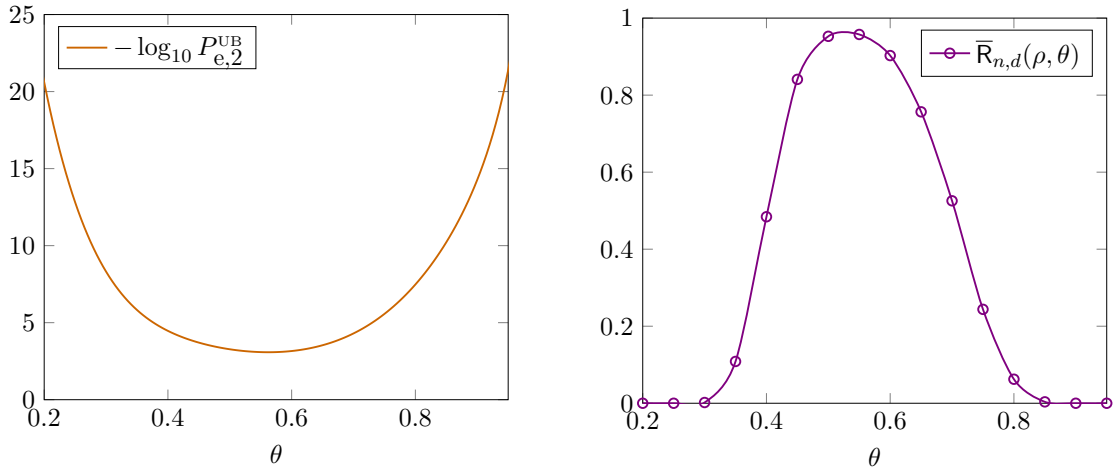

\section{Improved Partial Alignment Recovery} \label{Sec5}

The Threshold-and-Clean algorithm that was proposed in Section \ref{Sec4} is relatively efficient; its computational complexity grows like $\calO(n^{2})$, while the complexity of optimal full alignment (via the Hungarian algorithm) is $\calO(n^{3})$. In this section, we improve upon the Threshold-and-Clean algorithm by allowing an increased complexity; we propose a two-step algorithm that is based on the Hungarian algorithm. At the first step, we resort to full alignment recovery, even if we know a-priori that the two databases have only partial overlap. Then, after having an estimated full alignment, we order the $n$ empirical correlation coefficients in descending order, and consider only the $\mathsf{R}\%$ mostly correlated pairs as the final partial alignment recovery.   
We relate to this algorithm as `Maximum-Path'.
For $n=200$, $d=50$, and $\rho=0.6$, we illustrate this procedure in Figure \ref{fig:Improved_partial_alignment} below.

\begin{figure}[h!]	
	\centering 
	\begin{tikzpicture}[scale=0.9]
	\draw[black,thick,dash pattern={on 3pt off 2pt}] (4.425,0) -- (4.425,2.4);
	\draw[black,thick,dash pattern={on 3pt off 2pt}] (4.425,3.1) -- (4.425,4.8876);
	\draw[thick,<->] (0,4) -- (4.425,4);
	\node [anchor=center] at (2.2125,4.2) {$30\%$ pairs};
	
	\draw[black,thick,dash pattern={on 3pt off 2pt}] (7.425,0) -- (7.425,4.2228);
	\draw[thick,<->] (0,2.5) -- (7.425,2.5);
	\node [anchor=center] at (3.7125,2.7) {$50\%$ pairs};
	
	\draw[black,thick,dash pattern={on 3pt off 2pt}] (10.425,0) -- (10.425,3.624);
	\draw[thick,<->] (0,1) -- (10.425,1);
	\node [anchor=center] at (5.925,1.2) {$70\%$ pairs};

	\begin{axis}[
	disabledatascaling,
	x=0.075cm,
	y=12cm,
	scaled x ticks=false,
	xticklabel style={/pgf/number format/fixed,
		/pgf/number format/precision=3},
	scaled y ticks=false,
	yticklabel style={/pgf/number format/fixed,
		/pgf/number format/precision=3},
	xlabel={Pair number},
	ylabel={$\hat{\rho}$},
	xmin=1, xmax=200,
	ymin=0.25, ymax=0.85,
	legend pos=north west
	]
	
	\addplot+[
	only marks,
	scatter,
	mark=halfcircle*,
	mark size=2.9pt]
	table[row sep=crcr] 
	{
		1 0.8146 \\ 
		2 0.8093 \\ 
		3 0.7918 \\ 
		4 0.7815 \\ 
		5 0.768 \\ 
		6 0.7539 \\ 
		7 0.7534 \\ 
		8 0.7498 \\ 
		9 0.7483 \\ 
		10 0.7445 \\ 
		11 0.7423 \\ 
		12 0.7356 \\ 
		13 0.7346 \\ 
		14 0.7246 \\ 
		15 0.7233 \\ 
		16 0.7205 \\ 
		17 0.72 \\ 
		18 0.7144 \\ 
		19 0.714 \\ 
		20 0.7139 \\ 
		21 0.7135 \\ 
		22 0.7131 \\ 
		23 0.713 \\ 
		24 0.7125 \\ 
		25 0.7091 \\ 
		26 0.7076 \\ 
		27 0.7044 \\ 
		28 0.7034 \\ 
		29 0.6984 \\ 
		30 0.6968 \\ 
		31 0.6963 \\ 
		32 0.6907 \\ 
		33 0.6892 \\ 
		34 0.6874 \\ 
		35 0.6866 \\ 
		36 0.6858 \\ 
		37 0.6849 \\ 
		38 0.6821 \\ 
		39 0.682 \\ 
		40 0.6782 \\ 
		41 0.6769 \\ 
		42 0.6765 \\ 
		43 0.676 \\ 
		44 0.6755 \\ 
		45 0.674 \\ 
		46 0.6736 \\ 
		47 0.6718 \\ 
		48 0.6707 \\ 
		49 0.6699 \\ 
		50 0.6674 \\ 
		51 0.6666 \\ 
		52 0.6653 \\ 
		53 0.6652 \\ 
		54 0.6646 \\ 
		55 0.6636 \\ 
		56 0.6633 \\ 
		57 0.6624 \\ 
		58 0.6623 \\ 
		59 0.6595 \\ 
		60 0.6573 \\ 
		61 0.6567 \\ 
		62 0.6547 \\ 
		63 0.6519 \\ 
		64 0.6512 \\ 
		65 0.651 \\ 
		66 0.6508 \\ 
		67 0.6499 \\ 
		68 0.6485 \\ 
		69 0.6478 \\ 
		70 0.6456 \\ 
		71 0.644 \\ 
		72 0.6435 \\ 
		73 0.6431 \\ 
		74 0.6405 \\ 
		75 0.6402 \\ 
		76 0.6401 \\ 
		77 0.6356 \\ 
		78 0.6349 \\ 
		79 0.6349 \\ 
		80 0.6326 \\ 
		81 0.6288 \\ 
		82 0.6269 \\ 
		83 0.6257 \\ 
		84 0.6247 \\ 
		85 0.622 \\ 
		86 0.6197 \\ 
		87 0.6186 \\ 
		88 0.616 \\ 
		89 0.6159 \\ 
		90 0.6153 \\ 
		91 0.6142 \\ 
		92 0.6102 \\ 
		93 0.6085 \\ 
		94 0.6079 \\ 
		95 0.6073 \\ 
		96 0.6069 \\ 
		97 0.606 \\ 
		98 0.6051 \\ 
		99 0.6022 \\ 
		100 0.6019 \\ 
		101 0.6008 \\ 
		102 0.5996 \\ 
		103 0.5968 \\ 
		104 0.5965 \\ 
		105 0.5948 \\ 
		106 0.5944 \\ 
		107 0.5939 \\ 
		108 0.5929 \\ 
		109 0.5921 \\ 
		110 0.589 \\ 
		111 0.589 \\ 
		112 0.5887 \\ 
		113 0.5875 \\ 
		114 0.5865 \\ 
		115 0.5849 \\ 
		116 0.5845 \\ 
		117 0.5839 \\ 
		118 0.5816 \\ 
		119 0.5796 \\ 
		120 0.5792 \\ 
		121 0.5764 \\ 
		122 0.5759 \\ 
		123 0.5752 \\ 
		124 0.5724 \\ 
		125 0.5707 \\ 
		126 0.57 \\ 
		127 0.5691 \\ 
		128 0.5687 \\ 
		129 0.5673 \\ 
		130 0.5655 \\ 
		131 0.5634 \\ 
		132 0.561 \\ 
		133 0.5596 \\ 
		134 0.5569 \\ 
		135 0.5564 \\ 
		136 0.5563 \\ 
		137 0.5562 \\ 
		138 0.5559 \\ 
		139 0.5552 \\ 
		140 0.552 \\ 
		141 0.5484 \\ 
		142 0.5468 \\ 
		143 0.5464 \\ 
		144 0.5452 \\ 
		145 0.5445 \\ 
		146 0.5442 \\ 
		147 0.5421 \\ 
		148 0.5413 \\ 
		149 0.5397 \\ 
		150 0.5381 \\ 
		151 0.5378 \\ 
		152 0.5364 \\ 
		153 0.5337 \\ 
		154 0.5304 \\ 
		155 0.5267 \\ 
		156 0.5254 \\ 
		157 0.5247 \\ 
		158 0.5246 \\ 
		159 0.5221 \\ 
		160 0.5216 \\ 
		161 0.5175 \\ 
		162 0.5138 \\ 
		163 0.5137 \\ 
		164 0.5115 \\ 
		165 0.5081 \\ 
		166 0.5049 \\ 
		167 0.5032 \\ 
		168 0.5025 \\ 
		169 0.501 \\ 
		170 0.4991 \\ 
		171 0.4952 \\ 
		172 0.4948 \\ 
		173 0.4944 \\ 
		174 0.4911 \\ 
		175 0.4873 \\ 
		176 0.4855 \\ 
		177 0.4828 \\ 
		178 0.4779 \\ 
		179 0.4777 \\ 
		180 0.4777 \\ 
		181 0.473 \\ 
		182 0.4679 \\ 
		183 0.4609 \\ 
		184 0.4598 \\ 
		185 0.4476 \\ 
		186 0.4455 \\ 
		187 0.4439 \\ 
		188 0.4433 \\ 
		189 0.4334 \\ 
		190 0.4247 \\ 
		191 0.4245 \\ 
		192 0.4129 \\ 
		193 0.4097 \\ 
		194 0.393 \\ 
		195 0.3911 \\ 
		196 0.3803 \\ 
		197 0.363 \\ 
		198 0.3325 \\ 
		199 0.3126 \\ 
		200 0.2999 \\
	};
	\end{axis}
	\end{tikzpicture}
	\caption{An illustration for the improved partial alignment recovery.} 
	\label{fig:Improved_partial_alignment}
\end{figure}
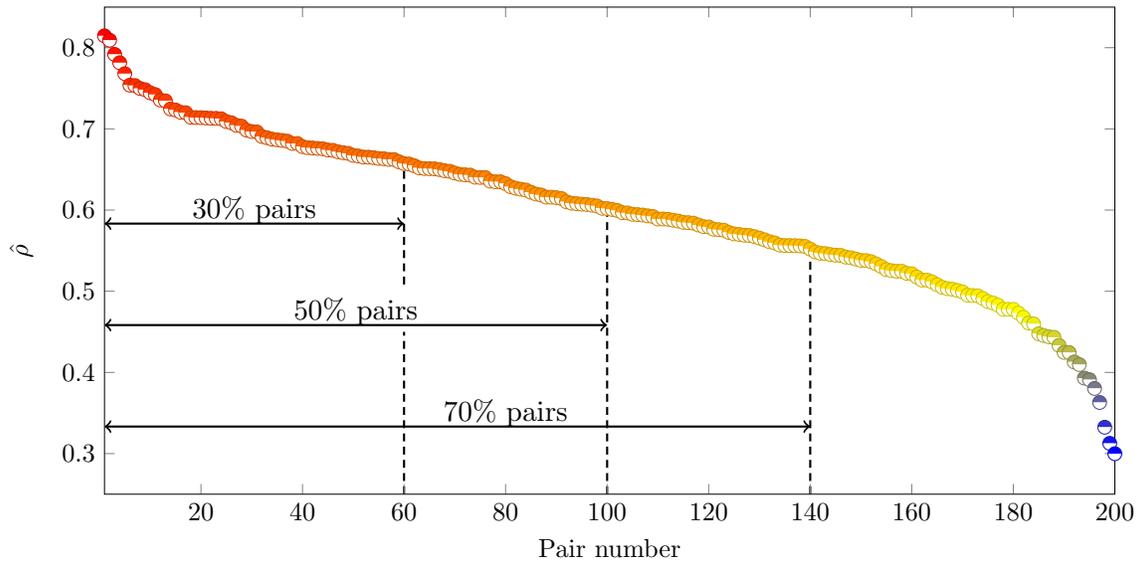

We now compare between three algorithms for $n=200$, $d=50$, and $\mathsf{R}=0.3$:
\begin{enumerate}
	\item Optimal full alignment recovery,
	\item Threshold-and-Clean partial alignment recovery,
	\item Maximum-Path partial alignment recovery.
\end{enumerate}

For the optimal full recovery as well as for the Maximum-Path algorithm, we rely on computer simulations in order to estimate their error probabilities, while for the Threshold-and-Clean algorithm, we use the error probability bound from Theorem \ref{Theorem_Alignment_Recovery}. As can be seen in Figure \ref{fig:Comparison} below, the reliability of the Threshold-and-Clean algorithm is about 10 times higher than the reliability of optimal full alignment recovery, and furthermore, the reliability of the Maximum-Path algorithm is about 10 times higher than the reliability of the Threshold-and-Clean algorithm.

\begin{figure}[h!]	
	\centering 
	\begin{tikzpicture}[scale=1.2]
	\begin{axis}[
	disabledatascaling,
	scaled x ticks=false,
	xticklabel style={/pgf/number format/fixed,
		/pgf/number format/precision=3},
	scaled y ticks=false,
	yticklabel style={/pgf/number format/fixed,
		/pgf/number format/precision=3},
	xlabel={$\rho$},
	ylabel={$-\log_{10}P_{\mbox{\small e}}$},
	xmin=0.35, xmax=0.6,
	ymin=0, ymax=3,
	legend pos=north west
	]
	
	\addplot[smooth,color=black!50!green,thick,mark=o]
	table[row sep=crcr]
	{
		0.30 0 \\
		0.32 0 \\
		0.34 0 \\
		0.36 0 \\
		0.38 0 \\
		0.4  0 \\
		0.42 0 \\
		0.44 0 \\
		0.46 0 \\
		0.48 0 \\
		0.5 0.017729 \\ 
		0.52 0.128427 \\
		0.54 0.39234 \\
		0.56 0.71533 \\
		0.58 1.056592 \\
		0.6 1.497027 \\  	
	};
	\legend{}
	\addlegendentry{$\overline{P}_{\mbox{\small e,ML}}$}
	
	\addplot[smooth,color=black!20!orange,thick,mark=square]
	table[row sep=crcr] 
	{
		0.44 0.0 \\ 
		0.46 0.046686 \\ 
		0.48 0.36381 \\ 	
		0.5 0.676993 \\ 
		0.52 1.024117 \\ 
		0.54 1.378391 \\ 
		0.56 1.754738 \\ 
		0.58 2.172845 \\ 
		0.6 2.600495 \\
	};
	\addlegendentry{$P_{\mbox{\small e,TC}}^{\mbox{\tiny UB}}$}
	
	\addplot[smooth,color=black!20!purple,thick,mark=pentagon]
	table[row sep=crcr] 
	{
		0.35 0.341989 \\ 
		0.4 0.744727 \\ 
		0.45 1.34 \\		
		0.5 1.98991 \\ 
		0.55 2.897338 \\ 
	};
	\addlegendentry{$\overline{P}_{\mbox{\small e,MP}}$}
	
	\end{axis}
	\end{tikzpicture} 
	\caption{Comparison between the error probabilities of three alignment recovery algorithms.} 
	\label{fig:Comparison}
\end{figure}
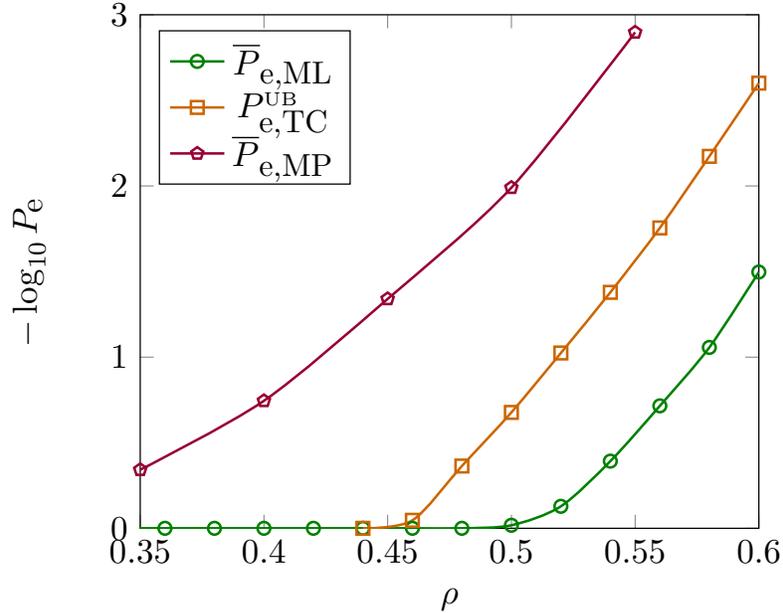

\section{Reduced Complexity Full Alignment Recovery} \label{Sec6}

The algorithms proposed in Sections \ref{Sec4} and \ref{Sec5} aim to find a partial alignment recovery. While the outputs of these algorithms are quite reliable, as proved in Theorem 2, it may be the case that one would like to have a full alignment recovery, and not just a partial recovery. For full alignment recovery, the optimal ML estimator can be realized via the Hungarian algorithm. As already mentioned before, the complexity of this algorithm is $\calO(n^{3})$, thus when $n$ is on the order of thousands or more, the amount of required computations may be quite demanding. 
Our current objective is to propose an improved algorithm, that allows to trade-off between reliability and complexity. 

\begin{figure}[h!]
	\centering
	\begin{tikzpicture}[scale=0.3]
	

	\draw[black,very thick] (0,0) -- (20,0);
	\draw[black,very thick] (0,2) -- (20,2);
	\draw[black,very thick] (0,4) -- (20,4);
	\draw[black,very thick] (0,6) -- (20,6);
	\draw[black,very thick] (0,8) -- (20,8);
	\draw[black,very thick] (0,10) -- (20,10);
	\draw[black,very thick] (0,12) -- (20,12);
	\draw[black,very thick] (0,14) -- (20,14);
	\draw[black,very thick] (0,16) -- (20,16);
	\draw[black,very thick] (0,18) -- (20,18);
	\draw[black,very thick] (0,20) -- (20,20);
	
	\draw[black,very thick] (0,0) -- (0,20);
	\draw[black,very thick] (2,0) -- (2,20);
	\draw[black,very thick] (4,0) -- (4,20);
	\draw[black,very thick] (6,0) -- (6,20);
	\draw[black,very thick] (8,0) -- (8,20);
	\draw[black,very thick] (10,0) -- (10,20);
	\draw[black,very thick] (12,0) -- (12,20);
	\draw[black,very thick] (14,0) -- (14,20);
	\draw[black,very thick] (16,0) -- (16,20);
	\draw[black,very thick] (18,0) -- (18,20);
	\draw[black,very thick] (20,0) -- (20,20);
	
	\foreach \x/\xtext in {1/1, 3/2, 5/3, 7/4, 9/5, 11/6, 13/7, 15/8, 17/9, 19/10}
	\draw[shift={(\x,0)}] (0pt,2pt) -- (0pt,-2pt) node[below] {$\bX_{\xtext}$};
	\foreach \y/\ytext in {1/1, 3/2, 5/3, 7/4, 9/5, 11/6, 13/7, 15/8, 17/9, 19/10}
	\draw[shift={(0,\y)}] (2pt,0pt) -- (-2pt,0pt) node[left] {$\bY_{\ytext}$};
	
	\node at (1,1) [circle,fill=blue!80] {};
	\node at (3,3) [circle,fill=blue!80] {};
	\node at (5,5) [circle,fill=blue!80] {};
	\node at (7,7) [circle,fill=blue!30] {};
	\node at (9,9) [circle,fill=blue!30] {};
	\node at (11,11) [circle,fill=blue!30] {};
	\node at (13,13) [circle,fill=blue!30] {};
	\node at (15,15) [circle,fill=blue!30] {};
	\node at (17,17) [circle,fill=blue!30] {};
	\node at (19,19) [circle,fill=blue!30] {};
	
	\node at (11,7) [circle,fill=red!30] {};
	\node at (17,9) [circle,fill=red!30] {};
	\node at (19,15) [circle,fill=red!30] {};
	\node at (13,19) [circle,fill=red!30] {};
	\node at (9,13) [circle,fill=red!30] {};
	
	\node at (10,22)[]{\underline{Step I: Threshold and clean:}};
	\end{tikzpicture}
	\hspace{0.15in}
	\begin{tikzpicture}[scale=0.3]
	
	\draw[fill=black!10!white] (0,0) -- (0,6) -- (20,6) -- (20,0) -- (0,0);
	\draw[fill=black!10!white] (0,0) -- (0,20) -- (6,20) -- (6,0) -- (0,0);

	\draw[black,very thick] (0,0) -- (20,0);
	\draw[black,very thick] (0,2) -- (20,2);
	\draw[black,very thick] (0,4) -- (20,4);
	\draw[black,very thick] (0,6) -- (20,6);
	\draw[black,very thick] (0,8) -- (20,8);
	\draw[black,very thick] (0,10) -- (20,10);
	\draw[black,very thick] (0,12) -- (20,12);
	\draw[black,very thick] (0,14) -- (20,14);
	\draw[black,very thick] (0,16) -- (20,16);
	\draw[black,very thick] (0,18) -- (20,18);
	\draw[black,very thick] (0,20) -- (20,20);
	
	\draw[black,very thick] (0,0) -- (0,20);
	\draw[black,very thick] (2,0) -- (2,20);
	\draw[black,very thick] (4,0) -- (4,20);
	\draw[black,very thick] (6,0) -- (6,20);
	\draw[black,very thick] (8,0) -- (8,20);
	\draw[black,very thick] (10,0) -- (10,20);
	\draw[black,very thick] (12,0) -- (12,20);
	\draw[black,very thick] (14,0) -- (14,20);
	\draw[black,very thick] (16,0) -- (16,20);
	\draw[black,very thick] (18,0) -- (18,20);
	\draw[black,very thick] (20,0) -- (20,20);
	
	\foreach \x/\xtext in {1/1, 3/2, 5/3, 7/4, 9/5, 11/6, 13/7, 15/8, 17/9, 19/10}
	\draw[shift={(\x,0)}] (0pt,2pt) -- (0pt,-2pt) node[below] {$\bX_{\xtext}$};
	\foreach \y/\ytext in {1/1, 3/2, 5/3, 7/4, 9/5, 11/6, 13/7, 15/8, 17/9, 19/10}
	\draw[shift={(0,\y)}] (2pt,0pt) -- (-2pt,0pt) node[left] {$\bY_{\ytext}$};
	
	\node at (7,7) [circle,fill=blue!80] {};
	\node at (9,9) [circle,fill=blue!80] {};
	\node at (11,11) [circle,fill=blue!80] {};
	\node at (13,13) [circle,fill=blue!80] {};
	\node at (15,15) [circle,fill=blue!80] {};
	\node at (17,17) [circle,fill=blue!80] {};
	\node at (19,19) [circle,fill=blue!80] {};
	
	\node at (10,22)[]{\underline{Step II: Optimal matching:}};
	\end{tikzpicture}
	\caption{An illustration for the efficient full alignment recovery.}\label{fig:Efficient_Full_Recovery}
\end{figure}
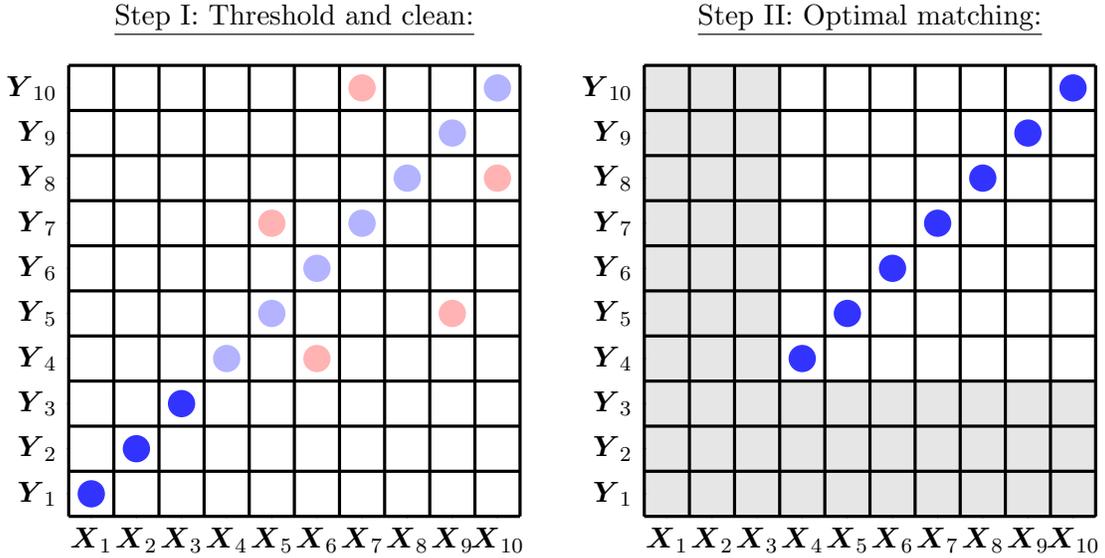

The proposed algorithm contains two steps. In the first step, we use the thresholding technique as in Section \ref{Sec4} to recover the relatively strong pairs. Then, for the second step, we create a new table by taking the original table, and erasing all rows and columns which contain a pair that has been recovered in the first step.  
We complete the alignment recovery by using the Hungarian algorithm in order to find the best match for the new (smaller) table. Since the size of the new table, to be denoted by $\tilde{n}$, is smaller than the size of the original table, then using the Hungarian algorithm will have a reduced computational cost, when compared to executing it on the original table. An illustration of this algorithm is given in Figure \ref{fig:Efficient_Full_Recovery}.

Concerning the computational complexity of this algorithm, it is obviously composed by two terms; for the first step, the complexity behaves like $\calO(n^{2})$, and for the second step, the complexity has the order of $\calO(\tilde{n}^{3})$. 
Assume that for a given $\rho$, we tune $\theta$ to attain a specific $\sR$. Then, we get that $\tilde{n}=(1-\sR)n$, which means that the computational complexity of the second step is given by $(1-\sR)^{3}\calO(n^{3})$. Thus, even if $\sR$ is not too large, the amount of computations may be reduced significantly. Specifically, the overall complexity of this algorithm is lower than the complexity of the optimal ML estimator (which is $\calO(n^{3})$ per the Hungarian algorithm).    

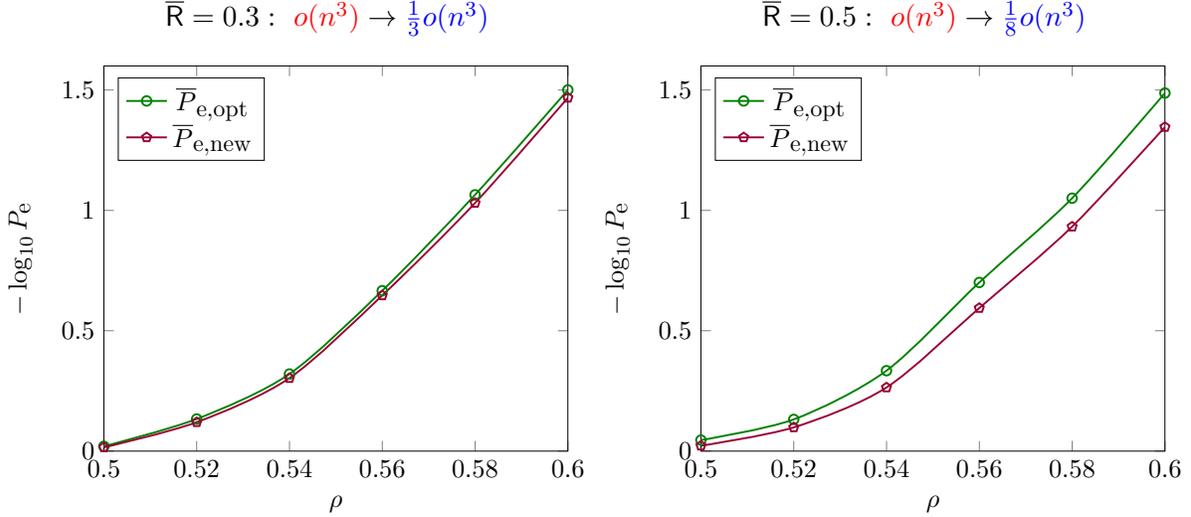
\begin{figure}[ht!]
	\centering
	\begin{subfigure}[b]{0.5\columnwidth}
		\centering 
		\begin{tikzpicture}[scale=0.9]
		\node at (3.3,6.4)[]{$\overline{\sR}=0.3:~
			\textcolor{red}{o(n^{3})} \to \textcolor{blue}{\frac{1}{3} o(n^{3})}$};
		\begin{axis}[
		disabledatascaling,
		scaled x ticks=false,
		xticklabel style={/pgf/number format/fixed,
			/pgf/number format/precision=3},
		scaled y ticks=false,
		yticklabel style={/pgf/number format/fixed,
			/pgf/number format/precision=3},
		xlabel={$\rho$},
		ylabel={$-\log_{10}P_{\mbox{\small e}}$},
		xmin=0.5, xmax=0.6,
		ymin=0, ymax=1.6,
		legend pos=north west
		]

		\addplot[smooth,color=black!50!green,thick,mark=o]
		table[row sep=crcr]
		{
			0.5 0.019997 \\ 
			0.52 0.133485 \\ 
			0.54 0.320269 \\ 
			0.56 0.666441 \\  
			0.58 1.064661 \\
			0.60 1.500313 \\	
		};
		\legend{}
		\addlegendentry{$\overline{P}_{\mbox{\small e,opt}}$}
		
		\addplot[smooth,color=black!20!purple,thick,mark=pentagon]
		table[row sep=crcr] 
		{
			0.5 0.015473 \\ 
			0.52 0.120066 \\ 
			0.54 0.303207 \\ 
			0.56 0.646746 \\
			0.58 1.030739 \\
			0.60 1.467883 \\
		};
		\addlegendentry{$\overline{P}_{\mbox{\small e,new}}$}
		
		\end{axis}
		\end{tikzpicture}
	\end{subfigure}%
	\begin{subfigure}[b]{0.5\columnwidth}
		\centering 
		\begin{tikzpicture}[scale=0.9]
		\node at (3.3,6.4)[]{$\overline{\sR}=0.5:~
			\textcolor{red}{o(n^{3})} \to \textcolor{blue}{\frac{1}{8} o(n^{3})}$};
		\begin{axis}[
		disabledatascaling,
		scaled x ticks=false,
		xticklabel style={/pgf/number format/fixed,
			/pgf/number format/precision=3},
		scaled y ticks=false,
		yticklabel style={/pgf/number format/fixed,
			/pgf/number format/precision=3},
		xlabel={$\rho$},
		ylabel={$-\log_{10}P_{\mbox{\small e}}$},
		xmin=0.5, xmax=0.6,
		ymin=0, ymax=1.6,
		legend pos=north west
		]

		\addplot[smooth,color=black!50!green,thick,mark=o]
		table[row sep=crcr]
		{
			0.50 0.045757 \\
			0.52 0.131672 \\
			0.54 0.334106 \\
			0.56 0.700581 \\
			0.58 1.05061 \\
			0.60 1.487449 \\  	
		};
		\legend{}
		\addlegendentry{$\overline{P}_{\mbox{\small e,opt}}$}
		
		\addplot[smooth,color=black!20!purple,thick,mark=pentagon]
		table[row sep=crcr] 
		{
			0.50 0.022276 \\
			0.52 0.098584 \\
			0.54 0.264268 \\
			0.56 0.594407 \\
			0.58 0.932433 \\
			0.60 1.346305 \\
		};
		\addlegendentry{$\overline{P}_{\mbox{\small e,new}}$}
		
		\end{axis}
		\end{tikzpicture}
	\end{subfigure}%
	\caption{Plots of $\overline{P}_{\mbox{\small e,new}}$ and $\overline{P}_{\mbox{\small e,opt}}$ as functions of $\rho$ for $n=200$ and $d=50$.}\label{fig:ErrorProbabilities}
\end{figure}

Of course, the savings in the amount of calculations comes with a price; the error probability of the new proposed algorithm, to be denoted by $P_{\mbox{\small e,new}}$, is not as low as the ML error probability (denoted here by $P_{\mbox{\small e,opt}}$).
Still, we show by computer simulations that $P_{\mbox{\small e,new}}$ is also not much higher than $P_{\mbox{\small e,opt}}$. In Figure \ref{fig:ErrorProbabilities}, we plot estimated values of $P_{\mbox{\small e,opt}}$ and $P_{\mbox{\small e,new}}$, as functions of $\rho$, for $n=200$, $d=50$, and for the two values $\sR=0.3$ (left) and $\sR=0.5$ (right). For each value of $\rho$, we tuned $\theta$ to attain the required $\sR$.  
As can be seen in Figure \ref{fig:ErrorProbabilities}, the error probability of the new proposed algorithm is not much lower than the ML error probability.

\section{Comparison with a Previous Work} \label{Sec7}
In this section, we show that for some different parameter choices, the new proposed tester in \eqref{Statistic_new}-\eqref{Test_new} achieves lower type-I and type-II error probabilities than the sum-of-inner-products tester in \eqref{Statistic}-\eqref{Test}, that was proposed in \cite{KN2022}. The fact that the new tester has lower error probabilities is not trivial, since the analysis in \cite{KN2022} is based on Chernoff's bound, which is known to be relatively tight in many different cases, while the analysis in the current manuscript follows other methods, that are able to accommodate the statistical dependencies between the indicator random variables in \eqref{Statistic_new}.     

In Figure \ref{fig:Diffrent_Parameters} below, we present plots of the trade-off between $-\log P_{\mbox{\tiny FA}}(\phi_{T})$ and $-\log P_{\mbox{\tiny MD}}(\phi_{T})$ (in solid line) and the trade-off between $-\log P_{\mbox{\tiny FA}}(\phi_{N})$ and $-\log P_{\mbox{\tiny MD}}(\phi_{N})$ (in dashed line), for the parameters $n=200$, $d=50$, and for two values of $\rho$.   
The quantity $Q(d,\theta)$ is calculated numerically using \eqref{Q_CALC} and $P(d,\rho,\theta)$ using \eqref{P_CALC} in Lemma \ref{Lemma_P_Bound}.
Since the numbers $\mathsf{B}(k)$ grow quite rapidly with $k$, when calculating the bounds in \eqref{FA_upper_bound}, we switched the infimum over $\naturals$ with a minimization over $\{1,2,\ldots,40\}$.
As can be seen in Figure \ref{fig:Diffrent_Parameters}, the new proposed statistical test has lower error probabilities for both $\rho=0.7$ and $\rho=0.4$.

\begin{figure}[h!]	
	\begin{subfigure}[b]{0.5\columnwidth}
		\centering 
		\begin{tikzpicture}[scale=0.9]
		\begin{axis}[
		disabledatascaling,
		scaled x ticks=false,
		xticklabel style={/pgf/number format/fixed,
			/pgf/number format/precision=3},
		scaled y ticks=false,
		yticklabel style={/pgf/number format/fixed,
			/pgf/number format/precision=3},
		xlabel={$-\log P_{\mbox{\tiny FA}}$},
		ylabel={$-\log P_{\mbox{\tiny MD}}$},
		xmin=0, xmax=20,
		ymin=0, ymax=20,
		legend pos=north east
		]
		
		\addplot[smooth,color=black!30!green,thick]
		table[row sep=crcr] 
		{
			0.0  16.8336  \\  
			0.0624  13.6414  \\  
			0.1247  12.457  \\  
			0.1868  11.6012  \\  
			0.2488  10.9144  \\  
			0.3106  10.3351  \\  
			0.3722  9.8316  \\  
			0.4338  9.385  \\  
			0.4951  8.9833  \\  
			0.5564  8.6179  \\  
			0.6174  8.2827  \\  
			0.6784  7.973  \\  
			0.7392  7.6854  \\  
			0.7998  7.4168  \\  
			0.8604  7.165  \\  
			0.9207  6.9282  \\  
			0.981  6.7047  \\  
			1.0411  6.4932  \\  
			1.1011  6.2926  \\  
			1.1609  6.1019  \\  
			1.2207  5.9202  \\  
			1.2802  5.7469  \\  
			1.3397  5.5812  \\  
			1.399  5.4226  \\  
			1.4582  5.2707  \\  
			1.5173  5.1248  \\  
			1.5763  4.9847  \\  
			1.6351  4.8499  \\  
			1.6938  4.7201  \\  
			1.7524  4.5951  \\  
			1.8109  4.4745  \\  
			1.8692  4.3581  \\  
			1.9274  4.2456  \\  
			1.9856  4.1369  \\  
			2.0435  4.0317  \\  
			2.1014  3.9299  \\  
			2.1592  3.8313  \\  
			2.2168  3.7358  \\  
			2.2744  3.6432  \\  
			2.3318  3.5533  \\  
			2.3891  3.4661  \\  
			2.4463  3.3814  \\  
			2.5034  3.2991  \\  
			2.5604  3.2191  \\  
			2.6172  3.1414  \\  
			2.674  3.0658  \\  
			2.7307  2.9923  \\  
			2.7872  2.9207  \\  
			2.8437  2.851  \\  
			2.9  2.7831  \\  
			2.9562  2.717  \\  
			3.0124  2.6526  \\  
			3.0684  2.5898  \\  
			3.1243  2.5285  \\  
			3.1802  2.4688  \\  
			3.2359  2.4106  \\  
			3.2915  2.3538  \\  
			3.3471  2.2983  \\  
			3.4025  2.2442  \\  
			3.4578  2.1914  \\  
			3.5131  2.1398  \\  
			3.5682  2.0894  \\  
			3.6233  2.0402  \\  
			3.6782  1.9921  \\  
			3.7331  1.9451  \\  
			3.7878  1.8992  \\  
			3.8425  1.8543  \\  
			3.8971  1.8105  \\  
			3.9516  1.7676  \\  
			4.006  1.7256  \\  
			4.0603  1.6846  \\  
			4.1145  1.6445  \\  
			4.1686  1.6053  \\  
			4.2226  1.5669  \\  
			4.2766  1.5294  \\  
			4.3304  1.4927  \\  
			4.3842  1.4567  \\  
			4.4379  1.4215  \\  
			4.4915  1.3871  \\  
			4.545  1.3534  \\  
			4.5984  1.3205  \\  
			4.6517  1.2882  \\  
			4.705  1.2566  \\  
			4.7582  1.2256  \\  
			4.8113  1.1954  \\  
			4.8643  1.1657  \\  
			4.9172  1.1367  \\  
			4.97  1.1082  \\  
			5.0228  1.0804  \\  
			5.0755  1.0531  \\  
			5.1281  1.0264  \\  
			5.1806  1.0003  \\  
			5.233  0.9747  \\  
			5.2854  0.9496  \\  
			5.3376  0.9251  \\  
			5.3898  0.901  \\  
			5.442  0.8775  \\  
			5.494  0.8544  \\  
			5.546  0.8318  \\  
			5.5979  0.8097  \\  
			5.6497  0.7881  \\  
			5.7014  0.7668  \\  
			5.7531  0.7461  \\  
			5.8047  0.7257  \\  
			5.8562  0.7058  \\  
			5.9076  0.6863  \\  
			5.959  0.6672  \\  
			6.0103  0.6485  \\  
			6.0615  0.6302  \\  
			6.1126  0.6123  \\  
			6.1637  0.5947  \\  
			6.2147  0.5775  \\  
			6.2657  0.5607  \\  
			6.3165  0.5442  \\  
			6.3673  0.5281  \\  
			6.418  0.5124  \\  
			6.4687  0.4969  \\  
			6.5192  0.4818  \\  
			6.5698  0.4671  \\  
			6.6202  0.4526  \\  
			6.6706  0.4385  \\  
			6.7209  0.4246  \\  
			6.7711  0.4111  \\  
			6.8213  0.3979  \\  
			6.8714  0.3849  \\  
			6.9214  0.3723  \\  
			6.9714  0.3599  \\  
			7.0213  0.3479  \\  
			7.0711  0.3361  \\  
			7.1209  0.3245  \\  
			7.1706  0.3132  \\  
			7.2202  0.3022  \\  
			7.2698  0.2915  \\  
			7.3193  0.281  \\  
			7.3688  0.2707  \\  
			7.4181  0.2607  \\  
			7.4674  0.251  \\  
			7.5167  0.2414  \\  
			7.5659  0.2322  \\  
			7.615  0.2231  \\  
			7.6641  0.2143  \\  
			7.7131  0.2057  \\  
			7.762  0.1973  \\  
			7.8109  0.1891  \\  
			7.8597  0.1811  \\  
			7.9085  0.1734  \\  
			7.9572  0.1659  \\  
			8.0058  0.1585  \\  
			8.0544  0.1514  \\  
			8.1029  0.1445  \\  
			8.1514  0.1377  \\  
			8.1998  0.1312  \\  
			8.2481  0.1248  \\  
			8.2964  0.1187  \\  
			8.3446  0.1127  \\  
			8.3928  0.1069  \\  
			8.4409  0.1013  \\  
			8.4889  0.0958  \\  
			8.5369  0.0906  \\  
			8.5849  0.0855  \\  
			8.6327  0.0806  \\  
			8.6806  0.0758  \\  
			8.7283  0.0712  \\  
			8.776  0.0668  \\  
			8.8237  0.0625  \\  
			8.8713  0.0584  \\  
			8.9188  0.0545  \\  
			8.9663  0.0507  \\  
			9.0137  0.047  \\  
			9.0611  0.0435  \\  
			9.1084  0.0402  \\  
			9.1557  0.037  \\  
			9.2029  0.034  \\  
			9.2501  0.0311  \\  
			9.2972  0.0283  \\  
			9.3442  0.0257  \\  
			9.3912  0.0232  \\  
			9.4381  0.0208  \\  
			9.485  0.0186  \\  
			9.5319  0.0165  \\  
			9.5787  0.0146  \\  
			9.6254  0.0128  \\  
			9.6721  0.0111  \\  
			9.7187  0.0095  \\  
			9.7653  0.0081  \\  
			9.8118  0.0068  \\  
			9.8583  0.0056  \\  
			9.9047  0.0045  \\  
			9.9511  0.0035  \\  
			9.9974  0.0027  \\  
			10.0437  0.002  \\  
			10.0899  0.0014  \\  
			10.136  0.0009  \\  
			10.1822  0.0005  \\  
			10.2282  0.0002  \\  
			10.2743  0.0001  \\  
		};
		\legend{}
		\addlegendentry{$\phi_{T}$}	
		
		\addplot[smooth,color=black!20!purple,thick,dash pattern={on 3pt off 2pt}]
		table[row sep=crcr]
		{
			-0.0  16.65  \\  
			0.6  16.483  \\  
			1.246  16.317  \\  
			1.636  16.15  \\  
			1.915  15.983  \\  
			2.134  15.817  \\  
			2.313  15.65  \\  
			2.465  15.483  \\  
			2.596  15.317  \\  
			2.713  15.15  \\  
			2.817  14.983  \\  
			2.911  14.817  \\  
			2.998  14.65  \\  
			3.077  14.483  \\  
			3.151  14.317  \\  
			3.219  14.15  \\  
			3.283  13.983  \\  
			3.344  13.817  \\  
			3.4  13.65  \\  
			3.475  13.483  \\  
			3.577  13.317  \\  
			3.674  13.15  \\  
			3.78  12.983  \\  
			3.912  12.817  \\  
			4.039  12.65  \\  
			4.161  12.483  \\  
			4.279  12.317  \\  
			4.41  12.15  \\  
			4.555  11.983  \\  
			4.695  11.817  \\  
			4.83  11.65  \\  
			4.961  11.483  \\  
			5.103  11.317  \\  
			5.256  11.15  \\  
			5.405  10.983  \\  
			5.55  10.817  \\  
			5.69  10.65  \\  
			5.837  10.483  \\  
			5.996  10.317  \\  
			6.152  10.15  \\  
			6.303  9.983  \\  
			6.451  9.817  \\  
			6.601  9.65  \\  
			6.765  9.483  \\  
			6.926  9.317  \\  
			7.083  9.15  \\  
			7.236  8.983  \\  
			7.388  8.817  \\  
			7.556  8.65  \\  
			7.721  8.483  \\  
			7.882  8.317  \\  
			8.04  8.15  \\  
			8.195  7.983  \\  
			8.365  7.817  \\  
			8.533  7.65  \\  
			8.697  7.483  \\  
			8.859  7.317  \\  
			9.018  7.15  \\  
			9.187  6.983  \\  
			9.358  6.817  \\  
			9.526  6.65  \\  
			9.691  6.483  \\  
			9.853  6.317  \\  
			10.021  6.15  \\  
			10.194  5.983  \\  
			10.365  5.817  \\  
			10.532  5.65  \\  
			10.697  5.483  \\  
			10.865  5.317  \\  
			11.04  5.15  \\  
			11.213  4.983  \\  
			11.383  4.817  \\  
			11.55  4.65  \\  
			11.718  4.483  \\  
			11.894  4.317  \\  
			12.069  4.15  \\  
			12.241  3.983  \\  
			12.41  3.817  \\  
			12.578  3.65  \\  
			12.756  3.483  \\  
			12.932  3.317  \\  
			13.105  3.15  \\  
			13.277  2.983  \\  
			13.446  2.686  \\  
			13.623  2.378  \\  
			13.8  2.088  \\  
			13.976  1.817  \\  
			14.149  1.565  \\  
			14.32  1.332  \\  
			14.496  1.117  \\  
			14.675  0.922  \\  
			14.851  0.745  \\  
			15.026  0.587  \\  
			15.199  0.448  \\  
			15.374  0.327  \\  
			15.553  0.226  \\  
			15.731  0.143  \\  
			15.907  0.079  \\  
			16.081  0.034  \\  
			16.256  0.008  \\ 
		};
		\addlegendentry{$\phi_{N}$}
		
		\end{axis}
		\end{tikzpicture}
		\caption{$\rho=0.7,~\theta=0.55$} 
		\label{fig:Different_Param_3}
	\end{subfigure}
	\begin{subfigure}[b]{0.5\columnwidth}
		\centering 
		\begin{tikzpicture}[scale=0.9]
		\begin{axis}[
		disabledatascaling,
		scaled x ticks=false,
		xticklabel style={/pgf/number format/fixed,
			/pgf/number format/precision=3},
		scaled y ticks=false,
		yticklabel style={/pgf/number format/fixed,
			/pgf/number format/precision=3},
		xlabel={$-\log P_{\mbox{\tiny FA}}$},
		ylabel={$-\log P_{\mbox{\tiny MD}}$},
		xmin=0, xmax=4,
		ymin=0, ymax=4,
		legend pos=north east
		]
		
		\addplot[smooth,color=black!30!green,thick]
		table[row sep=crcr] 
		{
			0.0  4.3588  \\  
			0.0624  3.2568  \\  
			0.1247  2.8518  \\  
			0.1868  2.5612  \\  
			0.2488  2.3297  \\  
			0.3106  2.1359  \\  
			0.3722  1.9687  \\  
			0.4338  1.8217  \\  
			0.4951  1.6906  \\  
			0.5564  1.5724  \\  
			0.6174  1.465  \\  
			0.6784  1.3668  \\  
			0.7392  1.2764  \\  
			0.7998  1.193  \\  
			0.8604  1.1157  \\  
			0.9207  1.0438  \\  
			0.981  0.9767  \\  
			1.0411  0.9141  \\  
			1.1011  0.8555  \\  
			1.1609  0.8005  \\  
			1.2207  0.7489  \\  
			1.2802  0.7003  \\  
			1.3397  0.6547  \\  
			1.399  0.6116  \\  
			1.4582  0.5711  \\  
			1.5173  0.5328  \\  
			1.5763  0.4967  \\  
			1.6351  0.4626  \\  
			1.6938  0.4303  \\  
			1.7524  0.3999  \\  
			1.8109  0.3712  \\  
			1.8692  0.3441  \\  
			1.9274  0.3184  \\  
			1.9856  0.2942  \\  
			2.0435  0.2714  \\  
			2.1014  0.2498  \\  
			2.1592  0.2295  \\  
			2.2168  0.2104  \\  
			2.2744  0.1923  \\  
			2.3318  0.1754  \\  
			2.3891  0.1594  \\  
			2.4463  0.1445  \\  
			2.5034  0.1304  \\  
			2.5604  0.1173  \\  
			2.6172  0.1051  \\  
			2.674  0.0936  \\  
			2.7307  0.083  \\  
			2.7872  0.0731  \\  
			2.8437  0.064  \\  
			2.9  0.0556  \\  
			2.9562  0.0479  \\  
			3.0124  0.0408  \\  
			3.0684  0.0344  \\  
			3.1243  0.0286  \\  
			3.1802  0.0233  \\  
			3.2359  0.0187  \\  
			3.2915  0.0146  \\  
			3.3471  0.0111  \\  
			3.4025  0.0081  \\  
			3.4578  0.0055  \\  
			3.5131  0.0035  \\  
			3.5682  0.002  \\  
			3.6233  0.0009  \\  
			3.6782  0.0002  \\  
			3.7331  -0.0  \\  
		};
		\legend{}
		\addlegendentry{$\phi_{T}$}	
		
		\addplot[smooth,color=black!20!purple,thick,dash pattern={on 3pt off 2pt}]
		table[row sep=crcr]
		{
			-0.0  3.28  \\  
			-0.0  3.215  \\  
			0.058  3.15  \\  
			0.448  3.086  \\  
			0.727  3.023  \\  
			0.945  2.96  \\  
			1.125  2.898  \\  
			1.276  2.837  \\  
			1.408  2.776  \\  
			1.524  2.716  \\  
			1.629  2.656  \\  
			1.723  2.598  \\  
			1.809  2.539  \\  
			1.889  2.482  \\  
			1.962  2.425  \\  
			2.031  2.369  \\  
			2.095  2.314  \\  
			2.155  2.259  \\  
			2.212  2.205  \\  
			2.266  2.151  \\  
			2.317  2.098  \\  
			2.365  2.046  \\  
			2.412  1.995  \\  
			2.456  1.944  \\  
			2.498  1.893  \\  
			2.539  1.844  \\  
			2.578  1.795  \\  
			2.616  1.747  \\  
			2.652  1.699  \\  
			2.687  1.652  \\  
			2.721  1.606  \\  
			2.753  1.56  \\  
			2.785  1.515  \\  
			2.816  1.471  \\  
			2.846  1.427  \\  
			2.874  1.384  \\  
			2.903  1.342  \\  
			2.93  1.3  \\  
			2.956  1.259  \\  
			2.982  1.219  \\  
			3.008  1.179  \\  
			3.032  1.14  \\  
			3.056  1.102  \\  
			3.08  1.064  \\  
			3.103  1.027  \\  
			3.125  0.991  \\  
			3.147  0.955  \\  
			3.168  0.92  \\  
			3.19  0.885  \\  
			3.21  0.852  \\  
			3.23  0.818  \\  
			3.25  0.786  \\  
			3.269  0.754  \\  
			3.288  0.723  \\  
			3.307  0.692  \\  
			3.325  0.663  \\  
			3.343  0.633  \\  
			3.361  0.605  \\  
			3.378  0.577  \\  
			3.395  0.55  \\  
			3.412  0.523  \\  
			3.429  0.497  \\  
			3.445  0.472  \\  
			3.461  0.448  \\  
			3.477  0.424  \\  
			3.492  0.4  \\  
			3.507  0.378  \\  
			3.522  0.356  \\  
			3.537  0.334  \\  
			3.552  0.314  \\  
			3.566  0.294  \\  
			3.58  0.275  \\  
			3.594  0.256  \\  
			3.608  0.238  \\  
			3.622  0.22  \\  
			3.635  0.204  \\  
			3.648  0.188  \\  
			3.661  0.172  \\  
			3.674  0.158  \\  
			3.687  0.144  \\  
			3.7  0.13  \\  
			3.712  0.117  \\  
			3.724  0.105  \\  
			3.736  0.094  \\  
			3.748  0.083  \\  
			3.76  0.073  \\  
			3.772  0.064  \\  
			3.783  0.055  \\  
			3.795  0.047  \\  
			3.806  0.039  \\  
			3.817  0.032  \\  
			3.828  0.026  \\  
			3.839  0.021  \\  
			3.85  0.016  \\  
			3.861  0.011  \\  
			3.871  0.008  \\  
			3.882  0.005  \\  
			3.892  0.003  \\  
			3.902  0.001  \\  
			3.912  0.0  \\    
		};
		\addlegendentry{$\phi_{N}$}
		
		\end{axis}
		\end{tikzpicture}
		\caption{$\rho=0.4,~\theta=0.6$} 
		\label{fig:Different_Param_4}
	\end{subfigure}
	\caption{Plots of the trade-offs between the exponential rates of decay of the two error probabilities for the sum-of-inner-products detector \eqref{Statistic}-\eqref{Test} (solid line) and for the sum-of-indicators detector \eqref{Statistic_new}-\eqref{Test_new} (dashed line).} 
	\label{fig:Diffrent_Parameters}
\end{figure}
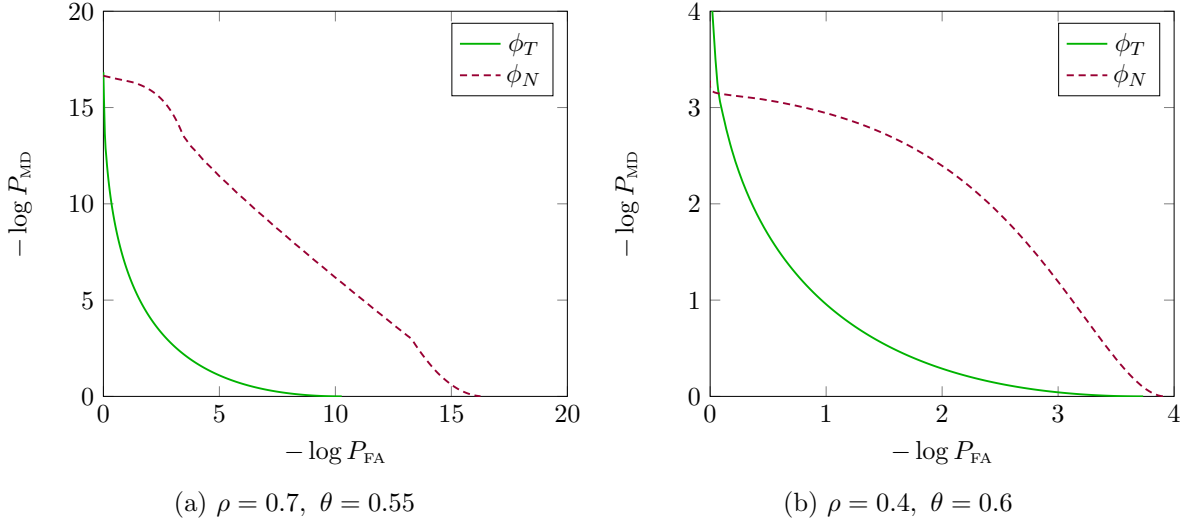

\section*{Appendix A - Proof of Theorem \ref{THM_FA_and_MD_probabilities}}
\renewcommand{\theequation}{A.\arabic{equation}}
\setcounter{equation}{0}

\subsection*{Analysis of False Alarms}
For $k \in \naturals$, consider the following
\begin{align}
P_{\mbox{\tiny FA}}(\phi_{N}) 
&= \pr_{0} \left\{N(\theta) \geq \beta nP \right\} \\
&= \pr_{0} \left\{\sum_{i=1}^{n} \sum_{j=1}^{n} \IND \left\{\tilde{\bX}_{i}^{\mathsf{T}} \tilde{\bY}_{j} \geq \theta \right\} \geq \beta nP \right\} \\
&= \pr_{0} \left\{\left(\sum_{i=1}^{n} \sum_{j=1}^{n} \IND\left\{\tilde{\bX}_{i}^{\mathsf{T}} \tilde{\bY}_{j} \geq \theta \right\}\right)^{k} \geq  \left(\beta nP\right)^{k} \right\} \\
\label{ref0}
&\leq \left(\beta nP\right)^{-k} \cdot \E_{0} \left[\left(\sum_{i=1}^{n} \sum_{j=1}^{n} \IND\left\{\tilde{\bX}_{i}^{\mathsf{T}} \tilde{\bY}_{j} \geq \theta \right\}\right)^{k}\right],
\end{align}
where \eqref{ref0} is due to Markov's inequality.
Since $k \in \naturals$ is arbitrary, it holds that
\begin{align} \label{ref3}
P_{\mbox{\tiny FA}}(\phi_{N}) 
\leq \inf_{k \in \naturals}  \left(\beta nP \right)^{-k} \cdot \E_{0} \left[\left(\sum_{i=1}^{n} \sum_{j=1}^{n} \IND\left\{\tilde{\bX}_{i}^{\mathsf{T}} \tilde{\bY}_{j} \geq \theta \right\}\right)^{k}\right],
\end{align}
so the main task is to evaluate the $k$-th moment of $N(\theta)$.

In order to upper-bound the $k$-th moment of $N(\theta)$, we will make use of the following result, which is proved in Appendix B.
\begin{lemma} \label{Lemma_Square_LD_Integers}
	Let $n \in \naturals$. Let $\bX_{1},\ldots,\bX_{n}$ and $\bY_{1},\ldots,\bY_{n}$ be two sets of IID random variables, taking values in $\calX$ and $\calY$, respectively. Let $J: \calX \times \calY \to \{0,1\}$ and assume that
	\begin{align}
	\pr(J(\bx,\bY)=1) &\leq Q, \\
	\pr(J(\bX,\by)=1) &\leq Q, \\
	\pr(J(\bX,\bY)=1) &\leq Q,
	\end{align}
	where $Q \in [0,1]$. Let $k \in \naturals$. Then,
	\begin{align}
	\label{Lemma_Square_LD_Integers_Result}
	\E \left[\left( \sum_{\ell=1}^{n} \sum_{m=1}^{n} J(\bX_{\ell},\bY_{m}) \right)^{k}\right]
	\leq k(k+1)\mathsf{B}(k) \cdot
	\left\{ 
	\begin{array}{l l}
	(n^{2}Q)^{k}   & \quad \text{if $n^{2}Q \geq 1$,} \\
	n^{2}Q   & \quad \text{if $n^{2}Q < 1$.} 
	\end{array} \right.
	\end{align} 
\end{lemma}

Continuing from \eqref{ref3} using the result of Lemma \ref{Lemma_Square_LD_Integers} yields
\begin{align} 
P_{\mbox{\tiny FA}}(\phi_{N}) 
&\leq \inf_{k \in \naturals}  
\frac{k(k+1)\mathsf{B}(k)\left[(n^{2}Q)^{k} \cdot \IND\{n^{2}Q \geq 1\} + n^{2}Q \cdot \IND\{n^{2}Q < 1\}\right]}{\left(\beta nP\right)^{k}},
\end{align}
which is exactly \eqref{FA_upper_bound} in Theorem \ref{THM_FA_and_MD_probabilities}.

\subsection*{Analysis of Miss-Detection}
Under $\HypB$, the expectation of $N(\theta)$ is given by 
\begin{align}
\Delta_{n,d}
=\E_{1}[N(\theta)]
&= \E_{1}\left[\sum_{i=1}^{n} \sum_{j=1}^{n} \IND\left\{\tilde{\bX}_{i}^{\mathsf{T}} \tilde{\bY}_{j} \geq \theta \right\}\right] \\
&= \sum_{i=1}^{n} \sum_{j=1}^{n} \pr_{1|\sigma} \left\{\tilde{\bX}_{i}^{\mathsf{T}} \tilde{\bY}_{j} \geq \theta \right\} \\
&= nP + n(n-1)Q.
\end{align}
Consider the following
\begin{align}
P_{\mbox{\tiny MD}}(\phi_{N}) 
&= \pr_{1|\sigma} \left\{N(\theta) \leq \beta nP \right\} \\
&\leq \pr_{1|\sigma} \left\{N(\theta) \leq \beta \left[nP+n(n-1)Q\right] \right\} \\
\label{ref2}
&= \pr_{1|\sigma} \left\{\sum_{i=1}^{n} \sum_{j=1}^{n} \IND\left\{\tilde{\bX}_{i}^{\mathsf{T}} \tilde{\bY}_{j} \geq \theta \right\} \leq \beta \Delta_{n,d} \right\}.
\end{align} 
Let us denote in short $\calI(i,j)=\IND\left\{\tilde{\bX}_{i}^{\mathsf{T}} \tilde{\bY}_{j} \geq \theta \right\}$.

In order to upper-bound \eqref{ref2}, we borrow one result from \cite{Janson} concerning large deviations behavior of sums of dependent indicator random variables.

Let $\{U_{\bk}\}_{\bk \in \calK}$, where $\calK$ is a set of multidimensional indexes, be a family of Bernoulli random variables. 
Let $G$ be a dependency graph for $\{U_{\bk}\}_{\bk \in \calK}$, i.e., a graph with vertex set $\calK$ such that if $\calA$ and $\calB$ are two disjoint subsets of $\calK$, and $G$ contains no edge between $\calA$ and $\calB$, then the families $\{U_{\bk}\}_{\bk \in \calA}$ and $\{U_{\bk}\}_{\bk \in \calB}$ are independent.   
Let $S = \sum_{\bk \in \calK} U_{\bk}$ and $\Delta = \mathbb{E}[S]$. Moreover, we write $\bi \sim \bj$ if $(\bi,\bj)$ is an edge in the dependency graph $G$. Let 
\begin{align}
\Omega =  \max_{\bi \in \calK} \sum_{\bj \in \calK,\bj \sim \bi} \mathbb{E} [U_{\bj}],
\end{align}
and
\begin{align}
\Theta = \frac{1}{2} \sum_{\bi \in \calK} \sum_{\bj \in \calK, \bj \sim \bi} \mathbb{E} [U_{\bi}U_{\bj}].
\end{align}
\begin{theorem}[\cite{Janson}, Theorem 10]
	With notations as above, then for any $0\leq \beta \leq 1$, 
	\begin{align}
	\prob \{S \leq \beta \Delta\} \leq \exp \left\{- \min \left(
	(1-\beta)^{2} \frac{\Delta^{2}}{8\Theta+2\Delta},
	(1-\beta) \frac{\Delta}{6\Omega}
	\right) \right\} .
	\end{align}
\end{theorem}

In our case, we have $\Delta = \Delta_{n,d}$, and it only remains to assess the quantities $\Theta$ and $\Omega$. One can easily check that the indicator random variables $\calI(i,j)$ and $\calI(k,\ell)$ are independent as long as $i \neq k$ and $j \neq \ell$. Thus, we define our dependency graph in a way that each vertex $(i,j)$ is connected to exactly $2n - 2$ vertices of the form $(i,\ell)$, $\ell \neq j$ or $(k,j)$, $k \neq i$. If the vertices $(i,j)$ and $(k,\ell)$ are connected, we denote it by $(i,j) \sim (k,\ell)$. 

Concerning $\Theta$ and $\Omega$, we get that
\begin{align}
\Theta &= \frac{1}{2} \sum_{(i,j) \in \Psi_{n}} \sum_{\substack{(k,\ell) \in \Psi_{n} \\ (k,\ell) \sim (i,j)}} \mathbb{E}[\calI(i,j)\calI(k,\ell)] \\
&= \frac{1}{2} \sum_{i=1}^{n} \sum_{\substack{(k,\ell) \in \Psi_{n} \\ (k,\ell) \sim (i,i)}} \mathbb{E}[\calI(i,i)\calI(k,\ell)]
+ \frac{1}{2} \sum_{i=1}^{n} \sum_{j \neq i} \sum_{\substack{(k,\ell) \in \Psi_{n} \\ (k,\ell) \sim (i,j)}} \mathbb{E}[\calI(i,j)\calI(k,\ell)] \\
&= \frac{1}{2} PQ (2n-2)n + \frac{1}{2}[2Q^{2}(n-2) + 2PQ]n(n-1) \\
&= PQ(n-1)n + [Q^{2}(n-2) + PQ]n(n-1) \\
&= 2PQn(n-1) + Q^{2}n(n-1)(n-2),
\end{align}
and 
\begin{align}
\Omega &=  \max_{(i,j) \in \Psi_{n}} \sum_{\substack{(k,\ell) \in \Psi_{n} \\ (k,\ell) \sim (i,j)}} \mathbb{E}[\calI(k,\ell)] \\
&= \max\left\{ \sum_{\substack{(k,\ell) \in \Psi_{n} \\ (k,\ell) \sim (i,i)}} \mathbb{E}[\calI(k,\ell)], 
\sum_{\substack{(k,\ell) \in \Psi_{n} \\ (k,\ell) \sim (i,j),~i \neq j}} \mathbb{E}[\calI(k,\ell)]
\right\} \\
&= \max\left\{(2n-2)Q, 2P + (2n-4)Q \right\} \\
&= 2P + (2n-4)Q.
\end{align}
Then,
\begin{align}
\label{RESULT1}
\frac{\Delta_{n,d}}{6\Omega}
&= \frac{nP + n(n-1)Q}{6[2P + (2n-4)Q]} \\
&= \frac{n[P + (n-1)Q]}{12[P + (n-2)Q]} \\
&\geq \frac{n[P + (n-2)Q]}{12[P + (n-2)Q]} \\
&= \frac{n}{12},
\end{align}
and,
\begin{align}
\label{RESULT2}
\frac{\Delta_{n,d}^{2}}{8\Theta + 2\Delta_{n,d}}
&= \frac{[nP + n(n-1)Q]^{2}}{8[2PQn(n-1) + Q^{2}n(n-1)(n-2)] + 2[nP + n(n-1)Q]} \\
&\geq \frac{[nP + n(n-1)Q]^{2}}{8[2PQn^{2} + 2Q^{2}n^{2}(n-1)] + 2[nP + n(n-1)Q]} \\
&= \frac{[nP + n(n-1)Q]^{2}}{16Qn[Pn + Qn(n-1)] + 2[nP + n(n-1)Q]} \\
&= \frac{nP + n(n-1)Q}{16Qn + 2} \\
&\geq \frac{nP}{16Qn + 2}.
\end{align}
Hence,
\begin{align} \label{DE_RESULT}
P_{\mbox{\tiny MD}}(\phi_{N})
&\leq \pr_{1|\sigma} \left\{\sum_{i=1}^{n} \sum_{j=1}^{n} \IND\left\{\tilde{\bX}_{i}^{\mathsf{T}} \tilde{\bY}_{j} \geq \theta \right\} \leq \beta \Delta_{n,d} \right\} \\ 
&\leq \exp \left\{- \min \left(
(1-\beta)^{2} \frac{\Delta_{n,d}^{2}}{8\Theta+2\Delta_{n,d}},
(1-\beta) \frac{\Delta_{n,d}}{6\Omega}
\right) \right\} \\
&\leq \exp \left\{- \min \left( (1-\beta)^{2} \frac{nP}{16nQ + 2}, (1-\beta) \frac{n}{12} \right) \right\},
\end{align}
which completes the proof of Theorem \ref{THM_FA_and_MD_probabilities}.

\section*{Appendix B - Proof of Lemma \ref{Lemma_Square_LD_Integers}}
\renewcommand{\theequation}{B.\arabic{equation}}
\setcounter{equation}{0}

For any $k \in \mathbb{N}$, let $\mathsf{S}(k,d)$ be the number of ways to partition a set of $k$ labeled objects into $d \in \{1,2, \dotsc, k\}$ nonempty unlabeled subsets, which is given by the Stirling numbers of the second kind \cite{RIORDAN}:
\begin{align}
\mathsf{S}(k,d) = \frac{1}{d!} \sum_{i=0}^{d} (-1)^{i} \binom{d}{i}(d-i)^{k}.
\end{align}
Obviously, $\mathsf{S}(k,1)=\mathsf{S}(k,k)=1$. 
Let us denote $\Psi_{n} = \{(m,m'):~m,m'\in \{1,2,\ldots,n\}\}$. Now,
\begin{align}
\mathbb{E} \left[ N^{k}  \right] 
&=\mathbb{E} \left[ \left( \sum_{(m,m') \in \Psi_{n}} J(\bX_{m},\bY_{m'}) \right)^{k} \right] \\
&= \sum_{(m_{1},m'_{1}) \in \Psi_{n}} \dots \sum_{(m_{k},m'_{k}) \in \Psi_{n}} \mathbb{E} \left[ J(\bX_{m_{1}},\bY_{m'_{1}}) \cdots J(\bX_{m_{k}},\bY_{m'_{k}}) \right] \\
\label{TERM2EXPh1}
&= \sum_{d=1}^{k} \sum_{\left\{\substack{(m_{i},m'_{i}) \in \Psi_{n},~1 \leq i \leq k,\\ \text{divided into $d$ subsets of identical pairs}} \right\}} \mathbb{E} \left[ J(\bX_{m_{1}},\bY_{m'_{1}})  \cdots J(\bX_{m_{k}},\bY_{m'_{k}}) \right] \\
\label{K_Summands}
&= \sum_{d=1}^{k} \mathsf{S}(k,d) \sum_{\left\{\substack{(m_{i},m'_{i}) \in \Psi_{n},~1 \leq i \leq d,\\ (m_{i},m'_{i}) \neq (m_{l},m'_{l}) ~ \forall i \neq l} \right\}} \mathbb{E} \left[ J(\bX_{m_{1}},\bY_{m'_{1}})  \cdots J(\bX_{m_{d}},\bY_{m'_{d}}) \right] ,
\end{align}
where in the inner summation in \eqref{TERM2EXPh1}, we sum over all possible $k$ pairs of codewords' indices, which are divided in any possible way into exactly $d$ subsets, all pairs in each subset are identical\footnote{Two pairs of indices $(m_{1},m'_{1})$ and $(m_{2},m'_{2})$ are said to be identical if and only if $m_{1}=m_{2}$ and $m'_{1}=m'_{2}$, otherwise, they said to be distinct.}. 
In \eqref{K_Summands}, we use the Stirling numbers of the second kind, and sum over exactly $d$ distinct pairs of codewords' indices, where all the identical pairs of indices in \eqref{TERM2EXPh1} have been merged together, using the trivial fact that multiplying any number of identical indicator random variables is equal to any one of them.  

Let us handle the inner sum of (\ref{K_Summands}). 
The idea is as follows. Instead of summing over the set $\{(m_{i},m'_{i}) \in \Psi_{n},~1 \leq i \leq d,~ (m_{i},m'_{i}) \neq (m_{l},m'_{l}) ~ \forall i \neq l\}$ of $d$ distinct pairs of indices of codewords, we represent each possible configuration of indices in this set as a \textit{graph} $G$, and sum over \textit{all different graphs} with exactly $d$ distinct edges.
In our graph representation, each codeword index $m_{i} \in \{1,2, \dotsc, n\}$ and $m'_{i} \in \{1,2, \dotsc, n\}$ is denoted by a vertex and each pair of indices $(m_{i},m'_{i})$, $m_{i} \neq m'_{i}$, is connected by an edge\footnote{Of course, indices that belong to distinct pairs may be joined together, e.g.,
	if $(m_{1},m'_{1})$ and $(m_{2},m'_{2})$ are distinct, then it may be that $m_{1}=m_{2}$ or $m'_{1}=m'_{2}$, but not both.}.
Hence, the number of edges is fixed, but the numbers of vertices and subgraphs (i.e., disconnected parts of the graph) are variable. 

Now, given $d \in \{1,2, \dotsc, k\}$, we sum over the \textit{set} $\calV(d) = \{(v_{x},v_{y})\}$ of pairs of integers, which consists of all possible pairs of vertices needed in order to support a graph with $d$ different edges. Of course, if all of the $d$ edges are disconnected, then we must have $(v_{x},v_{y}) = (d,d)$ vertices. 

\begin{figure}[h!]
	\centering
	\begin{tikzpicture}[scale=0.3]
	\draw[black,very thick,-latex] (0,0) -- (20,0);
	\draw[black,very thick,-latex] (0,0) -- (0,11);
	
	\foreach \x/\xtext in {1/1, 3/2, 5/3, 7/4, 9/5, 11/6, 13/7, 15/8, 17/9}
	\draw[shift={(\x,0)}] (0pt,2pt) -- (0pt,-2pt) node[below] {$\bX_{\xtext}$};
	\foreach \y/\ytext in {1/1, 3/2, 5/3, 7/4, 9/5}
	\draw[shift={(0,\y)}] (2pt,0pt) -- (-2pt,0pt) node[left] {$\bY_{\ytext}$};
	
	\node at (1,1) [circle,fill=black] {};
	\node at (3,1) [circle,fill=black] {};
	\node at (3,3) [circle,fill=black] {};
	\node at (5,1) [circle,fill=black] {};
	\node at (5,3) [circle,fill=black] {};
	\node at (7,5) [circle,fill=black] {};
	\node at (7,7) [circle,fill=black] {};
	\node at (9,7) [circle,fill=black] {};
	\node at (11,5) [circle,fill=black] {};
	\node at (11,7) [circle,fill=black] {};
	\node at (13,9) [circle,fill=black] {};
	\node at (15,9) [circle,fill=black] {};
	\node at (17,9) [circle,fill=black] {};
	\end{tikzpicture}
	\hspace{0.3in}
	\begin{tikzpicture}[scale=0.3]
	\draw[black,very thick,-latex] (0,0) -- (20,0);
	\draw[black,very thick,-latex] (0,0) -- (0,11);
	
	\foreach \x/\xtext in {1/1, 3/2, 5/3, 7/4, 9/5, 11/6, 13/7, 15/8, 17/9}
	\draw[shift={(\x,0)}] (0pt,2pt) -- (0pt,-2pt) node[below] {$\bX_{\xtext}$};
	\foreach \y/\ytext in {1/1, 3/2, 5/3, 7/4, 9/5}
	\draw[shift={(0,\y)}] (2pt,0pt) -- (-2pt,0pt) node[left] {$\bY_{\ytext}$};
	
	\node at (1,1) [circle,fill=black] {};
	\node at (3,5) [circle,fill=black] {};
	\node at (3,7) [circle,fill=black] {};
	\node at (5,3) [circle,fill=black] {};
	\node at (7,5) [circle,fill=black] {};
	\node at (7,7) [circle,fill=black] {};
	\node at (9,3) [circle,fill=black] {};
	\node at (9,7) [circle,fill=black] {};
	\node at (11,5) [circle,fill=black] {};
	\node at (13,7) [circle,fill=black] {};
	\node at (15,3) [circle,fill=black] {};
	\node at (15,7) [circle,fill=black] {};
	\node at (17,9) [circle,fill=black] {};
	\end{tikzpicture}
	\caption{Examples for two different graphs with $d=13$, $v_{x}=9$, $v_{y}=5$, and $s=3$.} 
	\label{fig:different_graphs}
\end{figure}
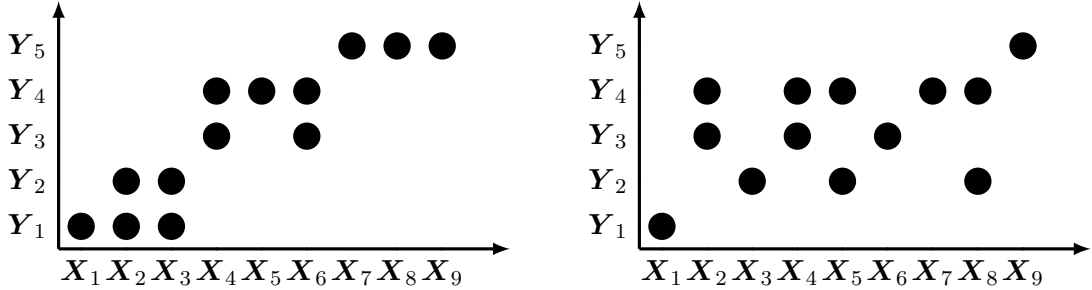

Next, given $d \in \{1,2, \dotsc, k\}$ and $(v_{x},v_{y}) \in \calV(d)$, we sum over the range of possible number of subgraphs. Let $\mathcal{S}_{\mbox{\tiny min}}(d,v_{x},v_{y})$ ($\mathcal{S}_{\mbox{\tiny max}}(d,v_{x},v_{y})$) be the minimal (maximal) number of subgraphs that a graph with $d$ edges and $(v_{x},v_{y})$ vertices can has. 
For a given quadruplet $(d,v_{x},v_{y},s)$, where $s \in [\mathcal{S}_{\mbox{\tiny min}}(d,v_{x},v_{y}),\mathcal{S}_{\mbox{\tiny max}}(d,v_{x},v_{y})]$ is the number of subgraphs within $G$, note that one can create many different graphs (see Figure \ref{fig:different_graphs}), and we have to take all of them into account. Hence, let $\mathsf{T}(d,v_{x},v_{y})$ ($\mathsf{T}(d,v_{x},v_{y},s)$) be the number of distinct ways to connect a graph with $d$ edges and $(v_{x},v_{y})$ vertices (and $s$ subgraphs). Finally, for any $1 \leq i \leq \mathsf{T}(d,v_{x},v_{y},s)$, let $\mathsf{G}_{i}(d,v_{x},v_{y},s)$ be the set of different graphs with $d$ edges, $(v_{x},v_{y})$ vertices, and $s$ subgraphs, that can be defined on the set $\Psi_{n}$ of pairs of codewords, as we explained before.  

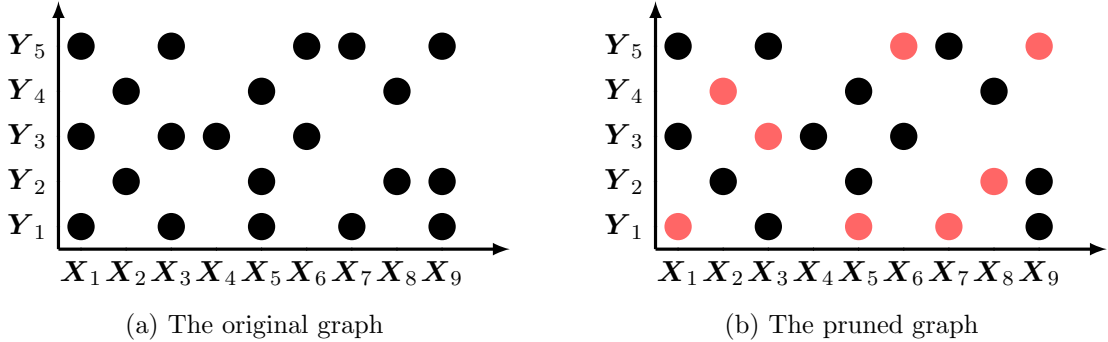
\begin{figure}[h!]
	\begin{subfigure}[b]{0.5\textwidth}
		\centering 
		\begin{tikzpicture}[scale=0.3]
		\draw[black,very thick,-latex] (0,0) -- (20,0);
		\draw[black,very thick,-latex] (0,0) -- (0,11);
		
		\foreach \x/\xtext in {1/1, 3/2, 5/3, 7/4, 9/5, 11/6, 13/7, 15/8, 17/9}
		\draw[shift={(\x,0)}] (0pt,2pt) -- (0pt,-2pt) node[below] {$\bX_{\xtext}$};
		\foreach \y/\ytext in {1/1, 3/2, 5/3, 7/4, 9/5}
		\draw[shift={(0,\y)}] (2pt,0pt) -- (-2pt,0pt) node[left] {$\bY_{\ytext}$};
		
		\node at (1,1) [circle,fill=black] {};
		\node at (1,5) [circle,fill=black] {};
		\node at (1,9) [circle,fill=black] {};
		\node at (3,3) [circle,fill=black] {};
		\node at (3,7) [circle,fill=black] {};
		\node at (5,1) [circle,fill=black] {};
		\node at (5,5) [circle,fill=black] {};
		\node at (5,9) [circle,fill=black] {};
		\node at (7,5) [circle,fill=black] {};
		\node at (9,1) [circle,fill=black] {};
		\node at (9,3) [circle,fill=black] {};
		\node at (9,7) [circle,fill=black] {};
		\node at (11,5) [circle,fill=black] {};
		\node at (11,9) [circle,fill=black] {};
		\node at (13,1) [circle,fill=black] {};
		\node at (13,9) [circle,fill=black] {};
		\node at (15,3) [circle,fill=black] {};
		\node at (15,7) [circle,fill=black] {};
		\node at (17,1) [circle,fill=black] {};
		\node at (17,3) [circle,fill=black] {};
		\node at (17,9) [circle,fill=black] {};
		\end{tikzpicture}
		\caption{The original graph} 
		\label{fig:NASA_Logo_Sub}
	\end{subfigure}%
	\begin{subfigure}[b]{0.5\textwidth}
		\centering 
		\begin{tikzpicture}[scale=0.3]
		\draw[black,very thick,-latex] (0,0) -- (20,0);
		\draw[black,very thick,-latex] (0,0) -- (0,11);
		
		\foreach \x/\xtext in {1/1, 3/2, 5/3, 7/4, 9/5, 11/6, 13/7, 15/8, 17/9}
		\draw[shift={(\x,0)}] (0pt,2pt) -- (0pt,-2pt) node[below] {$\bX_{\xtext}$};
		\foreach \y/\ytext in {1/1, 3/2, 5/3, 7/4, 9/5}
		\draw[shift={(0,\y)}] (2pt,0pt) -- (-2pt,0pt) node[left] {$\bY_{\ytext}$};
		
		\node at (1,1) [circle,fill=red!60] {};
		\node at (1,5) [circle,fill=black] {};
		\node at (1,9) [circle,fill=black] {};
		\node at (3,3) [circle,fill=black] {};
		\node at (3,7) [circle,fill=red!60] {};
		\node at (5,1) [circle,fill=black] {};
		\node at (5,5) [circle,fill=red!60] {};
		\node at (5,9) [circle,fill=black] {};
		\node at (7,5) [circle,fill=black] {};
		\node at (9,1) [circle,fill=red!60] {};
		\node at (9,3) [circle,fill=black] {};
		\node at (9,7) [circle,fill=black] {};
		\node at (11,5) [circle,fill=black] {};
		\node at (11,9) [circle,fill=red!60] {};
		\node at (13,1) [circle,fill=red!60] {};
		\node at (13,9) [circle,fill=black] {};
		\node at (15,3) [circle,fill=red!60] {};
		\node at (15,7) [circle,fill=black] {};
		\node at (17,1) [circle,fill=black] {};
		\node at (17,3) [circle,fill=black] {};
		\node at (17,9) [circle,fill=red!60] {};
		\end{tikzpicture}
		\caption{The pruned graph} 
		\label{fig:Orion_Logo_Sub}
	\end{subfigure}%
	\caption{Example for the process of graph pruning.} 
	\label{fig:graph_pruning}
\end{figure}

We now prove that the cardinality of the set $\mathsf{G}_{i}(d,v_{x},v_{y},s)$ is upper--bounded by an expression that depends only on the numbers of vertices $(v_{x},v_{y})$. 
First, let $N(v_{x})$ be the number of options to choose $v_{x}$ different vectors from a set of cardinality $n$. Then,
\begin{align}
N(v_{x}) = n \cdot (n-1) \cdot (n-2) \cdots (n-v_{x}+1) \leq n^{v_{x}}.
\end{align}
Thus,
\begin{align}
|\mathsf{G}_{i}(d,v_{x},v_{y},s)| 
&= N(v_{x}) \cdot N(v_{y}) \\
\label{TERM2USE2}
&\leq n^{v_{x}} \cdot n^{v_{y}}.
\end{align}
Let $\Theta(G)$ be an indicator random variable that equals one if and only if all of the pairs of codewords that are linked by the edges of $G$ have $J(\bX,\bY)=1$. 
Using the above definitions, 
the inner sum of (\ref{K_Summands}) can now be written as:   
\begin{align}
&\sum_{\{(m_{i},m'_{i}) \in \Psi_{n},~1 \leq i \leq d,~ (m_{i},m'_{i}) \neq (m_{l},m'_{l}) ~ \forall i \neq l\}}  
\mathbb{E} \left[ J(\bX_{m_{1}},\bY_{m'_{1}}) \cdots J(\bX_{m_{d}},\bY_{m'_{d}}) \right] \nonumber \\
\label{STEP0}
&~~~~\equiv \sum_{\{(m_{i},m'_{i}) \in \Psi_{n},~1 \leq i \leq d,~ (m_{i},m'_{i}) \neq (m_{l},m'_{l}) ~ \forall i \neq l\}}  
\mathbb{E} \left[ \prod_{i=1}^{d} J(\bX_{m_{i}},\bY_{m'_{i}}) \right] \\
\label{STEP1}
&~~~~= \sum_{(v_{x},v_{y}) \in \calV(d)} 
\sum_{s=\mathcal{S}_{\mbox{\tiny min}}(d,v_{x},v_{y})}^{\mathcal{S}_{\mbox{\tiny max}}(d,v_{x},v_{y})}
\sum_{i=1}^{\mathsf{T}(d,v_{x},v_{y},s)}
\sum_{G \in \mathsf{G}_{i}(d,v_{x},v_{y},s)}  \mathbb{E} \left[ \Theta(G) \right],
\end{align}
i.e., for any $d \in \{1,2, \dotsc, k\}$, we first sum over the numbers of vertices, then over the number of subgraphs, later on, for a fixed quadruplet $(d,v_{x},v_{y},s)$, over all possible $\mathsf{T}(d,v_{x},v_{y},s)$ topologies, and finally, over all specific graphs $G \in \mathsf{G}_{i}(d,v_{x},v_{y},s)$ with a given topology.
One should note, that all limits of the three outer sums in \eqref{STEP1} depend only on $k$, while $|\mathsf{G}_{i}(d,v_{x},v_{y},s)|$ is the only one that depends also on $n$. 
It turns out that the expectation of $\Theta(G)$ can be easily evaluated if all subgraphs of $G$ are trees (also known as a \textit{forest} in the terminology of graph theory). If at least one subgraph of $G$ contains loops, we apply a process of \textit{graph pruning}, in which we cut out the minimal amount of edges\footnote{In fact, this procedure is equivalent to upper--bounding some of the indicator functions in \eqref{STEP0} by one.}, while keeping all vertices intact, until we get a forest (for example, see Figure \ref{fig:graph_pruning}).   

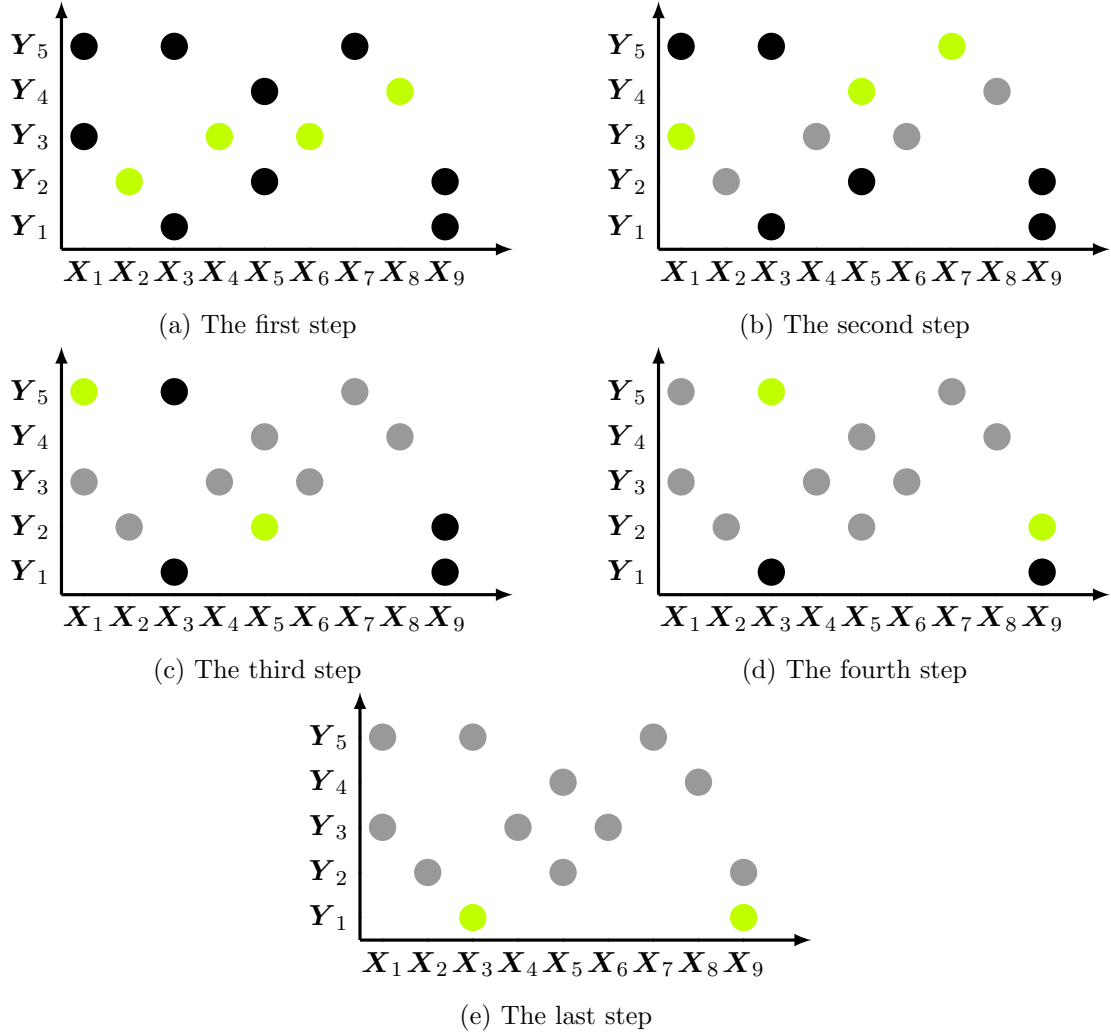
\begin{figure}[h!]
	\begin{subfigure}[b]{0.5\textwidth}
		\centering 
		\begin{tikzpicture}[scale=0.3]
		\draw[black,very thick,-latex] (0,0) -- (20,0);
		\draw[black,very thick,-latex] (0,0) -- (0,11);
		
		\foreach \x/\xtext in {1/1, 3/2, 5/3, 7/4, 9/5, 11/6, 13/7, 15/8, 17/9}
		\draw[shift={(\x,0)}] (0pt,2pt) -- (0pt,-2pt) node[below] {$\bX_{\xtext}$};
		\foreach \y/\ytext in {1/1, 3/2, 5/3, 7/4, 9/5}
		\draw[shift={(0,\y)}] (2pt,0pt) -- (-2pt,0pt) node[left] {$\bY_{\ytext}$};
		
		\node at (1,5) [circle,fill=black] {};
		\node at (1,9) [circle,fill=black] {};
		\node at (3,3) [circle,fill=lime] {};
		\node at (5,1) [circle,fill=black] {};
		\node at (5,9) [circle,fill=black] {};
		\node at (7,5) [circle,fill=lime] {};
		\node at (9,3) [circle,fill=black] {};
		\node at (9,7) [circle,fill=black] {};
		\node at (11,5) [circle,fill=lime] {};
		\node at (13,9) [circle,fill=black] {};
		\node at (15,7) [circle,fill=lime] {};
		\node at (17,1) [circle,fill=black] {};
		\node at (17,3) [circle,fill=black] {};
		\end{tikzpicture}
		\caption{The first step} 
		\label{fig:NASA_Logo_Sub}
	\end{subfigure}%
	\begin{subfigure}[b]{0.5\textwidth}
		\centering 
		\begin{tikzpicture}[scale=0.3]
		\draw[black,very thick,-latex] (0,0) -- (20,0);
		\draw[black,very thick,-latex] (0,0) -- (0,11);
		
		\foreach \x/\xtext in {1/1, 3/2, 5/3, 7/4, 9/5, 11/6, 13/7, 15/8, 17/9}
		\draw[shift={(\x,0)}] (0pt,2pt) -- (0pt,-2pt) node[below] {$\bX_{\xtext}$};
		\foreach \y/\ytext in {1/1, 3/2, 5/3, 7/4, 9/5}
		\draw[shift={(0,\y)}] (2pt,0pt) -- (-2pt,0pt) node[left] {$\bY_{\ytext}$};
		
		\node at (1,5) [circle,fill=lime] {};
		\node at (1,9) [circle,fill=black] {};
		\node at (3,3) [circle,fill=black!40] {};
		\node at (5,1) [circle,fill=black] {};
		\node at (5,9) [circle,fill=black] {};
		\node at (7,5) [circle,fill=black!40] {};
		\node at (9,3) [circle,fill=black] {};
		\node at (9,7) [circle,fill=lime] {};
		\node at (11,5) [circle,fill=black!40] {};
		\node at (13,9) [circle,fill=lime] {};
		\node at (15,7) [circle,fill=black!40] {};
		\node at (17,1) [circle,fill=black] {};
		\node at (17,3) [circle,fill=black] {};
		\end{tikzpicture}
		\caption{The second step} 
		\label{fig:Orion_Logo_Sub}
	\end{subfigure}%
	\\
	\centering
	\begin{subfigure}[b]{0.5\textwidth}
		\centering 
		\begin{tikzpicture}[scale=0.3]
		\draw[black,very thick,-latex] (0,0) -- (20,0);
		\draw[black,very thick,-latex] (0,0) -- (0,11);
		
		\foreach \x/\xtext in {1/1, 3/2, 5/3, 7/4, 9/5, 11/6, 13/7, 15/8, 17/9}
		\draw[shift={(\x,0)}] (0pt,2pt) -- (0pt,-2pt) node[below] {$\bX_{\xtext}$};
		\foreach \y/\ytext in {1/1, 3/2, 5/3, 7/4, 9/5}
		\draw[shift={(0,\y)}] (2pt,0pt) -- (-2pt,0pt) node[left] {$\bY_{\ytext}$};
		
		\node at (1,5) [circle,fill=black!40] {};
		\node at (1,9) [circle,fill=lime] {};
		\node at (3,3) [circle,fill=black!40] {};
		\node at (5,1) [circle,fill=black] {};
		\node at (5,9) [circle,fill=black] {};
		\node at (7,5) [circle,fill=black!40] {};
		\node at (9,3) [circle,fill=lime] {};
		\node at (9,7) [circle,fill=black!40] {};
		\node at (11,5) [circle,fill=black!40] {};
		\node at (13,9) [circle,fill=black!40] {};
		\node at (15,7) [circle,fill=black!40] {};
		\node at (17,1) [circle,fill=black] {};
		\node at (17,3) [circle,fill=black] {};
		\end{tikzpicture}
		\caption{The third step} 
		\label{fig:Orion_Logo_Sub}
	\end{subfigure}%
	\begin{subfigure}[b]{0.5\textwidth}
		\centering 
		\begin{tikzpicture}[scale=0.3]
		\draw[black,very thick,-latex] (0,0) -- (20,0);
		\draw[black,very thick,-latex] (0,0) -- (0,11);
		
		\foreach \x/\xtext in {1/1, 3/2, 5/3, 7/4, 9/5, 11/6, 13/7, 15/8, 17/9}
		\draw[shift={(\x,0)}] (0pt,2pt) -- (0pt,-2pt) node[below] {$\bX_{\xtext}$};
		\foreach \y/\ytext in {1/1, 3/2, 5/3, 7/4, 9/5}
		\draw[shift={(0,\y)}] (2pt,0pt) -- (-2pt,0pt) node[left] {$\bY_{\ytext}$};
		
		\node at (1,5) [circle,fill=black!40] {};
		\node at (1,9) [circle,fill=black!40] {};
		\node at (3,3) [circle,fill=black!40] {};
		\node at (5,1) [circle,fill=black] {};
		\node at (5,9) [circle,fill=lime] {};
		\node at (7,5) [circle,fill=black!40] {};
		\node at (9,3) [circle,fill=black!40] {};
		\node at (9,7) [circle,fill=black!40] {};
		\node at (11,5) [circle,fill=black!40] {};
		\node at (13,9) [circle,fill=black!40] {};
		\node at (15,7) [circle,fill=black!40] {};
		\node at (17,1) [circle,fill=black] {};
		\node at (17,3) [circle,fill=lime] {};
		\end{tikzpicture}
		\caption{The fourth step} 
		\label{fig:Orion_Logo_Sub}
	\end{subfigure}%
	\\
	\centering
	\begin{subfigure}[b]{0.5\textwidth}
		\centering 
		\begin{tikzpicture}[scale=0.3]
		\draw[black,very thick,-latex] (0,0) -- (20,0);
		\draw[black,very thick,-latex] (0,0) -- (0,11);
		
		\foreach \x/\xtext in {1/1, 3/2, 5/3, 7/4, 9/5, 11/6, 13/7, 15/8, 17/9}
		\draw[shift={(\x,0)}] (0pt,2pt) -- (0pt,-2pt) node[below] {$\bX_{\xtext}$};
		\foreach \y/\ytext in {1/1, 3/2, 5/3, 7/4, 9/5}
		\draw[shift={(0,\y)}] (2pt,0pt) -- (-2pt,0pt) node[left] {$\bY_{\ytext}$};
		
		\node at (1,5) [circle,fill=black!40] {};
		\node at (1,9) [circle,fill=black!40] {};
		\node at (3,3) [circle,fill=black!40] {};
		\node at (5,1) [circle,fill=lime] {};
		\node at (5,9) [circle,fill=black!40] {};
		\node at (7,5) [circle,fill=black!40] {};
		\node at (9,3) [circle,fill=black!40] {};
		\node at (9,7) [circle,fill=black!40] {};
		\node at (11,5) [circle,fill=black!40] {};
		\node at (13,9) [circle,fill=black!40] {};
		\node at (15,7) [circle,fill=black!40] {};
		\node at (17,1) [circle,fill=lime] {};
		\node at (17,3) [circle,fill=black!40] {};
		\end{tikzpicture}
		\caption{The last step} 
		\label{fig:Orion_Logo_Sub}
	\end{subfigure}%
	\caption{Example for the process of graph reduction.} 
	\label{fig:graph_reduction}
\end{figure}

Denote by $\mathcal{P}(G)$ the pruned graph of $G$. Notice that the expectation of $\Theta(G)$ is upper-bounded\footnote{It follows from the fact that $\Theta(G) \leq\Theta(\mathcal{P}(G))$ with probability one.} by the expectation of $\Theta(\mathcal{P}(G))$, which can be evaluated in a simple iterative process of \textit{graph reduction}. 
In the first step, we take the expectation with respect to all vectors that are labels of leaf vertices in $\mathcal{P}(G)$,
while we condition on the realizations of all other vectors, those corresponding to inner vertices in $\mathcal{P}(G)$; afterwards, we erase leaf vector vertices and corresponding edges.
Successive steps are identical to the first one on the remaining (unerased) graph, continuing until all vectors that are attributed to $G$ have been considered (for example, see Figure \ref{fig:graph_reduction}). 

In each step of the graph reduction, the expectation with respect to each of the leaf vectors is given by 
$\prob \left\{J(\bx,\bY_{\mbox{\tiny leaf}})=1 \right\} \leq Q$,
since we condition on the realizations of vectors that are attributed to the inner vertices.
For any graph $G \in \mathsf{G}_{i}(d,v_{x},v_{y},s)$, we conclude that the expectation of $\Theta(\mathcal{P}(G))$ is upper-bounded by $Q^{\calE(G)}$, where $\calE(G)$ is the number of edges in $\calP(G)$. 
Of course, if $G$ is already a forest, then $\mathcal{E}(G) = d$.
For any $G \in \mathsf{G}_{i}(d,v_{x},v_{y},s)$ which is not a forest, we find $\mathcal{E}(G)$ as follows. Assume that $v_{x}(1),v_{x}(2), \dotsc , v_{x}(s)$ and $v_{y}(1),v_{y}(2), \dotsc , v_{y}(s)$ are the numbers of vertices in each of the $s$ subgraphs of $G$. Then, in the process of graph pruning, each of the $j \in \{1,2, \dotsc,s\}$ subgraphs of $G$ will transform into a tree with exactly $v_{x}(j) + v_{y}(j) - 1$ edges. Hence,
\begin{align}
\label{TERM2USE1}
\mathcal{E}(G) = \sum_{j=1}^{s} (v_{x}(j) + v_{y}(j) - 1) = v_{x} + v_{y} - s.
\end{align}  
The innermost sum of (\ref{STEP1}) can be treated as follows:  
\begin{align}
\sum_{G \in \mathsf{G}_{i}(d,v_{x},v_{y},s)}  \mathbb{E} \left[ \Theta(G) \right] 
&\leq \sum_{G \in \mathsf{G}_{i}(d,v_{x},v_{y},s)}  \mathbb{E} \left[ \Theta(\mathcal{P}(G)) \right] \\
&\leq \sum_{G \in \mathsf{G}_{i}(d,v_{x},v_{y},s)} Q^{\calE(G)} \\
\label{TERM2EXPh2}
&= \sum_{G \in \mathsf{G}_{i}(d,v_{x},v_{y},s)}  Q^{v_{x}+v_{y}-s} \\
&= |\mathsf{G}_{i}(d,v_{x},v_{y},s)| \cdot Q^{v_{x}+v_{y}-s} \\
\label{TERM2EXPh3}
&\leq n^{v_{x}} \cdot n^{v_{y}} \cdot Q^{v_{x}+v_{y}-s},
\end{align}
where \eqref{TERM2EXPh2} follows from \eqref{TERM2USE1} and \eqref{TERM2EXPh3} is due to \eqref{TERM2USE2}.  
Next, we substitute \eqref{TERM2EXPh3} back into (\ref{STEP1}) and get that
\begin{align}
&\sum_{s=\mathcal{S}_{\mbox{\tiny min}}(d,v_{x},v_{y})}^{\mathcal{S}_{\mbox{\tiny max}}(d,v_{x},v_{y})}
\sum_{i=1}^{\mathsf{T}(d,v_{x},v_{y},s)} n^{v_{x}} \cdot n^{v_{y}} \cdot Q^{v_{x}+v_{y}-s} \nn \\
&~~~~~~= \sum_{s=\mathcal{S}_{\mbox{\tiny min}}(d,v_{x},v_{y})}^{\mathcal{S}_{\mbox{\tiny max}}(d,v_{x},v_{y})}
\mathsf{T}(d,v_{x},v_{y},s) \cdot n^{v_{x}} \cdot n^{v_{y}} \cdot Q^{v_{x}+v_{y}-s} \\
\label{TERM2EXPh4}
&~~~~~~\leq \left( \sum_{s=\mathcal{S}_{\mbox{\tiny min}}(d,v_{x},v_{y})}^{\mathcal{S}_{\mbox{\tiny max}}(d,v_{x},v_{y})}
\mathsf{T}(d,v_{x},v_{y},s) \right) \cdot n^{v_{x}} \cdot n^{v_{y}} \cdot Q^{v_{x}+v_{y}-\mathcal{S}_{\mbox{\tiny max}}(d,v_{x},v_{y})} \\
\label{TERM2EXPh5}
&~~~~~~= \mathsf{T}(d,v_{x},v_{y}) \cdot n^{v_{x}} \cdot n^{v_{y}} \cdot Q^{v_{x}+v_{y}-\mathcal{S}_{\mbox{\tiny max}}(d,v_{x},v_{y})}, 
\end{align}
where \eqref{TERM2EXPh4} is true since $Q \in [0,1]$, such that replacing $s$ by its maximal value $\mathcal{S}_{\mbox{\tiny max}}(d,v_{x},v_{y})$ provides an upper bound. The passage \eqref{TERM2EXPh5} follows from the definitions of $\mathsf{T}(d,v_{x},v_{y})$ and $\mathsf{T}(d,v_{x},v_{y},s)$. 
Before we substitute it back into \eqref{STEP1} and then into (\ref{K_Summands}), we summarize some minor results that will be needed in the sequel. Let us define $d^{*} = \max\{v_{x},v_{y}\}$.
\begin{lemma} \label{HELPER_LEMMA}
	We have the following.
	\begin{enumerate}
		\item For fixed $(v_{x},v_{y})$, $\mathcal{S}_{\mbox{\tiny max}}(d,v_{x},v_{y})$ is a non--increasing sequence of $d$.
		\item For any $(v_{x},v_{y})$, we have that $\mathcal{S}_{\mbox{\tiny max}}(d^{*},v_{x},v_{y})=\min\{v_{x},v_{y}\}$.
	\end{enumerate}
\end{lemma}
\textbf{Proof:} 
\begin{enumerate}
	\item For fixed $(v_{x},v_{y})$, if we add to the graph an edge, we have only two options. On the one hand, we may connect vertices that belong to the same subgraph, such that the number of subgraphs remains the same. On the other hand, we can connect vertices that belong to different subgraphs, and then, the number of subgraphs decreases.  
	\item Without loss of generality, assume that $v_{x} \leq v_{y}$. Then $d^{*}=v_{y}$, hence it follows by definition that each column contains exactly one edge, such that the set of edges in each row provides a subgraph. Thus, the number of subgraphs equals exactly $v_{x}$. 
\end{enumerate}

Now, we have that
\begin{align}
\mathbb{E} \left[ N^{k}  \right]
&\leq \sum_{d=1}^{k} \mathsf{S}(k,d) \sum_{(v_{x},v_{y}) \in \calV(d)} 
\mathsf{T}(d,v_{x},v_{y}) \cdot n^{v_{x}} \cdot n^{v_{y}} \cdot Q^{v_{x}+v_{y}-\mathcal{S}_{\mbox{\tiny max}}(d,v_{x},v_{y})} \\
\label{Summation_Set}
&= \sum_{d=1}^{k} \sum_{(v_{x},v_{y}) \in \calV(d)}
\mathsf{S}(k,d) \cdot \mathsf{T}(d,v_{x},v_{y}) \cdot n^{v_{x}} \cdot n^{v_{y}} \cdot Q^{v_{x}+v_{y}-\mathcal{S}_{\mbox{\tiny max}}(d,v_{x},v_{y})} \\
\label{CHANGING_ORDER}
&= \sum_{v_{x}=1}^{k} \sum_{v_{y}=1}^{k} \sum_{d=\max\{v_{x},v_{y}\}}^{\min\{k,v_{x} \cdot v_{y}\}}  
\mathsf{S}(k,d) \cdot \mathsf{T}(d,v_{x},v_{y}) \cdot n^{v_{x}} \cdot n^{v_{y}} \cdot Q^{v_{x}+v_{y}-\mathcal{S}_{\mbox{\tiny max}}(d,v_{x},v_{y})} \\
\label{TERM2EXPh6}
&\leq \sum_{v_{x}=1}^{k} \sum_{v_{y}=1}^{k} 
\left( \sum_{d=\max\{v_{x},v_{y}\}}^{\min\{k,v_{x} \cdot v_{y}\}}  
\mathsf{S}(k,d) \cdot \mathsf{T}(d,v_{x},v_{y}) \right) \cdot n^{v_{x}} \cdot n^{v_{y}}  
\cdot Q^{v_{x}+v_{y}-\mathcal{S}_{\mbox{\tiny max}}(d^{*},v_{x},v_{y})},
\end{align}
where in \eqref{CHANGING_ORDER} we changed the order of summation, and \eqref{TERM2EXPh6} is true since $Q \in [0,1]$, and thus, according to the first point in Lemma \ref{HELPER_LEMMA}, we attain an upper bound by substituting $d=d^{*}$.
Moving forward, we use the trivial bound
\begin{align}
\mathsf{T}(d,v_{x},v_{y}) \leq \binom{v_{x}v_{y}}{d} \leq \frac{(v_{x}v_{y})^{d}}{d!},
\end{align} 
and arrive at
\begin{align}
\mathbb{E} \left[ N^{k}  \right]
\label{TERM2EXPh7}
&\leq \sum_{v_{x}=1}^{k} \sum_{v_{y}=1}^{k} 
\left( \sum_{d=\max\{v_{x},v_{y}\}}^{\min\{k,v_{x} \cdot v_{y}\}}  
\mathsf{S}(k,d) \cdot \frac{(v_{x}v_{y})^{d}}{d!} \right) \cdot n^{v_{x}} \cdot n^{v_{y}} \cdot   
Q^{v_{x}+v_{y} - \min\{v_{x},v_{y}\}} \\
&\leq \sum_{v_{x}=1}^{k} \sum_{v_{y}=1}^{k}
\left( \sum_{d=1}^{k}  
\mathsf{S}(k,d) \cdot \frac{(v_{x}v_{y})^{d}}{d!} \right) \cdot n^{v_{x}} \cdot n^{v_{y}} \cdot   
Q^{\max\{v_{x},v_{y}\}} \\
&\leq \sum_{v_{x}=1}^{k} \sum_{v_{y}=1}^{k}
\left( \sum_{d=1}^{k}  
\mathsf{S}(k,d) \cdot \frac{k^{2d}}{d!} \right) \cdot n^{v_{x}} \cdot n^{v_{y}} \cdot   
Q^{\max\{v_{x},v_{y}\}} \\
\label{TERM2EXPh8}
&= \sum_{v_{x}=1}^{k} \sum_{v_{y}=1}^{k}
\mathsf{B}(k) \cdot n^{v_{x}} \cdot n^{v_{y}} \cdot  Q^{\max\{v_{x},v_{y}\}},
\end{align}
where \eqref{TERM2EXPh7} follows from the second point in Lemma \ref{HELPER_LEMMA} and \eqref{TERM2EXPh8} follows from the definition in \eqref{DEF_Bk}.
Finally,
\begin{align}
\mathbb{E} \left[ N^{k} \right]
&\leq \sum_{v_{x}=1}^{k} \sum_{v_{y}=1}^{v_{x}} 
\mathsf{B}(k) \cdot n^{v_{x}} \cdot n^{v_{y}} \cdot   
Q^{\max\{v_{x},v_{y}\}} \nn \\
&~~~~~~+ \sum_{v_{y}=1}^{k} \sum_{v_{x}=1}^{v_{y}} 
\mathsf{B}(k) \cdot n^{v_{x}} \cdot n^{v_{y}} \cdot   
Q^{\max\{v_{x},v_{y}\}} \\
&= \sum_{v_{x}=1}^{k} \sum_{v_{y}=1}^{v_{x}} 
\mathsf{B}(k) \cdot n^{v_{x}} \cdot n^{v_{y}} \cdot   
Q^{v_{x}} \nn \\
&~~~~~~+ \sum_{v_{y}=1}^{k} \sum_{v_{x}=1}^{v_{y}} 
\mathsf{B}(k) \cdot n^{v_{x}} \cdot n^{v_{y}} \cdot   
Q^{v_{y}} \\
&\leq \sum_{v_{x}=1}^{k} v_{x} \cdot  
\mathsf{B}(k) \cdot (n^{2}Q)^{v_{x}} + \sum_{v_{y}=1}^{k} v_{y} \cdot
\mathsf{B}(k) \cdot (n^{2}Q)^{v_{y}} \\
\label{ref1}
&= 2 \mathsf{B}(k) \sum_{\ell=1}^{k}  
\ell \cdot (n^{2}Q)^{\ell}.
\end{align}

Now, if $n^{2}Q \geq 1$, then
\begin{align}
\mathbb{E} \left[ N^{k} \right]
&\leq \left( 2\mathsf{B}(k) \sum_{\ell=1}^{k}  
\ell \right) \cdot (n^{2}Q)^{k} \\
&\leq k(k+1)\mathsf{B}(k) \cdot (n^{2}Q)^{k},
\end{align}
and otherwise, if $n^{2}Q < 1$, then
\begin{align}
\mathbb{E} \left[ N^{k} \right]
&\leq k(k+1)\mathsf{B}(k) \cdot n^{2}Q.
\end{align}
Thus, Lemma \ref{Lemma_Square_LD_Integers} is proved.

\section*{Appendix C - Proof of Theorem \ref{Theorem_Alignment_Recovery}}
\renewcommand{\theequation}{C.\arabic{equation}}
\setcounter{equation}{0}

\subsection*{Erroneous Recovery Error Probability}

Define the error events:
\begin{align}
\calE_{1} = \bigcup_{i=1}^{n} \calA_{i,i} = \bigcup_{i=1}^{n} \left\{\tilde{\bX}_{i}^{\mathsf{T}} \tilde{\bY}_{i} \leq \theta \right\},
\end{align}
and 
\begin{align}
\calE_{2} = \bigcup_{i=1}^{n} \bigcup_{j \neq i} \calA_{i,j} = \bigcup_{i=1}^{n} \bigcup_{j \neq i} \left\{\tilde{\bX}_{i}^{\mathsf{T}} \tilde{\bY}_{j} \geq \theta \right\}.
\end{align}
Then, the probability of alignment error is given by
\begin{align}
P_{\mbox{\scriptsize e,ER}}(\psi_{\mbox{\tiny e}})
= \pr_{1|\sigma}\left\{\calE_{1} \cup \calE_{2} \right\},
\end{align}
and can be upper-bounded using the clipped union bound as 
\begin{align}
P_{\mbox{\scriptsize e,ER}}(\psi_{\mbox{\tiny e}})
&= \pr_{1|\sigma}\left\{\bigcup_{i=1}^{n} \left\{\tilde{\bX}_{i}^{\mathsf{T}} \tilde{\bY}_{i} \leq \theta \right\} \cup \bigcup_{i=1}^{n} \bigcup_{j \neq i} \left\{\tilde{\bX}_{i}^{\mathsf{T}} \tilde{\bY}_{j} \geq \theta \right\}\right\} \\
&\leq \min\left[1, \sum_{i=1}^{n} \pr_{1|\sigma}\left\{ \tilde{\bX}_{i}^{\mathsf{T}} \tilde{\bY}_{i} \leq \theta \right\} 
+ \sum_{i=1}^{n} \sum_{j \neq i} \pr_{1|\sigma}\left\{ \tilde{\bX}_{i}^{\mathsf{T}} \tilde{\bY}_{j} \geq \theta \right\}\right] \\
&= \min\left[1,\sum_{i=1}^{n} \left[1 - \pr_{1|\sigma}\left\{ \tilde{\bX}_{i}^{\mathsf{T}} \tilde{\bY}_{i} \geq \theta \right\}\right] 
+ \sum_{i=1}^{n} \sum_{j \neq i} \pr_{1|\sigma}\left\{ \tilde{\bX}_{i}^{\mathsf{T}} \tilde{\bY}_{j} \geq \theta \right\}\right] \\
\label{ref37}
&= \min\left[1,n(1-P) + n(n-1)Q\right],  
\end{align}
which proves \eqref{ER_Error1}.

In order to derive a lower bound on $P_{\mbox{\scriptsize e,ER}}(\psi_{\mbox{\tiny e}})$, we first invoke de Caen's bound \cite{DeCaen}.  
\begin{proposition}
	Let $\{A_{i}\}_{i \in \calI}$ be any finite set of events in a probability space $(\Omega,\calF,\pr)$. The probability of the union $\pr\{\cup_{i\in\calI}A_{i}\}$ is lower-bounded by
	\begin{align}
	\pr\left\{\bigcup_{i\in\calI}A_{i}\right\} 
	\geq \sum_{i\in\calI}\frac{\pr\{A_{i}\}^{2}}{\sum_{j\in\calI}\pr\{A_{i}\cap A_{j}\}}.
	\end{align}
\end{proposition}

Now,
\begin{align}
P_{\mbox{\scriptsize e,ER}}(\psi_{\mbox{\tiny e}})
&= \pr_{1|\sigma}\left\{\bigcup_{i=1}^{n} \bigcup_{j=1}^{n} \calA_{i,j}\right\} \\
&\geq \sum_{i=1}^{n}\sum_{j=1}^{n}\frac{\pr\{\calA_{i,j}\}^{2}}{ \sum_{i'=1}^{n}\sum_{j'=1}^{n}\pr\{\calA_{i,j}\cap \calA_{i',j'}\}} \\
&= \sum_{i=1}^{n}\sum_{j=1}^{n}\frac{\pr\{\calA_{i,j}\}^{2}}{ \pr\{\calA_{i,j}\} + \sum_{(i',j')\neq(i,j)}\pr\{\calA_{i,j}\cap \calA_{i',j'}\}} \\
&= \sum_{i=1}^{n}\sum_{j=1}^{n}\frac{\pr\{\calA_{i,j}\}^{2}}{ \pr\{\calA_{i,j}\} + \sum_{(i',j')\neq(i,j)}\pr\{\calA_{i,j}\}\pr\{\calA_{i',j'}\}} \\
&= \sum_{i=1}^{n}\sum_{j=1}^{n}\frac{\pr\{\calA_{i,j}\}}{ 1 + \sum_{(i',j')\neq(i,j)}\pr\{\calA_{i',j'}\}} \\
&= \sum_{i=1}^{n}\frac{\pr\{\calA_{i,i}\}}{ 1 + \sum_{(i',j')\neq(i,i)}\pr\{\calA_{i',j'}\}} 
+
\sum_{i=1}^{n}\sum_{j \neq i}\frac{\pr\{\calA_{i,j}\}}{ 1 + \sum_{(i',j')\neq(i,j)}\pr\{\calA_{i',j'}\}}.
\end{align}
Substituting $\pr\{\calA_{i,i}\}=1-P$ and $\pr\{\calA_{i,j}\}=Q$ yields that
\begin{align}
P_{\mbox{\scriptsize e,ER}}(\psi_{\mbox{\tiny e}})
&\geq \sum_{i=1}^{n}\frac{1-P}{1 + (n-1)(1-P)+n(n-1)Q} 
+
\sum_{i=1}^{n}\sum_{j \neq i}\frac{Q}{1 + n(1-P)+[n(n-1)-1]Q} \\
&= \frac{n(1-P)}{P + n(1-P) + n(n-1)Q} 
+ \frac{n(n-1)Q}{1-Q + n(1-P)+n(n-1)Q} \\
&\geq \frac{n(1-P)+n(n-1)Q}{\max\{P,1-Q\} + n(1-P) + n(n-1)Q}.
\end{align}
The proof of \eqref{ER_Error2} is now complete.

\subsection*{Mismatch-Undetected Error Probability}
Define the events
\begin{align}
\calB_{k,\ell} = \{(k,\ell)\text{ is an undetected error}\}.
\end{align}
It follows by the clipped union bound that
\begin{align}
P_{\mbox{\scriptsize e,MU}}(\psi_{\mbox{\tiny e}})
&= \pr\left\{\bigcup_{k=1}^{n}\bigcup_{\ell \neq k} \calB_{k,\ell} \right\} \\ 
&\leq \min\left[1,\sum_{k=1}^{n}\sum_{\ell \neq k} \pr\left\{\calB_{k,\ell}\right\}\right].
\end{align}
Now, for a specific pair of sequences to cause an undetected error, say $(\bX_{k},\bY_{\ell})$, the rest of the $k$-th column and the $\ell$-th row must be empty of points; see Figure \ref{fig:Undetected_Error} below. 

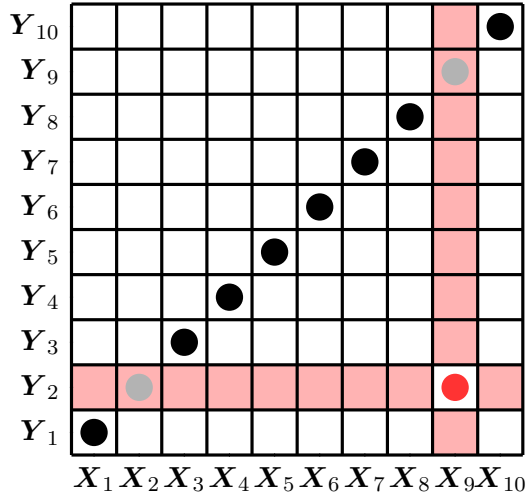
\begin{figure}[h!]
	\centering
	\begin{tikzpicture}[scale=0.3]
	
	\draw[fill=red!30!white] (16,4) -- (16,20) -- (18,20) -- (18,4) -- (16,4);
	\draw[fill=red!30!white] (16,0) -- (16,2) -- (18,2) -- (18,0) -- (16,0);
	
	\draw[fill=red!30!white] (0,2) -- (0,4) -- (16,4) -- (16,2) -- (0,2);
	\draw[fill=red!30!white] (18,2) -- (18,4) -- (20,4) -- (20,2) -- (18,2);
	
	\draw[black,very thick] (0,0) -- (20,0);
	\draw[black,very thick] (0,2) -- (20,2);
	\draw[black,very thick] (0,4) -- (20,4);
	\draw[black,very thick] (0,6) -- (20,6);
	\draw[black,very thick] (0,8) -- (20,8);
	\draw[black,very thick] (0,10) -- (20,10);
	\draw[black,very thick] (0,12) -- (20,12);
	\draw[black,very thick] (0,14) -- (20,14);
	\draw[black,very thick] (0,16) -- (20,16);
	\draw[black,very thick] (0,18) -- (20,18);
	\draw[black,very thick] (0,20) -- (20,20);
	
	\draw[black,very thick] (0,0) -- (0,20);
	\draw[black,very thick] (2,0) -- (2,20);
	\draw[black,very thick] (4,0) -- (4,20);
	\draw[black,very thick] (6,0) -- (6,20);
	\draw[black,very thick] (8,0) -- (8,20);
	\draw[black,very thick] (10,0) -- (10,20);
	\draw[black,very thick] (12,0) -- (12,20);
	\draw[black,very thick] (14,0) -- (14,20);
	\draw[black,very thick] (16,0) -- (16,20);
	\draw[black,very thick] (18,0) -- (18,20);
	\draw[black,very thick] (20,0) -- (20,20);
	
	\foreach \x/\xtext in {1/1, 3/2, 5/3, 7/4, 9/5, 11/6, 13/7, 15/8, 17/9, 19/10}
	\draw[shift={(\x,0)}] (0pt,2pt) -- (0pt,-2pt) node[below] {$\bX_{\xtext}$};
	\foreach \y/\ytext in {1/1, 3/2, 5/3, 7/4, 9/5, 11/6, 13/7, 15/8, 17/9, 19/10}
	\draw[shift={(0,\y)}] (2pt,0pt) -- (-2pt,0pt) node[left] {$\bY_{\ytext}$};
	
	\node at (17,3) [circle,fill=red!80] {};
	
	\node at (1,1) [circle,fill=black] {};
	\node at (3,3) [circle,fill=gray!60] {};
	\node at (5,5) [circle,fill=black] {};
	\node at (7,7) [circle,fill=black] {};
	\node at (9,9) [circle,fill=black] {};
	\node at (11,11) [circle,fill=black] {};
	\node at (13,13) [circle,fill=black] {};
	\node at (15,15) [circle,fill=black] {};
	\node at (17,17) [circle,fill=gray!60] {};
	\node at (19,19) [circle,fill=black] {};
	
	\end{tikzpicture}
	\caption{An example of a table with an undetected error caused by the pair $(\bX_{9},\bY_{2})$.} 
	\label{fig:Undetected_Error}
\end{figure}

Let us denote in short $\calI_{i,j}=\IND\left\{\tilde{\bX}_{i}^{\mathsf{T}} \tilde{\bY}_{j} \geq \theta \right\}$. Thus,
\begin{align}
\pr\left\{\calB_{k,\ell}\right\}
&= \pr\left\{\calI_{k,\ell}\cap\bigcap_{i \neq k} \calI_{i,\ell}^{\mbox{\scriptsize c}}\cap\bigcap_{j \neq \ell} \calI_{k,j}^{\mbox{\scriptsize c}} \right\} \\
&= \int \pr\left\{\calI_{k,\ell}\cap\bigcap_{i \neq k} \calI_{i,\ell}^{\mbox{\scriptsize c}}\cap\bigcap_{j \neq \ell} \calI_{k,j}^{\mbox{\scriptsize c}} \middle| \bX_{k}=\bx, \bY_{\ell}=\by \right\} f_{\bX_{k}}(\bx) f_{\bY_{\ell}}(\by) \dint\bx \dint\by \\
&= \int \IND\left\{\tilde{\bx}^{\mathsf{T}} \tilde{\by} \geq \theta \right\} (1-P)^{2} (1-Q)^{2n-4} f_{\bX_{k}}(\bx) f_{\bY_{\ell}}(\by) \dint\bx \dint\by \\
&= \pr\left\{\tilde{\bX}_{k}^{\mathsf{T}} \tilde{\bY}_{\ell} \geq \theta \right\} (1-P)^{2} (1-Q)^{2n-4} \\
\label{ref41}
&= Q (1-P)^{2} (1-Q)^{2n-4}.
\end{align}
Substituting back yields
\begin{align}
P_{\mbox{\scriptsize e,MU}}(\psi_{\mbox{\tiny e}}) 
&\leq \min\left[1,\sum_{k=1}^{n}\sum_{\ell \neq k} Q (1-P)^{2} (1-Q)^{2n-4}\right] \\
&= \min\left[1,n(n-1) Q (1-P)^{2} (1-Q)^{2n-4}\right],
\end{align}
which is exactly \eqref{MU_Error1}.

\section*{Appendix D - Proof of Lemma \ref{Lemma_P_Bound}}
\renewcommand{\theequation}{D.\arabic{equation}}
\setcounter{equation}{0}

Observe that under the assumption of  
\begin{align}
(\bX,\bY) \sim \calN^{\otimes d}\left(\left[\begin{matrix} 0 \\ 0
\end{matrix}\right], \left[\begin{matrix} 1 & \rho \\ \rho & 1 \end{matrix}\right] \right),
\end{align}
it holds that $\bY = \rho\bX + \sqrt{1-\rho^{2}}\bZ$ with $\bZ \sim \calN(\boldsymbol{0}_{d},\bI_{d})$ and independent of $\bX$.
Consider the following for some $\theta \in (0,1)$
\begin{align}
\pr\left\{\tilde{\bX}^{\mathsf{T}} \tilde{\bY} \geq \theta \right\} 
&= \pr\left\{\frac{\bX^{\mathsf{T}}\bY}{\|\bX\|\|\bY\|} \geq \theta \right\} \\
&= \pr\left\{\frac{\bX^{\mathsf{T}}\left(\rho\bX + \sqrt{1-\rho^{2}}\bZ\right)}{\|\bX\|\left\|\rho\bX + \sqrt{1-\rho^{2}}\bZ\right\|} \geq \theta \right\} \\
\label{ref42}
&= \pr\left\{\frac{\rho\|\bX\|^{2} + \sqrt{1-\rho^{2}}\bX^{\mathsf{T}}\bZ}
{\|\bX\| \sqrt{\rho^{2}\|\bX\|^{2} + 2\rho\sqrt{1-\rho^{2}}\bX^{\tr}\bZ + (1-\rho^{2})\|\bZ\|^{2}}} \geq \theta \right\}.
\end{align}
Since $\theta > 0$, then the event in \eqref{ref42} can occur only if $\rho\|\bX\|^{2} + \sqrt{1-\rho^{2}}\bX^{\mathsf{T}}\bZ \geq 0$. Let us denote
\begin{align}
\calF_{\rho}(\bX,\bZ) = \{\rho\|\bX\|^{2} + \sqrt{1-\rho^{2}}\bX^{\mathsf{T}}\bZ \geq 0\}.
\end{align}  
Then,
\begin{align}
&\pr\left\{\tilde{\bX}^{\mathsf{T}} \tilde{\bY} \geq \theta \right\} \nn \\ 
&= \pr\left\{\frac{\rho\|\bX\|^{2} + \sqrt{1-\rho^{2}}\bX^{\mathsf{T}}\bZ}
{\|\bX\| \sqrt{\rho^{2}\|\bX\|^{2} + 2\rho\sqrt{1-\rho^{2}}\bX^{\tr}\bZ + (1-\rho^{2})\|\bZ\|^{2}}} \geq \theta , \calF_{\rho}(\bX,\bZ) \right\} \\
&= \pr\left\{\frac{\rho^{2}\|\bX\|^{4} + 2 \rho \sqrt{1-\rho^{2}} \|\bX\|^{2} \bX^{\mathsf{T}}\bZ
+(1-\rho^{2})\left(\bX^{\mathsf{T}}\bZ\right)^{2}}
{\rho^{2}\|\bX\|^{4} + 2\rho\sqrt{1-\rho^{2}}\|\bX\|^{2}\bX^{\tr}\bZ + (1-\rho^{2})\|\bX\|^{2}\|\bZ\|^{2}} \geq \theta^{2} , \calF_{\rho}(\bX,\bZ) \right\} \\
&= \pr\left\{\frac{\rho^{2}\|\bX\|^{4} + 2 \rho \sqrt{1-\rho^{2}} \|\bX\|^{3}\|\bZ\|\cdot\cos(\Phi_{\txz})
	+(1-\rho^{2})\|\bX\|^{2}\|\bZ\|^{2}\cdot\cos^{2}(\Phi_{\txz})}
{\rho^{2}\|\bX\|^{4} + 2\rho\sqrt{1-\rho^{2}}\|\bX\|^{3}\|\bZ\|\cdot\cos(\Phi_{\txz}) + (1-\rho^{2})\|\bX\|^{2}\|\bZ\|^{2}} \geq \theta^{2} , \calF_{\rho}(\bX,\bZ) \right\} \\
\label{ref40}
&= \pr\left\{\frac{\rho^{2}\|\bX\|^{2} + 2 \rho \sqrt{1-\rho^{2}} \|\bX\|\|\bZ\|\cdot\cos(\Phi_{\txz})
	+(1-\rho^{2})\|\bZ\|^{2}\cdot\cos^{2}(\Phi_{\txz})}
{\rho^{2}\|\bX\|^{2} + 2\rho\sqrt{1-\rho^{2}}\|\bX\|\|\bZ\|\cdot\cos(\Phi_{\txz}) + (1-\rho^{2})\|\bZ\|^{2}} \geq \theta^{2} , \calF_{\rho}(\bX,\bZ) \right\}.
\end{align}
Rearranging, we get that
\begin{align}
&\pr\left\{\tilde{\bX}^{\mathsf{T}} \tilde{\bY} \geq \theta \right\} \nn \\
&= \pr\left\{ (1-\rho^{2})\|\bZ\|^{2}\cdot\cos^{2}(\Phi_{\txz}) + 2 \rho \sqrt{1-\rho^{2}} (1-\theta^{2}) \|\bX\|\|\bZ\|\cdot\cos(\Phi_{\txz}) \right. \nn \\
&~~~~~~~~~~~~~~~~\left. + \rho^{2}(1-\theta^{2})\|\bX\|^{2} 
- (1-\rho^{2})\theta^{2}\|\bZ\|^{2} \geq 0 , \calF_{\rho}(\bX,\bZ) \right\} \\
&= \pr\left\{ (1-\rho^{2})\cdot\cos^{2}(\Phi_{\txz}) + 2 \rho \sqrt{1-\rho^{2}} (1-\theta^{2}) \frac{\|\bX\|}{\|\bZ\|}\cdot\cos(\Phi_{\txz}) \right. \nn \\
&~~~~~~~~~~~~~~~~\left. + \rho^{2}(1-\theta^{2})\frac{\|\bX\|^{2}}{\|\bZ\|^{2}} 
- (1-\rho^{2})\theta^{2} \geq 0 , \calF_{\rho}(\bX,\bZ) \right\} \\
&\dfn \pr\left\{\cos^{2}(\Phi_{\txz}) + \frac{2\rho  (1-\theta^{2})}{\sqrt{1-\rho^{2}}} \sqrt{U} \cdot\cos(\Phi_{\txz}) + \frac{\rho^{2}(1-\theta^{2})}{1-\rho^{2}} U - \theta^{2} \geq 0 , \calF_{\rho}(\bX,\bZ) \right\} \\
&= \pr\left\{\left(\cos(\Phi_{\txz}) + \frac{\rho  (1-\theta^{2})}{\sqrt{1-\rho^{2}}}\sqrt{U} \right)^{2} 
+ \frac{\rho^{2}\theta^{2}(1-\theta^{2})}{1-\rho^{2}} U - \theta^{2} \geq 0 , \calF_{\rho}(\bX,\bZ) \right\} \\
&= \pr\left\{\left(\cos(\Phi_{\txz}) + \frac{\rho  (1-\theta^{2})}{\sqrt{1-\rho^{2}}}\sqrt{U} \right)^{2} \geq \theta^{2} \left[1 - \frac{\rho^{2}(1-\theta^{2})}{1-\rho^{2}} U\right], \calF_{\rho}(\bX,\bZ) \right\}.
\end{align}
Note that $\rho\|\bX\|^{2} + \sqrt{1-\rho^{2}}\bX^{\mathsf{T}}\bZ \geq 0$ is equivalent to 
\begin{align}
\cos(\Phi_{\txz}) \geq -\frac{\rho}{\sqrt{1-\rho^{2}}}\sqrt{U},
\end{align} 
so,
\begin{align}
&\pr\left\{\tilde{\bX}^{\mathsf{T}} \tilde{\bY} \geq \theta \right\} \nn \\
\label{ref43}
&= \pr\left\{\left(\cos(\Phi_{\txz}) + \frac{\rho  (1-\theta^{2})}{\sqrt{1-\rho^{2}}}\sqrt{U} \right)^{2} \geq \theta^{2} \left[1 - \frac{\rho^{2}(1-\theta^{2})}{1-\rho^{2}} U\right], \cos(\Phi_{\txz}) \geq -\frac{\rho}{\sqrt{1-\rho^{2}}}\sqrt{U}\right\}.
\end{align}
Now, the left-hand-side event in \eqref{ref43} becomes trivial if and only if 
\begin{align}
1 - \frac{\rho^{2}(1-\theta^{2})}{1-\rho^{2}} U \leq 0 
~\iff~
U \geq \frac{1-\rho^2}{\rho^2(1-\theta^2)} \dfn \beta(\rho,\theta),
\end{align}
while the right-hand-side event in \eqref{ref43} becomes trivial if and only if
\begin{align}
\frac{\rho}{\sqrt{1-\rho^{2}}} \sqrt{U} \geq 1 
~\iff~
U \geq \frac{1-\rho^2}{\rho^2} \dfn \alpha(\rho).
\end{align}
Define the functions
\begin{align}
F_{1}(u,\rho,\theta) &= -\frac{\rho  (1-\theta^{2})}{\sqrt{1-\rho^{2}}}\sqrt{u} - \theta\sqrt{1 - \frac{\rho^{2}(1-\theta^{2})}{1-\rho^{2}} u} \\
F_{2}(u,\rho,\theta) &= -\frac{\rho  (1-\theta^{2})}{\sqrt{1-\rho^{2}}}\sqrt{u} + \theta\sqrt{1 - \frac{\rho^{2}(1-\theta^{2})}{1-\rho^{2}} u} \\
F_{3}(u,\rho) &= - \frac{\rho }{\sqrt{1-\rho^{2}}}\sqrt{u},  
\end{align}
such that
\begin{align}
\pr\left\{\tilde{\bX}^{\mathsf{T}} \tilde{\bY} \geq \theta \right\} 
= \pr\left\{\cos(\Phi_{\txz}) \in [-1,F_{1}(U,\rho,\theta)]\cup [F_{2}(U,\rho,\theta),1], \cos(\Phi_{\txz}) \in [F_{3}(U,\rho),1] \right\}.
\end{align}

\begin{figure}[h!]
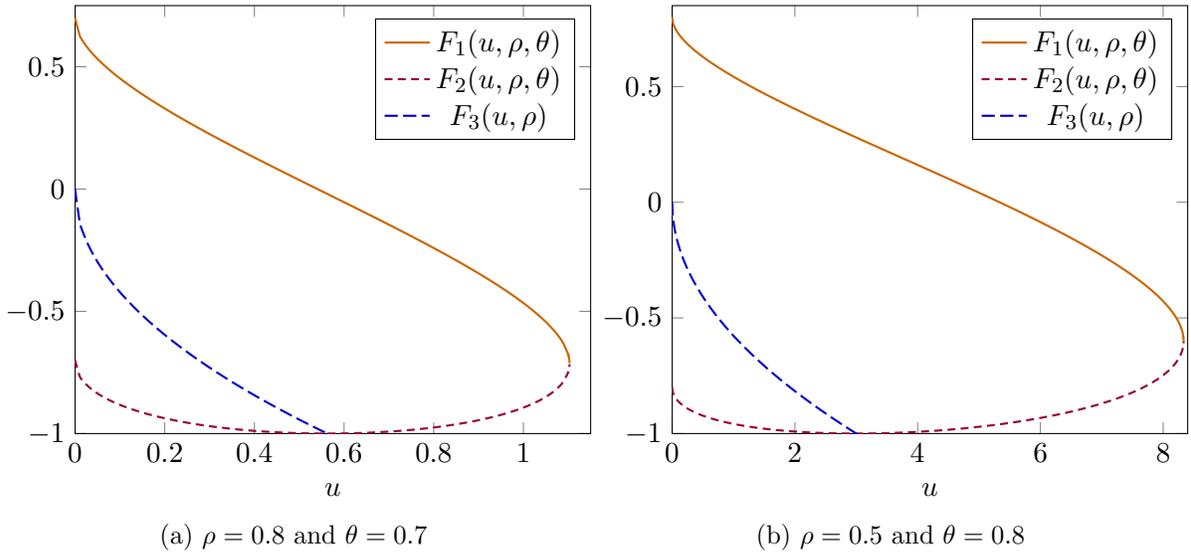
	
	\begin{subfigure}[b]{0.5\columnwidth}
		\centering 

		\caption{$\rho=0.8$ and $\theta=0.7$} 
		\label{fig:Bound2}
	\end{subfigure}%
	\begin{subfigure}[b]{0.5\columnwidth}
		\centering 
	%
		\caption{$\rho=0.5$ and $\theta=0.8$} 
		\label{fig:Bound1}
	\end{subfigure}
	\caption{Plots of $F_{1}(u,\rho,\theta)$, $F_{2}(u,\rho,\theta)$, and $F_{3}(u,\rho)$.}\label{fig:Boundaries}
\end{figure}

Note that both $\|\bX\|^{2}$ and $\|\bZ\|^{2}$ follow a $\chi^{2}$ distribution with $d$ degrees of freedom, hence, the random variable given by the ratio 
\begin{align}
U = \frac{\|\bX\|^{2}}{\|\bZ\|^{2}}
\end{align}   
will have the $F$-distribution with $(d,d)$ degrees of freedom. The PDF of a general $F$-distributed random variable with $(d_{1},d_{2})$ degrees of freedom is given by
\begin{align}
f_{U}(u;d_{1},d_{2}) = \frac{1}{B\left(\frac{d_{1}}{2},\frac{d_{2}}{2}\right)} \left(\frac{d_{1}}{d_{2}}\right)^{d_{1}/2} u^{d_{1}/2-1}\left(1+\frac{d_{1}}{d_{2}}u\right)^{-(d_{1}+d_{2})/2},
\end{align}
and for $d_{1}=d_{2}=d$, we denote it by
\begin{align}
f_{U}(u) = \frac{1}{B\left(\frac{d}{2},\frac{d}{2}\right)}  u^{d/2-1}\left(1+u\right)^{-d}.
\end{align}
Now,
\begin{align}
&\pr\left\{\tilde{\bX}^{\mathsf{T}} \tilde{\bY} \geq \theta \right\} \nn \\
&= \int_{0}^{\infty} \pr\left\{\cos(\Phi_{\txz}) \in [-1,F_{1}(u)]\cup [F_{2}(u),1], \cos(\Phi_{\txz}) \in [F_{3}(u),1] \right\} f_{U}(u) \dint u \\
\label{ref44}
&= \int_{0}^{\alpha(\rho)} \pr\left\{\cos(\Phi_{\txz}) \in [F_{2}(u),1] \right\} f_{U}(u) \dint u \nn \\ 
&~~~+ \int_{\alpha(\rho)}^{\beta(\rho,\theta)} \pr\left\{\cos(\Phi_{\txz}) \in [-1,F_{1}(u)]\cup [F_{2}(u),1] \right\} f_{U}(u) \dint u + \int_{\beta(\rho,\theta)}^{\infty} f_{U}(u) \dint u. 
\end{align}
In order to derive the first two integrals in \eqref{ref44}, consider the following result.
\begin{lemma}
	Let $d \in \naturals$ and $\bX,\bY \sim \calN(\boldsymbol{0}_{d},\bI_{d})$ be two independent random vectors. Let $\tilde{\bX} = \tfrac{\bX}{\|\bX\|},~\tilde{\bY} = \tfrac{\bY}{\|\bY\|}$ and define $S=\cos(\Phi) = \tilde{\bX}^{\tr}\tilde{\bY}$. Then,
	\begin{align}
	g_{S}(s) = \frac{(1-s^{2})^{\frac{d-3}{2}}}{B(\frac{d-1}{2},\frac{1}{2})},
	\end{align} 
	where the complete beta function 
	\begin{align}
	B(a,b) = \int_{0}^{1} t^{a-1}(1-t)^{b-1}\dint t.
	\end{align}
\end{lemma}
Then, finally,
\begin{align}
&\pr\left\{\tilde{\bX}^{\mathsf{T}} \tilde{\bY} \geq \theta \right\} \nn \\
&= \int_{0}^{\alpha(\rho)} \left[\int_{F_{2}(u)}^{1} g_{S}(s) \dint s \right] f_{U}(u) \dint u \nn \\ 
&~~~+ \int_{\alpha(\rho)}^{\beta(\rho,\theta)} \left[\int_{-1}^{F_{1}(u)} g_{S}(s) \dint s + \int_{F_{2}(u)}^{1} g_{S}(s) \dint s\right] f_{U}(u) \dint u + \int_{\beta(\rho,\theta)}^{\infty} f_{U}(u) \dint u,
\end{align}
which completes the proof of Lemma \ref{Lemma_P_Bound}.


\end{document}